\documentclass[a4paper, 11pt]{article}
\usepackage[left=2cm, right=1cm, top=0.5cm, bottom=1cm, includeheadfoot]{geometry}
\usepackage[english]{babel}
\usepackage[utf8]{inputenc}
\usepackage[authoryear,round,longnamesfirst]{natbib}
\usepackage{framed}
\usepackage{amsmath,amsfonts,amssymb,mathtools,bbm}
\usepackage{booktabs, multirow}
\usepackage{adjustbox}
\usepackage{float}
\usepackage{listings}
\usepackage{color}
\usepackage{url}
\usepackage{hyperref}
\usepackage{xpatch}
\usepackage{xcolor}
\usepackage{listings}
\usepackage{realboxes}
\usepackage{marginnote}
\usepackage{changepage}
\usepackage{fancyhdr}

\newcommand{\orcid}[1]{\href{https://orcid.org/#1}{\includegraphics[height=\fontcharht\font`A]{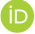}}}

\hypersetup{
	colorlinks=true,
	linkcolor=blue,
	citecolor=magenta,   
	urlcolor=cyan,
	pdfpagemode=FullScreen,
}

\let\pkg\textbf

\DeclareMathOperator{\EX}{\mathbb{E}}
\definecolor{applegreen}{rgb}{0.55, 0.71, 0.0}
\definecolor{darkolivegreen}{rgb}{0.33, 0.42, 0.18}
\definecolor{codecolors}{RGB}{ 219,225,226}
\definecolor{codecolorsinline}{RGB}{ 230.1000,  234.3000 , 235.0000}

\newcommand{\consolein}{\noindent\par\reversemarginpar\marginnote{\noindent\textcolor{applegreen}{In [~]}:}[0.2cm]}
\newcommand{\consoleout}{\noindent\par\reversemarginpar\marginnote{\noindent\textcolor{gray}{Out [~]}:}[0.2cm]}

\lstdefinestyle{codeinput}{
	backgroundcolor = \color{codecolors},
	language = python,
	mathescape=true,
	keywordstyle=\color{blue}\bf,
	commentstyle=\color{darkolivegreen},
	stringstyle=\color{darkolivegreen},
	emphstyle=\color{magenta},	
	basicstyle=\footnotesize\ttfamily,
	frame=single
}

\lstdefinestyle{codeinputinline}{
	backgroundcolor = \color{codecolorsinline},
	language = python,
	mathescape=true,
	keywordstyle=\color{blue}\bf,
	commentstyle=\color{green},
	stringstyle=\color{darkolivegreen},
	emphstyle=\color{magenta},	
	basicstyle=\normalsize\ttfamily,
}

\lstdefinestyle{codeoutput}{
	backgroundcolor = \color{codecolors},
	basicstyle=\footnotesize\ttfamily,
	 mathescape=true,
	 frame=single,
}

\newcommand{\code}{\lstinline[style=codeinputinline]}
\makeatletter
\xpretocmd\lstinline{\Colorbox{codecolorsinline}\bgroup\appto\lst@DeInit{\egroup}}{}{}
\makeatother

\newcommand{\class}[1]{`\code{#1}'}

\usepackage{pgfplots}
\usepackage{tikz}
\usetikzlibrary{angles, quotes, math, shapes, snakes, arrows.meta, decorations.pathreplacing, fit}
\usepgflibrary{arrows}
\definecolor{mycolor}{RGB}{202,230,223}
\definecolor{mycolor3}{RGB}{10,103,79}
\definecolor{mycolor4}{RGB}{193,17,17}
\definecolor{mycolor2}{RGB}{171,131,174}
\definecolor{y_box}{RGB}{ 171,131,174}
\definecolor{g_box}{RGB}{56,168,83}
\definecolor{gray_box}{rgb}{0.85, 0.85, 0.85}
\tikzset{
	my_latent/.style={circle, draw=mycolor3, thick, fill=mycolor},
	my_obs/.style={rectangle, draw=mycolor3, thick, fill=mycolor},
	my_obs_exo/.style={regular polygon, regular polygon sides=3,inner sep=0pt,text width=3mm, draw = mycolor4, thick, fill=mycolor},
	my_manif/.style={},
	my_arrow/.style={-{Latex[length=2mm]}, mycolor3, thick},
	my_arrow_mp/.style={-{Latex[length=1.5mm]}, shorten <=2pt, black},
	my_covariance/.style={{Latex[length=2mm]}-{Latex[length=2mm]}, 
		mycolor2, thick},
	my_covariance_mp/.style={{Latex[length=1.5mm]}-{Latex[length=1.5mm]}, 
		mycolor2, thick},
	style1/.style={color=blue,
		mark=square*,
		mark size=4pt,
	},
	style2/.style={color=green,
		mark=*,
		mark size=4pt,
	},
	style3/.style={color=orange,
		mark=triangle*,
		mark size=4pt,
	}
}

\title{\textbf{semopy 2}: A Structural Equation Modeling Package with Random Effects in Python}
\author{
	\begin{tabular}{c}
		Georgy Meshcheryakov \orcid{0000-0003-0751-8286} \\ \href{mailto:iam@georgy.top}{\texttt{iam@georgy.top}}
	\end{tabular}\\
	\begin{tabular}{cc}
	Anna A. Igolkina \orcid{0000-0001-8851-9621} &  Maria G. Samsonova \\
	\href{mailto:igolkinaanna11@gmail.com}{\texttt{igolkinaanna11@gmail.com}} & \href{mailto:m.samsonova@spbstu.ru}{\texttt{m.samsonova@spbstu.ru}}
 	\end{tabular}
}
\begin{document}
\date{June 1, 2021}
\maketitle
\thispagestyle{empty}
\begin{abstract}
Structural Equation Modeling (SEM) is an umbrella term that includes numerous multivariate statistical techniques that are employed throughout a plethora of research areas, ranging from social to natural sciences. Until recently, SEM software was either commercial or restricted to niche languages, and the lack of SEM packages compatible with more mainstream programming languages was dire. To combat that, we introduced a \textit{Python} package \textbf{semopy v1} that surpassed other state-of-the-art software in terms of performance and estimation accuracy. Yet, it was lacking in functionality and its usage was burdened with unnecessary boilerplate code. Here, we introduce a complete overhaul of \textbf{semopy} that improves upon the previous results and comes with lots of new capabilities. Furthermore, we propose a novel SEM model that combines in itself a notion of random effects from linear mixed models (LMMs) to model numerous phenomena, such as spatial data, time series or population stratification in genetics.
\end{abstract}

\section{Introduction}\label{sec:intro}
Over the past years, structural equation modelling software has grown in number, but neither of the numerous SEM tools has been developed natively in any of the mainstream languages, with \texttt{R} being the most common choice. Notable examples are \texttt{R} packages \textbf{lavaan} \citep{Rosseel:2012}, \textbf{openMX} \citep{Neale:2016}, \textbf{sem} \citep{Fox:2016}, with the former being the most popular one. According to PYPL \citep{Carbonelle:2021} and TIOBE \citep{Tiobe:2021} indices, \texttt{Python} is easily among the most popular languages, whereas \texttt{R} popularity is stagnating at best. We believe that the lack of SEM packages in \texttt{Python} prevents or makes it harder for some researchers to contribute to the area of SEM. 

Previously, we created the first \texttt{Python} package for SEM \textbf{semopy} that outperformed the most popular package \textbf{lavaan} in terms of performance and estimates accuracy \citep{Igolkina:2020}. However, at the time, we didn't anticipate significant interest in \textbf{semopy}, and the first versions of the package were more of an ad-hoc solution for our own specific research interests in bioinformatics, namely, in applying SEM to genome-wide association studies \citep{IgolkinaMesh:2020} than a package that is designated to a wider audience. We were proved wrong by vast unexpected feedback, and the presence of a public desire for a better and easier to use \texttt{Python} package became clear. In \textbf{semopy 2}, we aim to bring together the most requested functionalities that were lacking previously, while maintaining an easy-to-use-and-extend codebase.

Also, we introduce a notion of random effects to SEM that comes from the area of linear mixed models (LMMs). Random effects are a powerful technique that helps to model different types of dependencies in data. For example, those dependencies can be caused by observations being separated into groups (e.g. pupils polled from different schools), time series, spatial data and genetic kinship. Although some of the cases listed are studied and implemented as separate SEM techniques, such as latent curve analysis \citep{Meredith:1990} or spatial SEM \citep{Liu:2005}, no one has provided a generalized framework or an easy-to-use solution. Here, we propose a novel generalized SEM model that can be used to take into account any of the above phenomena.

This article aims to serve as a tutorial on package \textbf{semopy} with a strong emphasis on algebra behind its methods. The latest version \textbf{semopy} is available at PyPi repository (\url{https://pypi.org/projects/semopy}). A comprehensive guide on its usage can be found at \textbf{semopy} website (\url{https://semopy.com}), albeit without rigorous mathematical details. The rest part of the article is structured as follows: first, to better highlight the power of SEM technique and some strengths of the \textbf{semopy} package, we provide a short overview of the SEM history; second, we explain in detail four mathematical models behind the package and their use cases; third, we describe secondary features of \textbf{semopy}; fourth, we provide results of numerical experiments that showcase that the 2.0+ version of the package improves upon results of the previous 1.3.1 version in Section~\ref{sect:experiments}. Finally, a reader can find some tedious derivations of formulae in appendices, alongside tabular data of numerical experiments.
\section{Briefly on SEM history}
\begin{figure}[t!]
	\label{fig:pa}
	\centering
	\begin{tikzpicture}
	\begin{scope}[transform shape]

		\node[my_obs] (x1) at (0,0) {$x_1$};
		\node[my_obs] (x2) at (1,0) {$x_2$};
		\node[my_obs] (x3) at (2,0) {$x_3$};
		\node[my_obs] (x4) at (3,1) {$x_4$};
		\node[my_obs] (x5) at (3,-1) {$x_5$};
		\node[my_obs] (x6) at (4, 1) {$x_6$};
		\node[my_obs] (x7) at (4,-1) {$x_7$};

		\draw[my_arrow] (x1) -- (x2);
		\draw[my_arrow] (x2) -- (x3);
		\draw[my_arrow] (x3) -- (x4);
		\draw[my_arrow] (x3) -- (x5);
		\draw[my_arrow] (x4) -- (x6);
		\draw[my_arrow] (x5) -- (x7);
		
	\end{scope}
\end{tikzpicture}
	\vspace{-1em}
	\begin{tabular}{rlrlrl}
			\begin{tikzpicture}
				\node[my_obs] (x) at (0,0) {$x$};
			\end{tikzpicture} & -- observed variable;
		&
      \lapbox[\width]{1em}{\begin{tikzpicture}
		\node (a1) at (0,0) {};
		\node (a2) at (0.8,0) {};
		\draw[my_arrow] (a1) -- (a2);
	\end{tikzpicture}} & -- loading;
\end{tabular}
	\caption{PA model example; $x_k$ are observed variables whereas arrows are regressions. To fit PA model, or SEM model, to data, is to find regression coefficients on those arrows, and, possibly, variances of $x_k$.}
\end{figure}
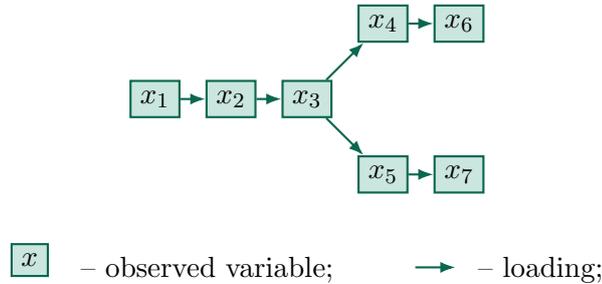

SEM is not a new field and its roots date back to works of geneticist Sewall Green Wright and his technique of path analysis (PA) \citep{Wright:1921}. PA is best explained by Equation~\ref{eq:pa}:

\begin{equation}\label{eq:pa}
	x = B x + \epsilon,
\end{equation}
where $x$ is a $n_{x} \times 1$ vector of observed variables, $\epsilon$ is a random vector of errors and $B$ is a parameterized loadings matrix. Notice that Equation~\ref{eq:pa} can be reformulated as a special case of a linear regression model by inferring $x$: 
$$x = \underbrace{(I_{n_{x}} - B)^{-1}}_{C} \epsilon = C \epsilon = \widehat{\epsilon}$$
If we further assume that $\epsilon \sim \mathcal{N}(0, \Psi)$, where $\Psi$ is a covariance matrix of shape $n_{x} \times n_{x}$, then $x \sim \mathcal{N}(C \mu, C \Psi C^T)$. For brevity, we also assume that data is centred. Usually, $\Psi$ and $B$/$C$ are unknown and estimated via maximum likelihood scheme, although at the times of Wright such a computational technique was unfeasible and was not considered in his original works \citep{Wright:1934}. Also, often some kind of structure is imposed on $\Psi$. In the context of PA, it is natural to assume it to be diagonal. An example of the PA model can be seen in Figure~\ref{fig:pa}.

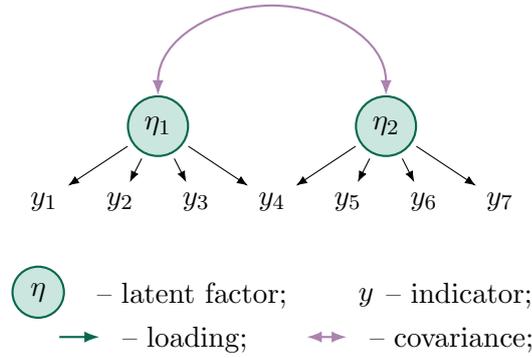
\begin{figure}[t!]
	\label{fig:fa}
	\centering
	\begin{tikzpicture}
	\begin{scope}[transform shape]
		\node[my_latent] (eta1) at (0.5,0) {$\eta_1$};
		\node[my_manif] (y1) at (-1, -1) {$y_1$};
		\node[my_manif] (y2) at (0, -1) {$y_2$};
		\node[my_manif] (y3) at (1, -1) {$y_3$};
		\node[my_manif] (y4) at (2, -1) {$y_4$};
		\node[my_manif] (y5) at (3, -1) {$y_5$};
		\node[my_manif] (y6) at (4, -1) {$y_6$};
		\node[my_manif] (y7) at (5, -1) {$y_7$};
		\node[my_latent] (eta2) at (3.5,0) {$\eta_2$};
		
		\draw[my_arrow_mp] (eta1) -- (y1);
		\draw[my_arrow_mp] (eta1) -- (y2);
		\draw[my_arrow_mp] (eta1) -- (y3);
		\draw[my_arrow_mp] (eta1) -- (y4);
		\draw[my_arrow_mp] (eta2) -- (y4);
		\draw[my_arrow_mp] (eta2) -- (y5);
		\draw[my_arrow_mp] (eta2) -- (y6);
		\draw[my_arrow_mp] (eta2) -- (y7);
		
			\draw[my_covariance] (eta1) to [out= 90, in=180] (2, 1.6) to [out=0, in= 90] (eta2);
	\end{scope}
\end{tikzpicture}
	\vspace{-1em}
	\begin{tabular}{rlrlrl}
		\raisebox{-0.3\height}{\begin{tikzpicture}
			\node[my_latent] (eta) at (0,0) {$\eta$};
		\end{tikzpicture}} & -- latent factor;
	   &
	   \lapbox[\width]{1em}{\raisebox{-0.35\height}{\begin{tikzpicture}
			\node[my_manif] (y) at (0,0) {$y$};
		\end{tikzpicture}}} & -- indicator;
	\end{tabular}
\\
\begin{tabular}{rlrl}
      \lapbox[\width]{1em}{\begin{tikzpicture}
		\node (a1) at (0,0) {};
		\node (a2) at (0.8,0) {};
		\draw[my_arrow] (a1) -- (a2);
	\end{tikzpicture}} & -- loading;
 \lapbox[\width]{1em}{\begin{tikzpicture}
	\node (a1) at (0,0) {};
	\node (a2) at (0.8,0) {};
	\draw[my_covariance] (a1) -- (a2);
\end{tikzpicture}} & -- covariance;
\end{tabular}
	\caption{CFA model example; $y_k$ are indicators/observed variables, $\eta_k$ are latent factors, black arrows are regressions and bidirectional arrows are parameterized covariances. }
\end{figure}

Some phenomena can not be observed, or, at least, were not at the time of data collection, yet the researcher seeks to incorporate its effects into a model via the introduction of a latent variable (or, in other terms, a factor). The first task is to confirm if there is an underlying latent factor in the first place, as designed by a researcher. The most common approach is to fit a linear model with latent factors in it defined explicitly as in Equation~\ref{eq:fa}
\begin{equation}\label{eq:fa}
	y = \Lambda \eta + \delta,
\end{equation}
where $y$ is an $n_y \times 1$ vector of \textit{indicators}/observed variables (also known as \textit{manifest} variables), $\delta$ is a random vector of errors, $\eta$ is a $n_\eta \times 1$ vector of latent factors and $\Lambda$ is a parameterized loading of factors onto indicators matrix. Per usual, it is assumed that $\delta \sim \mathcal{N}(0, \Theta)$ and $\eta \sim \mathcal{N}(0, \Psi)$, where $\Theta$ is usually restricted to diagonal matrix, however extra non-diagonal covariances can be introduced at researcher's disposal. This approach, named confirmatory factor analysis (CFA), is due to \cite{Joreskog:1967}. An example of the CFA model can be seen in Figure~\ref{fig:fa}. Please notice, that despite the fact indicator and observed variables are effectively synonymous, at the moment we are using different notation for them both in formulae and on figures. The reason for this is historic and will be discussed in Section~\ref{sect:model}.

The model from Equation~\ref{eq:fa} is not identifiable, though. This comes from an observation that we can imagine latent variables of an arbitrary magnitude that will still fit the data exactly the same. Indeed, we can write down
$$y = \Lambda \eta + \delta = \Lambda \frac{1}{\alpha} \alpha \eta + \delta = \left(\frac{1}{\alpha} \Lambda \right) (\alpha \eta) + \delta = \widehat{\Lambda} \widehat{\eta} + \delta,$$
where $\alpha$ is an arbitrary scalar. This issue is dealt with by fixing some of the loading or variance parameters to a certain value; the most common approach (and the one followed by \textbf{semopy}) is to fix a first loading for each of the latent variables to $1.0$ (for example, at Figure~\ref{fig:fa} two loadings are fixed: the loading between $y_1$ and $\eta_1$ and between $y_4$ and $\eta_2$).

CFA models, however, do not allow for casual interactions between observed variables or latent factors. SEM, on the other hand, can model arbitrary interactions between variables. Although there are no strict definition of SEM and it is rather a broad term, here we shall assume that by "SEM model" we mean a hybrid of PA and CFA, with \textbf{LISREL} \citep{Joreskog:1972} being the notable example and one of pioneers of SEM. Following the idea of SEM = PA + CFA, we get the Equation~\ref{eq:sem}:

\begin{equation}\label{eq:sem}
	\begin{cases}
		\begin{bmatrix}
			\eta \\ x
		\end{bmatrix} = \omega = B \omega + \epsilon, ~~~\epsilon \sim \mathcal{N}(0, \Psi) \\
		\begin{bmatrix}y \\ x \end{bmatrix} = z = \Lambda \omega + \delta, ~~~\delta \sim \mathcal{N}(0, \Theta)
	\end{cases},
\end{equation}

where $\eta, x, y$ are the same vectors from Equation~\ref{eq:pa} and Equation~\ref{eq:fa}, $w$ is a $n_\omega \times 1$ (where $n_\omega = n_\eta + n_x$) vector of latent variables $\eta$ and observed variables $x$, $z$ is a $n_z \times 1$ (where $n_z = n_y + n_x$) vector of all observed variables (including indicators $y$), $\epsilon$  and $\delta$ are random vectors of shapes $n_\omega \times 1$ and $n_z \times 1$ respectively. The vector $x$ is separated into subvectors $x^{(1)}, x^{(2)}$ where $x^{(1)}$ is a vector of endogenous $x_i$ and $x^{(2)}$ is a vector of exogenous $x_i$. Covariances between $x^{(2)}$ in $\Psi$ are fixed to their sample values. This model is also used by \textbf{lavaan}, for instance.

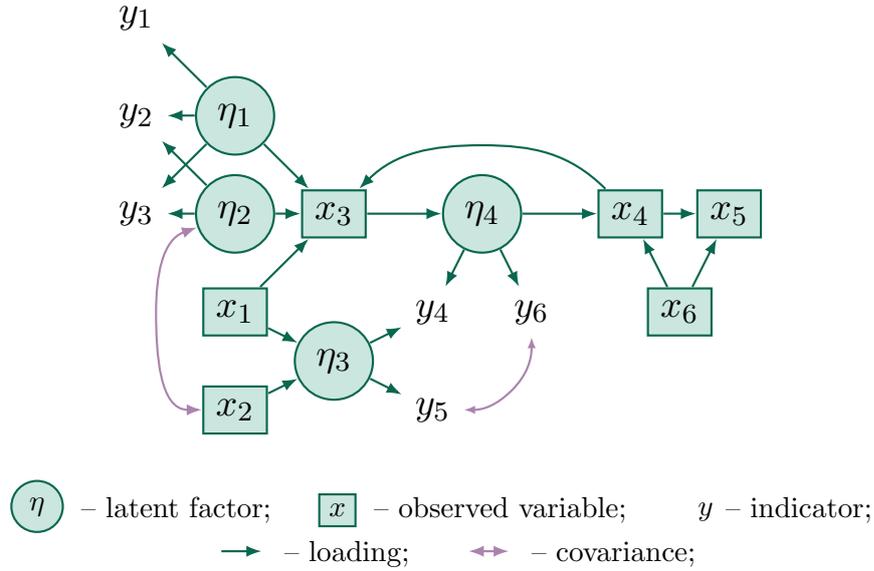
\begin{figure}[t!]
	\label{fig:sem}
	\centering
	\begin{tikzpicture}
	\begin{scope}[scale=1.3, transform shape]
		
		\node[my_manif] (y1) at (1,10) {$y_1$};
		\node[my_manif] (y2) at (1,9) {$y_2$};
		\node[my_manif] (y3) at (1,8) {$y_3$};
		\node[my_manif] (y4) at (4,7) {$y_4$};
		\node[my_manif] (y5) at (4,6) {$y_5$};
		\node[my_manif] (y6) at (5,7) {$y_6$};
		
		\node[my_obs] (x1) at (2,7) {$x_1$};
		\node[my_obs] (x2) at (2,6) {$x_2$};
		\node[my_obs] (x3) at (3,8) {$x_3$};
		\node[my_obs] (x4) at (6,8) {$x_4$};
		\node[my_obs] (x5) at (7,8) {$x_5$};
		\node[my_obs] (x6) at (6.5,7) {$x_6$};

		\node[my_latent] (eta1) at (2,9) {$\eta_1$};
		\node[my_latent] (eta2) at (2,8) {$\eta_2$};
		\node[my_latent] (eta3) at (3,6.5) {$\eta_3$};
		\node[my_latent] (eta4) at (4.5, 8) {$\eta_4$};
		
		\draw[my_arrow] (x6) -- (x5);
		\draw[my_arrow] (x6) -- (x4);
		\draw[my_arrow] (eta1) -- (x3) ; 
		\draw[my_arrow] (eta2) -- (x3); 
		\draw[my_arrow] (x1) -- (x3); 
		\draw[my_arrow] (x1) -- (eta3); 
		\draw[my_arrow] (x2) -- (eta3) ; 
		\draw[my_arrow] (x3) -- (eta4) ; 
		\draw[my_arrow] (eta4) -- (x4); 
		\draw[my_arrow] (x4) -- (x5) ; 
		\draw[my_arrow] (x4) to [out=135, in=0] (4.5, 8.7) to [out=180, in=45] (x3); 
		\draw[my_covariance] (eta2) to [out= 200, in=90] (1.2, 7) to [out=-90, in= 180] (x2);
		\draw[my_covariance_mp] (y5) to [out= 0, in=225] (4.8, 6.2) to [out=45, in= -90]   (y6);
		
		\draw[my_arrow] (eta1) -- (y1) ; 
		\draw[my_arrow] (eta1) -- (y2); 
		\draw[my_arrow] (eta1) -- (y3); 
		\draw[my_arrow] (eta2) -- (y3) ; 
		\draw[my_arrow] (eta2) -- (y2);
		\draw[my_arrow] (eta3) -- (y4) ;
		\draw[my_arrow] (eta3) -- (y5) ; 
		\draw[my_arrow] (eta4) -- (y4);
		\draw[my_arrow] (eta4) -- (y6);
		
	\end{scope}
\end{tikzpicture}
	\vspace{-1em}
	\begin{tabular}{rlrlrlrl}
		\lapbox[\width]{0.5em}{\raisebox{-0.3\height}{\begin{tikzpicture}
			\node[my_latent] (eta) at (0,0) {$\eta$};
	\end{tikzpicture}}} & -- latent factor;
	&
	\lapbox[\width]{0.5em}{\raisebox{-0.3\height}{\begin{tikzpicture}
				\node[my_obs] (x) at (0,0) {$x$};
	\end{tikzpicture}}} & -- observed variable;
	&
	\lapbox[\width]{1em}{\raisebox{-0.35\height}{\begin{tikzpicture}
				\node[my_manif] (y) at (0,0) {$y$};
	\end{tikzpicture}}} & -- indicator;

\end{tabular}
\\
\begin{tabular}{rlrl}
	\lapbox[\width]{1em}{\begin{tikzpicture}
			\node (a1) at (0,0) {};
			\node (a2) at (0.8,0) {};
			\draw[my_arrow] (a1) -- (a2);
	\end{tikzpicture}} & -- loading;
	\lapbox[\width]{1em}{\begin{tikzpicture}
			\node (a1) at (0,0) {};
			\node (a2) at (0.8,0) {};
			\draw[my_covariance] (a1) -- (a2);
	\end{tikzpicture}} & -- covariance;
\end{tabular}
	\caption{SEM model example; $y_k$ are indicators/observed variables, $\eta_k$ are latent factors, $x_k$ are observed variables (but not indicators) unidirectional arrows are regressions and bidirectional arrows are parameterized covariances. }
\end{figure}

One can infer covariance matrix $\Sigma$ of shape $n_z \times n_z$ from Equation~\ref{eq:sem}:
\begin{equation}\label{eq:sigma}
	\Sigma = \EX[z z^T] = \Lambda C \Psi C^T \Lambda^T + \Theta,
\end{equation}
and then fit to a sample covariance matrix $S$ using either maximum likelihood (ML) or weighted least squares (WLS) \citep{Hoyle:2015}.

\section[semopy models]{\textbf{semopy} models}

\textbf{semopy} cornerstones are \textit{model} classes.  At the moment, there are 4 models present: \class{Model}, \class{ModelMeans}, \class{ModelEffects} and \\ \class{ModelGeneralizedEffects}. Each of them treats data and estimation differently, has it is own distinct set of hyperparameters, but they all share the core interface, specifically, they all posses key methods \code{fit} (fits model to data), \code{inspect} (produces a dataframe with parameters estimates) , \code{predict} (predicts observed variables values from given data), \\ \code{predict_factors} (estimates factor scores/latent variables values). Next, we describe each of the 4 models in detail.

\subsection[Model]{Conventional SEM: \class{Model}}\label{sect:model}
This model is closest to the classical SEM and is almost the same for an end user as one present in Equation~\ref{eq:sem}. However, there are some technical differences that result in a better performance and a possibility to define a broader class of models. The proposed model is:

\begin{equation}\label{eq:model}
	\begin{cases}
		\begin{bmatrix}
			\eta \\ x
		\end{bmatrix} = \omega = B \omega + \epsilon, ~~~\epsilon \sim \mathcal{N}(0, \Psi) \\
		\begin{bmatrix}
			y \\ x^{(1)}
		\end{bmatrix} = z = \Lambda \omega + \delta, ~~~\delta \sim \mathcal{N}(0, \Theta)
	\end{cases},
\end{equation}
the biggest change here is that $y$ is not a vector of indicator variables, it is a vector of \textit{output} observed variables -- those that depend on other variables but do not cast any regression arrows to other variables. In fact, there is no concept of "indicators"/"manifest" variables in \textbf{semopy} unlike in other SEM software. We think that the sole reason such a separation of observed variables exists is due to historical leftovers from CFA. Although not crucial, having a class of indicator variables over plain output variables has three drawbacks:

\begin{enumerate}
	\item Creates confusion among new SEM researchers, as there is no apparent reason to move out variables that are loaded onto only by latent variables to a class of indicators;
	\item Once a variable is specified as an indicator, it is impossible to regress onto it by any observable variable;
	\item The missing opportunity to rule out output variables from $x$ vector result in a larger size of $C = (I - B)^{-1}$ matrix, and its inversion is numerically expensive.
\end{enumerate}

The other change in Equation~\ref{eq:model} is reduced size of $z$ as it now consists of $x^{(1)}$ instead of $x$. This change is natural as covariances of exogenous variables are usually not parameterized and hence, a part of $\Sigma$ that corresponds to exogenous variables is static and therefore not of interest. This results in further performance gains as in some objective functions $\Sigma$ has to be inverted.
\begin{leftbar}
	It may appear that with \class{Model} one can't define a covariance parameter between an exogenous variable and some other variable. Although the authors of \textbf{semopy} can't think of a reason to do that, it is still possible, but such exogenous variable is moved out of the class of "exogenous" variables $x^{(2)}$ to "endogenous" variables $x^{(1)}$. It should not affect an end-user experience other than a slightly longer estimation time.
\end{leftbar}

\begin{figure}[t!]
	\label{fig:model}
	\centering
	\begin{tikzpicture}
	\begin{scope}[scale=1.3, transform shape]
		
		\node[my_obs] (y1) at (1,10) {$y_1$};
		\node[my_obs] (y2) at (1,9) {$y_2$};
		\node[my_obs] (y3) at (1,8) {$y_3$};
		\node[my_obs] (y4) at (4,7) {$y_4$};
		\node[my_obs] (y5) at (4,6) {$y_{5}$};
		\node[my_obs] (y6) at (5,7) {$y_{6}$};
		
		\node[my_obs] (x1) at (2,7) {$x_1$};
		\node[my_obs] (x2) at (2,6) {$x_2$};
		\node[my_obs] (x3) at (3,8) {$x_3$};
		\node[my_obs] (x4) at (6,8) {$x_4$};
		\node[my_obs] (x5) at (7,8) {$y_7$};
		\node[my_obs] (x6) at (6.5,7) {$x_6$};

		\node[my_latent] (eta1) at (2,9) {$\eta_1$};
		\node[my_latent] (eta2) at (2,8) {$\eta_2$};
		\node[my_latent] (eta3) at (3,6.5) {$\eta_3$};
		\node[my_latent] (eta4) at (4.5, 8) {$\eta_4$};
		
		\draw[my_arrow] (x6) -- (x5);
		\draw[my_arrow] (x6) -- (x4);
		\draw[my_arrow] (eta1) -- (x3) ; 
		\draw[my_arrow] (eta2) -- (x3); 
		\draw[my_arrow] (x1) -- (x3); 
		\draw[my_arrow] (x1) -- (eta3); 
		\draw[my_arrow] (x2) -- (eta3) ; 
		\draw[my_arrow] (x3) -- (eta4) ; 
		\draw[my_arrow] (eta4) -- (x4); 
		\draw[my_arrow] (x4) -- (x5) ; 
		\draw[my_arrow] (x4) to [out=135, in=0] (4.5, 8.7) to [out=180, in=45] (x3); 
		\draw[my_covariance] (eta2) to [out= 200, in=90] (1.2, 7) to [out=-90, in= 180] (x2);
		\draw[my_covariance_mp] (y5) to [out= 0, in=225] (4.8, 6.2) to [out=45, in= -90]   (y6);
		
		\draw[my_arrow] (eta1) -- (y1) ; 
		\draw[my_arrow] (eta1) -- (y2); 
		\draw[my_arrow] (eta1) -- (y3); 
		\draw[my_arrow] (eta2) -- (y3) ; 
		\draw[my_arrow] (eta2) -- (y2);
		\draw[my_arrow] (eta3) -- (y4) ;
		\draw[my_arrow] (eta3) -- (y5) ; 
		\draw[my_arrow] (eta4) -- (y4);
		\draw[my_arrow] (eta4) -- (y6);
		
	\end{scope}
\end{tikzpicture}
	\vspace{-1em}
	\begin{tabular}{rlrlrlrl}
		\lapbox[\width]{0.5em}{\raisebox{-0.3\height}{\begin{tikzpicture}
			\node[my_latent] (eta) at (0,0) {$\eta$};
	\end{tikzpicture}}} & -- latent factor;
	&
	\lapbox[\width]{0.5em}{\raisebox{-0.3\height}{\begin{tikzpicture}
				\node[my_obs] (x) at (0,0) {$x$};
	\end{tikzpicture}}} & -- observed variable;

\end{tabular}
\\
\begin{tabular}{rlrl}
	\lapbox[\width]{1em}{\begin{tikzpicture}
			\node (a1) at (0,0) {};
			\node (a2) at (0.8,0) {};
			\draw[my_arrow] (a1) -- (a2);
	\end{tikzpicture}} & -- loading;
	\lapbox[\width]{1em}{\begin{tikzpicture}
			\node (a1) at (0,0) {};
			\node (a2) at (0.8,0) {};
			\draw[my_covariance] (a1) -- (a2);
	\end{tikzpicture}} & -- covariance;
\end{tabular}
	\caption{\class{Model} representation of the same SEM model presented previously. Notice, that there is no disambiguation between $y_k$ and $x_k$ now. }
\end{figure}
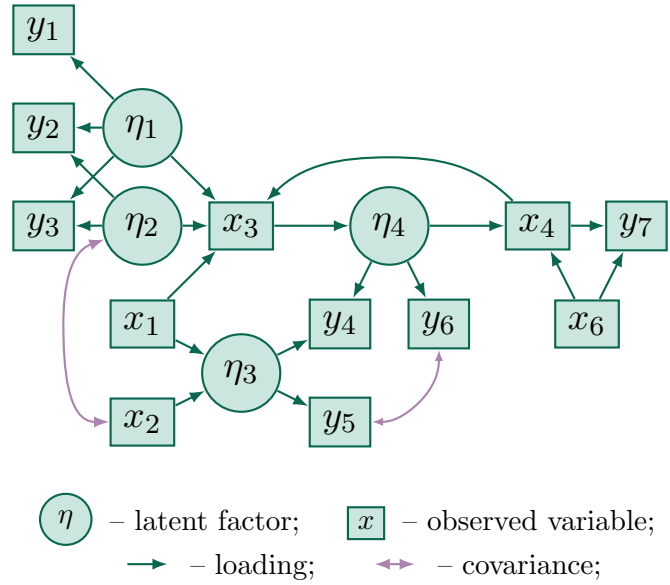

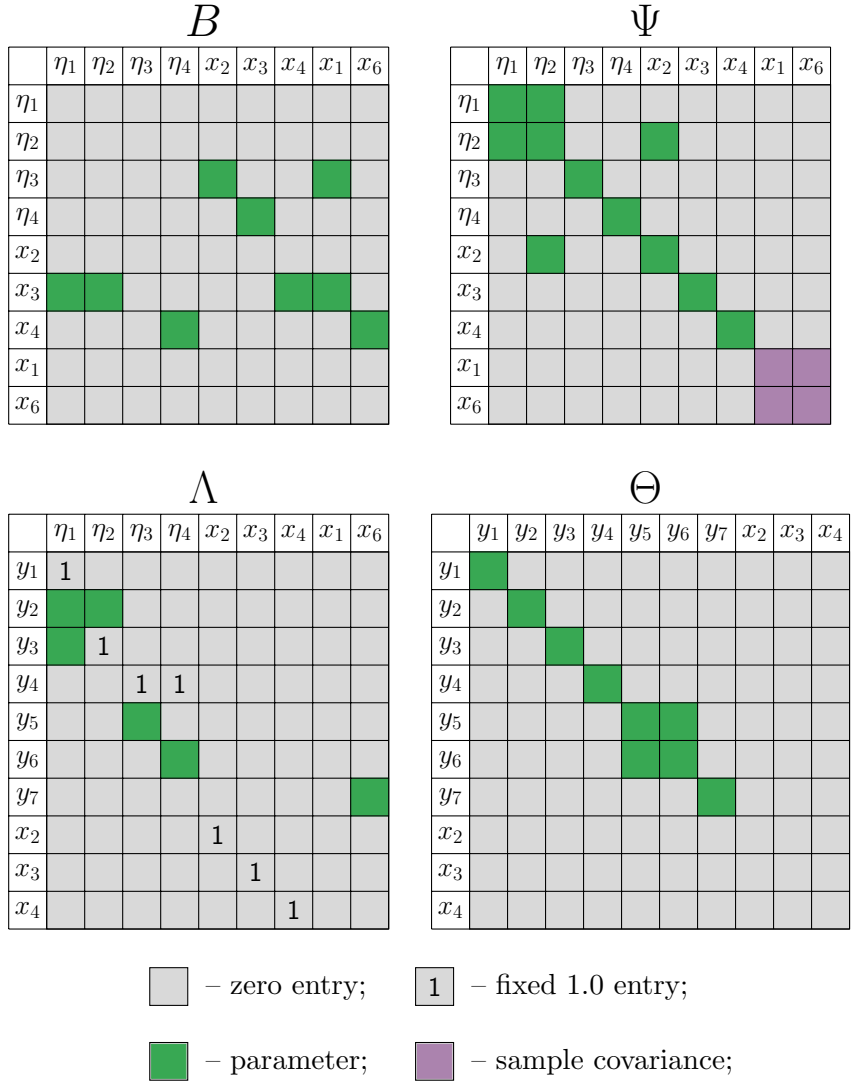
\begin{figure}[t!]
	\label{fig:model_matrices}
	\centering
	\begin{tabular}{cc}
		{\LARGE$B$} & {\LARGE$\Psi$}
		\\
		\begin{tikzpicture}[font=\sffamily\huge]
	\begin{scope}[scale=0.5, transform shape]

	\draw[color=black, fill=gray_box, step=1cm] (1, -1) rectangle(10, 8);

	\draw[color=g_box, fill=g_box, step=1cm](2, 3) rectangle(1, 2);
	\draw[color=g_box, fill=g_box, step=1cm](2, 3) rectangle(3, 2);
	\draw[color=g_box, fill=g_box, step=1cm](5, 2) rectangle(4, 1);
	\draw[color=g_box, fill=g_box, step=1cm](6, 6) rectangle(5, 5);
	\draw[color=g_box, fill=g_box, step=1cm](7, 5) rectangle(6, 4);
	\draw[color=g_box, fill=g_box, step=1cm](8, 3) rectangle(7, 2);
	\draw[color=g_box, fill=g_box, step=1cm](9, 3) rectangle(8, 2);
	\draw[color=g_box, fill=g_box, step=1cm](9, 6) rectangle(8, 5);
	\draw[color=g_box, fill=g_box, step=1cm](10, 2) rectangle(9, 1);

	\draw[step=1cm, color=black] (0, -1) grid (10,9);
	\node at (1.5, 8.5) {$\eta_1$};
	\node at (2.5, 8.5) {$\eta_2$};
	\node at (3.5, 8.5) {$\eta_3$};
	\node at (4.5, 8.5) {$\eta_4$};
	\node at (5.5, 8.5) {$x_2$};
	\node at (6.5, 8.5) {$x_3$};
	\node at (7.5, 8.5) {$x_4$};
	\node at (8.5, 8.5) {$x_1$};
	\node at (9.5, 8.5) {$x_6$};
	
	\node at (0.5, 7.5) {$\eta_1$};
	\node at (0.5, 6.5) {$\eta_2$};
	\node at (0.5, 5.5) {$\eta_3$};
	\node at (0.5, 4.5) {$\eta_4$};
	\node at (0.5, 3.5) {$x_2$};
	\node at (0.5, 2.5) {$x_3$};
	\node at (0.5, 1.5) {$x_4$};
	\node at (0.5, 0.5) {$x_1$};
	\node at (0.5, -0.5) {$x_6$};
	\end{scope}
	\end{tikzpicture}
		&
		\begin{tikzpicture}[font=\sffamily\huge]
	\begin{scope}[scale=0.5, transform shape]
		\draw[color=black, fill=gray_box, step=1cm] (1, -1) rectangle(10, 8);

		\draw[color=g_box, fill=g_box, step=1cm](2, 8) rectangle(1, 7);
		\draw[color=g_box, fill=g_box, step=1cm](8, 1) rectangle(7, 2);
		\draw[color=g_box, fill=g_box, step=1cm](3, 4) rectangle(2, 3);
		\draw[color=g_box, fill=g_box, step=1cm](6, 7) rectangle(5, 6);
		\draw[color=g_box, fill=g_box, step=1cm](3, 8) rectangle(2, 7);
		\draw[color=g_box, fill=g_box, step=1cm](2, 7) rectangle(1, 6);
		\draw[color=g_box, fill=g_box, step=1cm](3, 7) rectangle(2, 6);
		\draw[color=g_box, fill=g_box, step=1cm](4, 6) rectangle(3, 5);
		\draw[color=g_box, fill=g_box, step=1cm](5, 5) rectangle(4, 4);
		\draw[color=g_box, fill=g_box, step=1cm](6, 4) rectangle(5, 3);
		\draw[color=g_box, fill=g_box, step=1cm](7, 3) rectangle(6, 2);
		
		\draw[color=y_box, fill=y_box, step=1cm](10,1) rectangle(8, -1);
		\draw[step=1cm, color=black] (0, -1) grid (10,9);
		\node at (1.5, 8.5) {$\eta_1$};
		\node at (2.5, 8.5) {$\eta_2$};
		\node at (3.5, 8.5) {$\eta_3$};
		\node at (4.5, 8.5) {$\eta_4$};
		\node at (5.5, 8.5) {$x_2$};
		\node at (6.5, 8.5) {$x_3$};
		\node at (7.5, 8.5) {$x_4$};
		\node at (8.5, 8.5) {$x_1$};
		\node at (9.5, 8.5) {$x_6$};
		
		\node at (0.5, 7.5) {$\eta_1$};
		\node at (0.5, 6.5) {$\eta_2$};
		\node at (0.5, 5.5) {$\eta_3$};
		\node at (0.5, 4.5) {$\eta_4$};
		\node at (0.5, 3.5) {$x_2$};
		\node at (0.5, 2.5) {$x_3$};
		\node at (0.5, 1.5) {$x_4$};
		\node at (0.5, 0.5) {$x_1$};
		\node at (0.5, -0.5) {$x_6$};
	\end{scope}
\end{tikzpicture} 
		\\
		\\
		{\LARGE$\Lambda$} & {\LARGE$\Theta$}
		\\
		\begin{tikzpicture}[font=\sffamily\huge]
	\begin{scope}[scale=0.5, transform shape]
		\draw[color=black, fill=gray_box, step=1cm] (1, -2) rectangle(10, 8);

		\draw[color=g_box, fill=g_box, step=1cm](2, 7) rectangle(1, 6);

		\node at (1.5, 7.5) {1};
		\draw[color=g_box, fill=g_box, step=1cm](2, 7) rectangle(1, 6);
		\draw[color=g_box, fill=g_box, step=1cm](3, 7) rectangle(2, 6);
		\draw[color=g_box, fill=g_box, step=1cm](2, 6) rectangle(1, 5);
		\node at (2.5, 5.5) {1};
		\node at (3.5, 4.5) {1};
		\draw[color=g_box, fill=g_box, step=1cm](3, 3) rectangle(4, 4);
		\node at (4.5, 4.5) {1};
		\draw[color=g_box, fill=g_box, step=1cm](4, 2) rectangle(5, 3);
		\node at (5.5,  0.5) {1};
		\node at (6.5,  -0.5) {1};

		\draw[color=g_box, fill=g_box, step=1cm](10, 1) rectangle(9, 2);

		\node at (7.5,  -1.5) {1};

		\draw[step=1cm, color=black] (0, -2) grid (10, 9);
		\node at (1.5, 8.5) {$\eta_1$};
		\node at (2.5, 8.5) {$\eta_2$};
		\node at (3.5, 8.5) {$\eta_3$};
		\node at (4.5, 8.5) {$\eta_4$};
		\node at (5.5, 8.5) {$x_2$};
		\node at (6.5, 8.5) {$x_3$};
		\node at (7.5, 8.5) {$x_4$};
		\node at (8.5, 8.5) {$x_1$};
		\node at (9.5, 8.5) {$x_6$};
		\node at (0.5, 7.5) {$y_1$};
		\node at (0.5, 6.5) {$y_2$};
		\node at (0.5, 5.5) {$y_3$};
		\node at (0.5, 4.5) {$y_4$};
		\node at (0.5, 3.5) {$y_5$};
		\node at (0.5, 2.5) {$y_6$};
		\node at (0.5, 1.5) {$y_7$};
		\node at (0.5, 0.5) {$x_2$};
		\node at (0.5, -0.5) {$x_3$};
		\node at (0.5, -1.5) {$x_4$};

	\end{scope}
\end{tikzpicture} 
		&
		\begin{tikzpicture}[font=\sffamily\huge]
	\begin{scope}[scale=0.5, transform shape]
		\draw[color=black, fill=gray_box, step=1cm] (1, -2) rectangle(11, 8);
		\draw[color=g_box, fill=g_box, step=1cm](2, 8) rectangle(1, 7);
		\draw[color=g_box, fill=g_box, step=1cm](3, 7) rectangle(2, 6);
		\draw[color=g_box, fill=g_box, step=1cm](4, 6) rectangle(3, 5);
		\draw[color=g_box, fill=g_box, step=1cm](5, 5) rectangle(4, 4);
		\draw[color=g_box, fill=g_box, step=1cm](6, 4) rectangle(5, 3);
		\draw[color=g_box, fill=g_box, step=1cm](7, 3) rectangle(6, 2);
		\draw[color=g_box, fill=g_box, step=1cm](8, 2) rectangle(7, 1);

		\draw[color=g_box, fill=g_box, step=1cm](7, 4) rectangle(6, 3);
		\draw[color=g_box, fill=g_box, step=1cm](6, 3) rectangle(5, 2);
		\draw[step=1cm, color=black] (0, -2) grid (11,9);
		\node at (1.5, 8.5) {$y_1$};
		\node at (2.5, 8.5) {$y_2$};
		\node at (3.5, 8.5) {$y_3$};
		\node at (4.5, 8.5) {$y_4$};
		\node at (5.5, 8.5) {$y_5$};
		\node at (6.5, 8.5) {$y_6$};
		\node at (7.5, 8.5) {$y_7$};
		\node at (8.5, 8.5) {$x_2$};
		\node at (9.5, 8.5) {$x_3$};
		\node at (10.5, 8.5) {$x_4$};
		\node at (0.5, 7.5) {$y_1$};
		\node at (0.5, 6.5) {$y_2$};
		\node at (0.5, 5.5) {$y_3$};
		\node at (0.5, 4.5) {$y_4$};
		\node at (0.5, 3.5) {$y_5$};
		\node at (0.5, 2.5) {$y_6$};
		\node at (0.5, 1.5) {$y_7$};
		\node at (0.5, 0.5) {$x_2$};
		\node at (0.5, -0.5) {$x_3$};
		\node at (0.5, -1.5) {$x_4$};
	\end{scope} 
\end{tikzpicture}
\end{tabular}
	\begin{tabular}{rlrlrlrl}
	\lapbox[\width]{0.5em}{\raisebox{-0.3\height}{\begin{tikzpicture}[font=\sffamily\huge]
	\begin{scope}[scale=0.5, transform shape]
			\draw[color=black, fill=gray_box, step=1cm] (0, 0) rectangle(1, 1);
	\end{scope}
	\end{tikzpicture}}} & -- zero entry;
	&
	\lapbox[\width]{0.5em}{\raisebox{-0.3\height}{\begin{tikzpicture}[font=\sffamily\huge]
	\begin{scope}[scale=0.5, transform shape]
			\draw[color=black, fill=gray_box, step=1cm] (0, 0) rectangle(1, 1);
			\node at (0.5, 0.5) {1};
	\end{scope}
	\end{tikzpicture}}} & -- fixed $1.0$ entry;
\\

\\
\lapbox[\width]{0.5em}{\raisebox{-0.3\height}{\begin{tikzpicture}[font=\sffamily\huge]
	\begin{scope}[scale=0.5, transform shape]
			\draw[color=black, fill=gray_box, step=1cm] (0, 0) rectangle(1, 1);
			\draw[color=g_box, fill=g_box, step=1cm](0.05, 0.05) rectangle(0.95, 0.95);
	\end{scope}
	\end{tikzpicture}}} & -- parameter;
	&
	\lapbox[\width]{0.5em}{\raisebox{-0.3\height}{\begin{tikzpicture}[font=\sffamily\huge]
	\begin{scope}[scale=0.5, transform shape]
			\draw[color=black, fill=gray_box, step=1cm] (0, 0) rectangle(1, 1);
			\draw[color=y_box, fill=y_box, step=1cm](0.05, 0.05) rectangle(0.95, 0.95);
	\end{scope}
	\end{tikzpicture}}} & -- sample covariance;

\end{tabular}
	\caption{Matrices parameterized as implemented in \class{Model} class. Parameterized entries in symmetric matrices ($\Theta, \Psi$) have identical parameters in their lower and upper triangular parts. }
\end{figure}

To help reader understand how matrices are parameterized, compare Figure~\ref{fig:model} to Figure~\ref{fig:model_matrices} for clarification.

\begin{leftbar}
	In some SEM implementations covariances between "output" observed variables that are not indicators and "output" latent factors are parameterized too. To achieve this behavior in \textbf{semopy}, one can pass argument \code{mimic_lavaan=True} to the constructor of \class{Model}.
\end{leftbar}

\subsubsection{Estimation methods}
\class{Model} provides several loss functions to minimize for parameters estimation. Here and in other parts of the article, we shall often assume that parameterized matrices are parameterized by a certain vector $\theta$. Rigorously, we should always write $B(\theta), \Psi(\theta), \Sigma(\theta)$ etc, but we often omit explicit declaration of matrices as functions of parameter vector $\theta$ for brevity.

Maximum likelihood methods assume a normal distribution of variables:
\begin{itemize}
	\item \textbf{Wishart maximum likelihood}
	\begin{equation}\label{eq:mlw}
		F(\theta|S) = tr\{S \Sigma^{-1}(\theta)\} + \ln |\Sigma(\theta)|
	\end{equation}
	It is a maximum likelihood method for Wishart distribution. 
	\begin{leftbar}
		Wishart ML equivalent to multivariate normal ML, see that:
		\begin{equation*}
			\begin{split}
				l(\theta|Z) = \sum_{i=1}^n (z_{(i)} - \mu)^T \Sigma^{-1} (z_{(i)} - \mu) + ln|\Sigma| = tr\{\sum_{i=1}^n (z_{(i)} - \mu)^T \Sigma^{-1} (z_{(i)} - \mu)\} +\\+ n ln|\Sigma|  =\sum_{i=1}^n tr\{(z_{(i)} - \mu)^T \Sigma^{-1} (z_{(i)} - \mu) \} + n \ln|\Sigma| =
			\end{split},
		\end{equation*}
		\begin{equation*}
			\begin{split}
			\\=  tr\left\{\left(\sum_{i=1}^n (z_{(i)} - \mu)(z_{(i)} - \mu)^T \right)\Sigma^{-1} \right\} + n \ln|\Sigma|= tr\left\{ M M^T \Sigma^{-1} \right\} + n \ln |\Sigma| = \\= n tr \left\{S \Sigma^{-1} \right\} + n\ln|\Sigma| \propto tr \left\{S \Sigma^{-1} \right\} + \ln|\Sigma|
			\end{split},
		\end{equation*}
		where  $M = [z_{(1)}, z_{(2)}, \dots, z_{(n)}] - \mu \mathbbm{1}$ -- centered data matrix, $S = \frac{1}{n}M M^T$ is a biased sample covariance matrix. If we were to use unbiased $S = \frac{1}{n - 1}M M^T$, then 
		\begin{equation}\label{eq:mln_unbiased}
			F(\theta|Z) = (n - 1)tr\{S\Sigma^{-1}(\theta)\} - n ln|\Sigma|
		\end{equation}
	\end{leftbar}
	To estimate model parameters with Wishart ML, no extra actions are necessary as it is the default method in \class{Model}, but if that changes, you can supply \code{'MLW'} to argument \code{method} of \code{Model.fit}: \\ \code{model.fit(..., method='MLW')}.
	\item \textbf{Full information maximum likelihood (FIML)}

	All of the approaches above handle missing data naturally at the stage of computing $S$. However, one might consider FIML to be a more viable approach.

	FIML at its core is very similar to multivariate normal ML, but for each term in loglikelihood sum, we cull columns and rows in $\Sigma$ that correspond to missing variables. In case when all of the data is present, FIML degrades to multivariate normal as in Equation~\ref{eq:mln_unbiased}.
	
	To estimate model parameters with FIML, supply \code{'FIML'} to argument \code{method} of \code{Model.fit}: \code{model.fit(..., method='FIML')}.
\end{itemize}

The least-squares method might be more robust when the normality assumption is violated:
\begin{itemize}
	\item \textbf{Unweighted least squares (ULS)}
	\begin{equation}\label{eq:uls}
		F(\theta|S) = tr\{(\Sigma(\theta) - S)(\Sigma(\theta) - S)^T\} = tr\{(\Sigma(\theta) - S)^2\},
	\end{equation}
	where $S$ is a sample covariance matrix.
	
	To estimate model parameters with ULS, supply \code{'ULS'} to argument \code{method} of \code{Model.fit}:\\ \code{model.fit(..., method='ULS')}.
	
	\item \textbf{Generalized least squares (GLS)}
	\begin{equation}\label{eq:gls}
		F(\theta|S) = tr\{(I_{n_{z}} - \Sigma(\theta)S^{-1})^2\}
	\end{equation}
	It is the same as minimizing Mahalanobis distance.
	
	To estimate model parameters with GLS, supply \code{'GLS'} to argument \code{method} of \code{Model.fit}:\\ \code{model.fit(..., method='GLS')}.
	
	\item \textbf{Weighted least squares (WLS)}
	\begin{equation}\label{eq:wls}
		F(\theta|Z) = (vech(\Sigma(\theta) - vech(S))^T W^{-1} (vech(\Sigma(\theta) - vech(S)),
	\end{equation}
	where $vech$ is a half-vectorization operator (i.e. $vech(S)$ transforms $n_{z} \times n_{z}$ matrix to a $\frac{n_z(n_z + 1)}{2} \times 1$ vector of elements from an upper triangular part of $S$), $W$ is a weight matrix that is usually chosen to be an asymptotic covariance matrix for $\Sigma$. A user can pass any custom $W$ matrix to \code{Model.fit} as an \code{wls_w} argument, but by default it will use the fourth-moments matrix as proposed by \cite{Browne:1984}. In the latter case, it is also known as asymptotic distribution-free (ADL) estimator. 
	
	\begin{leftbar}
		Let ${z_k}$ be a $n \times 1$ vector of observations for $k$-th component of the $n_z \times 1$ vector $z$. The algorithm for computing is $W$ is following:
		\begin{enumerate}
			\item For each $\{{z_k}\}_{k=1}^{n_z}$ compute sample mean $\hat{z}_k = \frac{1}{N}\sum_{i=1}^n z_k^{(i)}$;
			\item Center each $\{{z_k}\}_{k=1}^{n_z}$: $\tilde{z_k} = z_k - \hat{z}_k$;
			\item Compute matrix $X$ of Cartesian Hadamard products for $\{\tilde{z}_k\}_{k=1}^{n_z}$:
			$$X = \left[\tilde{z}_1 \odot \tilde{z}_1, \tilde{z}_1 \odot \tilde{z}_2, \dots, \tilde{z}_1 \odot \tilde{z}_{n_z}, \tilde{z}_2 \odot \tilde{z}_1, \dots, \tilde{z}_{n_z} \odot \tilde{z}_{n_z} \right] $$
			\item Compute $W$ as a covariance matrix of $X$.
			
		\end{enumerate}
	\end{leftbar}
	
	WLS is often used when assumptions on data normality are violated.
	
	If $W$ can't be successfully inverted, then a warning is printed and a nearest positive-definite matrix of $W$ is used instead.
	
	To estimate model parameters with WLS, supply \code{'WLS'} to argument \code{method} of \code{Model.fit}:\\ \code{model.fit(..., method='WLS')}.
	
	\item \textbf{Diagonally weighted least squares (DWLS)}
	
	It is the same as WLS, but all non-diagonal elements of $W$ are zeroed. It might be helpful if $W$ is ill-conditioned or too big and inverting it is not an option. DWLS is also sometimes referred to as \textit{robust weighted least squares} \citep{DiStefano:2014}.
	
	To estimate model parameters with DWLS, supply \code{'DWLS'} to argument \code{method} of \code{Model.fit}:\\ \code{model.fit(..., method='DWLS')}.
\end{itemize}

If some of the variables are ordinal, then in all of the methods above it is possible to substitute sample covariance matrix $S$ with a heterogeneous correlation matrix $\widehat{S}$ computed from polychoric and polyserial correlations \citep{Drasgow:2004}: it can be done through model syntax as explained in Section~\ref{sect:syntax}. Sometimes, however, $\widehat{S}$ is degenerate: in that case, \textbf{semopy} finds a nearest positive-definite matrix and uses it instead of $\widehat{S}$. The latter also happens with the sample covariance matrix when ill data is passed. In either case, \textbf{semopy} will output a warning that informs the user of possible problems with data.

\subsubsection{Usage example}

\begin{table}[!t]
	\centering
	\begin{tabular}{r|p{9cm}}
		
		Module name & Description \\
		\hline
		{\texttt{univariate\_regression}} & Univariate regression wtih 1 independent varaible.\\
		{\texttt{univariate\_regression\_many}} & Univariate regression with 3 independent variables.\\
		{\texttt{multivariate\_regression}} & Multivariate regression with 5 independent and 3 dependent variables. \\
		{\texttt{example\_article}} & Toy model from Figure~\ref{fig:model}.\\
		{\texttt{political\_democracy}} & Political Democracy dataset that is frequently used as a testing dataset in SEM software \citep{Bollen:1980}. \\
		{\texttt{holzinger39}} & The classic Holzinger and Swineford dataset consists of mental ability test scores of seventh- and eighth-grade children from two different schools that is frequently used for showcasing in SEM software \citep{Joreskog:1969}.
	\end{tabular}
	\caption{Toy SEM models + dataset that are used in \textbf{semopy} for showcasing and testing. All modules reside in submodule \code{semopy.examples}, each of them has functions \code{get\_model()}, \code{get\_data()}, \code{get\_params()} that return string description of the model in \textbf{semopy} syntax (see Section~\ref{sect:syntax}), dataset and true parameter values respectively. The latter is obviously not returned for \code{political\_democracy} and \code{holzinger39} as they are real datasets and hence no true parameter estimates are available. A reader can check \textbf{semopy} results for those models in Appendix~\ref{app:toys}.}
	\label{table:examples}
\end{table}

\textbf{semopy} has some built-in SEM models to help its users dive into it: see list at Table~\ref{table:examples}. Throughout most of this article (namely, in Section~\ref{sect:model}, Section~\ref{sect:modelmeans}, Section~\ref{sect:modeleffects} and Section~\ref{sect:modelgeneralizedeffects}), we shall use only \code{example_article} as it is the model showcased at Figure~\ref{fig:model} (or Figure~\ref{fig:modelmeans}). The reason why we chose this model for showcasing is because it has all variety of interactions supported by \textbf{semopy}. Also, it is not identifiable with respect to some of its parameters -- usually, \textbf{semopy} provides hints to users that model is incorrect and some of parameter estimates should not be trusted (see below).

First, we get a built-in dataset and toy model:
	\consolein
\begin{lstlisting}[style=codeinput]
import semopy
ex = semopy.examples.example_article
desc, data = ex.get_model(), ex.get_data()
print(desc)
	\end{lstlisting}
	\consoleout
\begin{lstlisting}[style=codeoutput]
# Measurement part
eta1 =~ y1 + y2 + y3
eta2 =~ y3 + y2
eta3 =~ y4 + y5
eta4 =~ y4 + y6
# Structural part
eta3 ~ x2 + x1
eta4 ~ x3
x3 ~ eta1 + eta2 + x1
x4 ~ eta4 + x6
y7 ~ x4 + x6
# Additional covariances
y6 ~~ y5
x2 ~~ eta2		
	\end{lstlisting}
	\consolein
\begin{lstlisting}[style=codeinput]
print(data.head())
	\end{lstlisting}
	\consoleout
\begin{lstlisting}[style=codeoutput]
y1        y2        y3  ...        x4        x1        x6
0  0.729838 -0.781150 -0.473951  ...  1.984575 -1.187765 -0.025494
1 -1.895332  0.313026 -1.861669  ...  2.139032 -0.397323 -0.217159
2  0.771990 -2.019936 -0.452560  ...  1.908656  0.534365  0.058370
3 -0.956471 -0.374326  0.040394  ... -0.089787  0.091094 -0.603859
4  0.959640 -0.997909 -0.299834  ...  2.284195 -0.851540 -0.343206

[5 rows x 12 columns]
	\end{lstlisting}

Then, we instantiate a \class{Model} and fit it to data. We also print an optimization result:
	\consolein
\begin{lstlisting}[style=codeinput]
m = semopy.Model(desc)
r = m.fit(data)
print(r)
	\end{lstlisting}
	\consoleout
\begin{lstlisting}[style=codeoutput]
Name of objective: MLW
Optimization method: SLSQP
Optimization successful.
Optimization terminated successfully
Objective value: 0.091
Number of iterations: 58
Params: -0.488 -0.782 -0.183 1.225 1.444 -1.147 -1.344 1.223 1.071
-0.348 1.291
1.454 0.840 -0.388 -0.625 -0.106 1.252 -0.084 1.097
0.870 0.696 0.844 0.654 1.114 0.871 0.824 1.010 0.804 1.182 -0.499
1.264
	\end{lstlisting}

Finally, we can print a fancy table with parameter estimates and their p-values:

	\consolein
\begin{lstlisting}[style=codeinput]
ins = m.inspect()
print(ins)
	\end{lstlisting}
	\consoleout
\begin{lstlisting}[style=codeoutput]
WARNING:root:Fisher Information Matrix is not PD. Moore-Penrose inverse
will be used instead of Cholesky decomposition. See
10.1109/TSP.2012.2208105.

lval  op  rval  Estimate  Std. Err    z-value   p-value
0   eta3   ~    x2 -1.146663  0.065317  -17.55527       0.0
1   eta3   ~    x1 -1.344422  0.076917 -17.478884       0.0
2   eta4   ~    x3  1.222542  0.038071  32.112318       0.0
3     x3   ~  eta1  1.070822  0.287943   3.718868    0.0002
4     x3   ~  eta2 -0.347555  0.146593  -2.370895  0.017745
5     x3   ~    x1  1.291230  0.075725  17.051592       0.0
6     x4   ~  eta4  1.454421  0.041067   35.41557       0.0
7     x4   ~    x6  0.839923   0.06817  12.320923       0.0
8     y1   ~  eta1  1.000000         -          -         -
9     y2   ~  eta1 -0.488414  0.664931  -0.734533  0.462624
10    y2   ~  eta2 -0.781996  0.912859  -0.856646  0.391641
11    y3   ~  eta1 -0.182725  0.140074  -1.304484  0.192069
12    y3   ~  eta2  1.000000         -          -         -
13    y4   ~  eta3  1.000000         -          -         -
14    y4   ~  eta4  1.000000         -          -         -
15    y5   ~  eta3  1.224550  0.048392  25.304791       0.0
16    y6   ~  eta4  1.443567  0.040942  35.258544       0.0
17    y7   ~    x4 -0.387558   0.01444   -26.8399       0.0
18    y7   ~    x6 -0.624882     0.058 -10.773807       0.0
19    x2  ~~  eta2 -0.084431  0.087237  -0.967832  0.333128
20  eta3  ~~  eta3  0.869520  0.110941   7.837675       0.0
21    x3  ~~    x3  1.114065  0.566346   1.967111   0.04917
22    x4  ~~    x4  1.009523  0.136551   7.393021       0.0
23  eta4  ~~  eta4  0.803514  0.090644   8.864495       0.0
24  eta2  ~~  eta2  1.181504  0.855015   1.381853  0.167017
25  eta2  ~~  eta1 -0.498966  0.239579  -2.082676  0.037281
26  eta1  ~~  eta1  1.263544  0.456489   2.767959  0.005641
27    y6  ~~    y5 -0.105931  0.101857  -1.039999   0.29834
28    y6  ~~    y6  1.251659  0.151825   8.244097       0.0
29    y7  ~~    y7  1.096623  0.089539  12.247449       0.0
30    y1  ~~    y1  0.695780  0.435022   1.599413  0.109729
31    y3  ~~    y3  0.844282  0.961208   0.878355  0.379751
32    y4  ~~    y4  0.654485   0.11071   5.911725       0.0
33    y2  ~~    y2  0.871375  0.751912   1.158879  0.246505
34    y5  ~~    y5  0.823609  0.143472   5.740541       0.0
	\end{lstlisting}

Notice the warning -- \textbf{semopy} has managed to spot identification issues. Still, most parameters are correctly estimated. We can check mean absolute percentage error (MAPE) for true parameter values:

	\consolein
\begin{lstlisting}[style=codeinput]
import numpy as np
params = ex.get_params()
mape = np.mean(semopy.utils.compare_results(m, params))
print('MAPE: {:.2f}%'.format(mape * 100))
	\end{lstlisting}
	\consoleout
\begin{lstlisting}[style=codeoutput]
MAPE: 19.94%
	\end{lstlisting}

For an example of a clean working session, see Appendix~\ref{app:toys}.

\subsection[ModelMeans]{SEM with fixed effects: \class{ModelMeans}}\label{sect:modelmeans}
As it is evident from the Equation~\ref{eq:model}, \class{Model} has no support for modeling intercepts nor any kinds of fixed effects. The former is not crucial as it is rarely of any interest, the latter, however, might be a problem. \class{Model} analyzes sample covariance information only, and it will fail if a variation of a variable is too low to be captured effectively by a covariance function. Furthermore, it loses information on a relative scale of a variable: for instance, it will produce the same results for an ordinal variable encoded as $\{0, 1, 2\}$ and an ordinal variable encoded as $\{0, 1, 10\}$. Also, it will have problems with non-normal variables in the first place. In Equation~\ref{eq:modelmeans} we propose a model that lacks most of those downfalls and can be used with non-normal variables as long as they are exogenous:
\begin{equation}\label{eq:modelmeans}
	\begin{cases}
		\begin{bmatrix}
			\eta \\ x^{(1)}
		\end{bmatrix} = \omega = \Gamma_1 x^{(2)} + B \omega + \epsilon, ~~~\epsilon \sim \mathcal{N}(0, \Psi) \\
		\begin{bmatrix}
			y \\ x^{(1)}
		\end{bmatrix} = z = \Gamma_2 x^{(2)} + \Lambda \omega + \delta, ~~~\delta \sim \mathcal{N}(0, \Theta)
	\end{cases},
\end{equation}
where $\Gamma_1$ is a $n_{\omega} \times n_{x^{(2)}}$ loading matrix of exogenous variables $x^{(1)}$ onto latent factors $\eta$ and endogenous non-output variables $x^{(1)}$, and $\Gamma_2$ is a $n_{z} \times n_{x^{(2)}}$ loading matrix of exogenous variables onto output variables $y$. $\Gamma_1$ and $\Gamma_2$ are parameterized. By default, there is a fake $\mathbbm{1}$ exogenous variable that loads onto all endogenous observed variables to model intercepts, but if user seeks to reduce parameter space, he/she can pass \code{intercepts=False} argument to the \code{ModelMeans.fit} method. However, in that case, data has to be centered beforehand.

\begin{figure}[t!]
	\label{fig:modelmeans}
	\centering
	\begin{tikzpicture}
	\begin{scope}[scale=1.3, transform shape]
		
		\node[my_obs] (y1) at (1,10) {$y_1$};
		\node[my_obs] (y2) at (1,9) {$y_2$};
		\node[my_obs] (y3) at (1,8) {$y_3$};
		\node[my_obs] (y4) at (4,7) {$y_4$};
		\node[my_obs] (y5) at (4,6) {$y_{5}$};
		\node[my_obs] (y6) at (5,7) {$y_{6}$};
		
		\node[my_obs_exo] (x1) at (2,7) {$x_1$};
		\node[my_obs] (x2) at (2,6) {$x_2$};
		\node[my_obs] (x3) at (3,8) {$x_3$};
		\node[my_obs] (x4) at (6,8) {$x_4$};
		\node[my_obs] (x5) at (7,8) {$y_7$};
		\node[my_obs_exo] (x6) at (6.5,7) {$x_6$};

		\node[my_latent] (eta1) at (2,9) {$\eta_1$};
		\node[my_latent] (eta2) at (2,8) {$\eta_2$};
		\node[my_latent] (eta3) at (3,6.5) {$\eta_3$};
		\node[my_latent] (eta4) at (4.5, 8) {$\eta_4$};

		\draw[my_arrow] (x6) -- (x5);
		\draw[my_arrow] (x6) -- (x4);
		\draw[my_arrow] (eta1) -- (x3) ; 
		\draw[my_arrow] (eta2) -- (x3); 
		\draw[my_arrow] (x1) -- (x3); 
		\draw[my_arrow] (x1) -- (eta3); 
		\draw[my_arrow] (x2) -- (eta3) ; 
		\draw[my_arrow] (x3) -- (eta4) ; 
		\draw[my_arrow] (eta4) -- (x4); 
		\draw[my_arrow] (x4) -- (x5) ; 
		\draw[my_arrow] (x4) to [out=135, in=0] (4.5, 8.7) to [out=180, in=45] (x3); 
		\draw[my_covariance] (eta2) to [out= 200, in=90] (1.2, 7) to [out=-90, in= 180] (x2);
		\draw[my_covariance_mp] (y5) to [out= 0, in=225] (4.8, 6.2) to [out=45, in= -90]   (y6);
		
		\draw[my_arrow] (eta1) -- (y1) ; 
		\draw[my_arrow] (eta1) -- (y2); 
		\draw[my_arrow] (eta1) -- (y3); 
		\draw[my_arrow] (eta2) -- (y3);
		\draw[my_arrow] (eta2) -- (y2);
		\draw[my_arrow] (eta3) -- (y4) ;
		\draw[my_arrow] (eta3) -- (y5) ; 
		\draw[my_arrow] (eta4) -- (y4);
		\draw[my_arrow] (eta4) -- (y6);
		
	\end{scope}
\end{tikzpicture}
	\vspace{-1em}
	\begin{tabular}{rlrlrlrl}
		\lapbox[\width]{0.5em}{\raisebox{-0.3\height}{\begin{tikzpicture}
			\node[my_latent] (eta) at (0,0) {$\eta$};
	\end{tikzpicture}}} & -- latent factor;
	&
	\lapbox[\width]{0.5em}{\raisebox{-0.3\height}{\begin{tikzpicture}
				\node[my_obs] (x) at (0,0) {$x$};
	\end{tikzpicture}}} & -- observed variable;
	&
	\lapbox[\width]{0.5em}{\raisebox{-0.3\height}{\begin{tikzpicture}
				\node[my_obs_exo] (x) at (0,0) {$x$};
	\end{tikzpicture}}} & -- exogenous observed variable;

\end{tabular}
\\
\begin{tabular}{rlrl}
	\lapbox[\width]{1em}{\begin{tikzpicture}
			\node (a1) at (0,0) {};
			\node (a2) at (0.8,0) {};
			\draw[my_arrow] (a1) -- (a2);
	\end{tikzpicture}} & -- loading;
	\lapbox[\width]{1em}{\begin{tikzpicture}
			\node (a1) at (0,0) {};
			\node (a2) at (0.8,0) {};
			\draw[my_covariance] (a1) -- (a2);
	\end{tikzpicture}} & -- covariance;
\end{tabular}
	\caption{\class{ModelMeans} representation of the same SEM model presented previously. Notice that exogenous observed variables are now ruled out into a separate class. }
\end{figure}
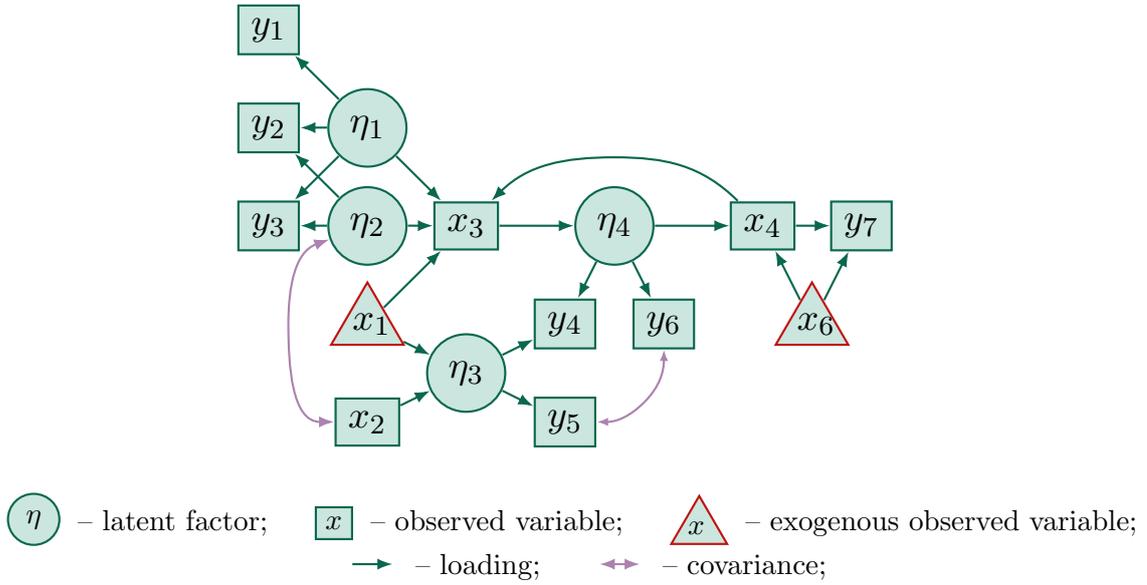
\begin{figure}[t!]
	\label{fig:modelmeans_matrices}
	\centering
	\begin{tabular}{ccc}
		{\LARGE$\Gamma_1$} & {\LARGE$B$} & {\LARGE$\Psi$}
		\\
		\begin{tikzpicture}[font=\sffamily\huge]
	\begin{scope}[scale=0.5, transform shape]

	\draw[color=black, fill=gray_box, step=1cm] (1, 1) rectangle(4, 8);

	\draw[color=g_box, fill=g_box, step=1cm](2, 3) rectangle(1, 2);
	\draw[color=g_box, fill=g_box, step=1cm](2, 6) rectangle(1, 5);
	\draw[color=g_box, fill=g_box, step=1cm](3, 2) rectangle(2, 1);
	\draw[color=g_box, fill=g_box, step=1cm](4, 4) rectangle(3, 1);

	\draw[step=1cm, color=black] (0, 1) grid (4,9);
	\node at (1.5, 8.5) {$x_1$};
	\node at (2.5, 8.5) {$x_6$};
	\node at (3.5, 8.5) {$\mathbbm{1}$};
	
	\node at (0.5, 7.5) {$\eta_1$};
	\node at (0.5, 6.5) {$\eta_2$};
	\node at (0.5, 5.5) {$\eta_3$};
	\node at (0.5, 4.5) {$\eta_4$};
	\node at (0.5, 3.5) {$x_2$};
	\node at (0.5, 2.5) {$x_3$};
	\node at (0.5, 1.5) {$x_4$};
	\end{scope}
\end{tikzpicture}
		&
		\begin{tikzpicture}[font=\sffamily\huge]
	\begin{scope}[scale=0.5, transform shape]

	\draw[color=black, fill=gray_box, step=1cm] (1, 1) rectangle(8, 8);

	\draw[color=g_box, fill=g_box, step=1cm](2, 3) rectangle(1, 2);
	\draw[color=g_box, fill=g_box, step=1cm](2, 3) rectangle(3, 2);
	\draw[color=g_box, fill=g_box, step=1cm](5, 2) rectangle(4, 1);
	\draw[color=g_box, fill=g_box, step=1cm](6, 6) rectangle(5, 5);
	\draw[color=g_box, fill=g_box, step=1cm](7, 5) rectangle(6, 4);
	\draw[color=g_box, fill=g_box, step=1cm](8, 3) rectangle(7, 2);

	\draw[step=1cm, color=black] (0, 1) grid (8,9);
	\node at (1.5, 8.5) {$\eta_1$};
	\node at (2.5, 8.5) {$\eta_2$};
	\node at (3.5, 8.5) {$\eta_3$};
	\node at (4.5, 8.5) {$\eta_4$};
	\node at (5.5, 8.5) {$x_2$};
	\node at (6.5, 8.5) {$x_3$};
	\node at (7.5, 8.5) {$x_4$};
	
	\node at (0.5, 7.5) {$\eta_1$};
	\node at (0.5, 6.5) {$\eta_2$};
	\node at (0.5, 5.5) {$\eta_3$};
	\node at (0.5, 4.5) {$\eta_4$};
	\node at (0.5, 3.5) {$x_2$};
	\node at (0.5, 2.5) {$x_3$};
	\node at (0.5, 1.5) {$x_4$};
	\end{scope}
\end{tikzpicture}
		&
		\begin{tikzpicture}[font=\sffamily\huge]
	\begin{scope}[scale=0.5, transform shape]
		\draw[color=black, fill=gray_box, step=1cm] (1, 1) rectangle(8, 8);

		\draw[color=g_box, fill=g_box, step=1cm](2, 8) rectangle(1, 7);
		\draw[color=g_box, fill=g_box, step=1cm](8, 1) rectangle(7, 2);
		\draw[color=g_box, fill=g_box, step=1cm](3, 4) rectangle(2, 3);
		\draw[color=g_box, fill=g_box, step=1cm](6, 7) rectangle(5, 6);
		\draw[color=g_box, fill=g_box, step=1cm](3, 8) rectangle(2, 7);
		\draw[color=g_box, fill=g_box, step=1cm](2, 7) rectangle(1, 6);
		\draw[color=g_box, fill=g_box, step=1cm](3, 7) rectangle(2, 6);
		\draw[color=g_box, fill=g_box, step=1cm](4, 6) rectangle(3, 5);
		\draw[color=g_box, fill=g_box, step=1cm](5, 5) rectangle(4, 4);
		\draw[color=g_box, fill=g_box, step=1cm](6, 4) rectangle(5, 3);
		\draw[color=g_box, fill=g_box, step=1cm](7, 3) rectangle(6, 2);
		
		\draw[color=y_box, fill=y_box, step=1cm](8,1) rectangle(8, 1);
		\draw[step=1cm, color=black] (0, 1) grid (8,9);
		\node at (1.5, 8.5) {$\eta_1$};
		\node at (2.5, 8.5) {$\eta_2$};
		\node at (3.5, 8.5) {$\eta_3$};
		\node at (4.5, 8.5) {$\eta_4$};
		\node at (5.5, 8.5) {$x_2$};
		\node at (6.5, 8.5) {$x_3$};
		\node at (7.5, 8.5) {$x_4$};
		
		\node at (0.5, 7.5) {$\eta_1$};
		\node at (0.5, 6.5) {$\eta_2$};
		\node at (0.5, 5.5) {$\eta_3$};
		\node at (0.5, 4.5) {$\eta_4$};
		\node at (0.5, 3.5) {$x_2$};
		\node at (0.5, 2.5) {$x_3$};
		\node at (0.5, 1.5) {$x_4$};
	\end{scope}
\end{tikzpicture} 
		\\
		\\
		{\LARGE$\Gamma_2$} & {\LARGE$\Lambda$} & {\LARGE$\Theta$}
		\\
		\begin{tikzpicture}[font=\sffamily\huge]
	\begin{scope}[scale=0.5, transform shape]
		\draw[color=black, fill=gray_box, step=1cm] (1, -2) rectangle(4, 8);

		\draw[color=g_box, fill=g_box, step=1cm](2, 1) rectangle(3, 2);
		\draw[color=g_box, fill=g_box, step=1cm](3, 1) rectangle(4, 8);

		\draw[step=1cm, color=black] (0, -2) grid (4, 9);
		\node at (1.5, 8.5) {$x_1$};
		\node at (2.5, 8.5) {$x_6$};
		\node at (3.5, 8.5) {$\mathbbm{1}$};
		\node at (0.5, 7.5) {$y_1$};
		\node at (0.5, 6.5) {$y_2$};
		\node at (0.5, 5.5) {$y_3$};
		\node at (0.5, 4.5) {$y_4$};
		\node at (0.5, 3.5) {$y_5$};
		\node at (0.5, 2.5) {$y_6$};
		\node at (0.5, 1.5) {$y_7$};
		\node at (0.5, 0.5) {$x_2$};
		\node at (0.5, -0.5) {$x_3$};
		\node at (0.5, -1.5) {$x_4$};
		
	\end{scope}
\end{tikzpicture} 
		&
		\begin{tikzpicture}[font=\sffamily\huge]
	\begin{scope}[scale=0.5, transform shape]
		\draw[color=black, fill=gray_box, step=1cm] (1, -2) rectangle(8, 8);

		\draw[color=g_box, fill=g_box, step=1cm](2, 7) rectangle(1, 6);

		\node at (1.5, 7.5) {1};
		\draw[color=g_box, fill=g_box, step=1cm](2, 7) rectangle(1, 6);
		\draw[color=g_box, fill=g_box, step=1cm](3, 7) rectangle(2, 6);
		\draw[color=g_box, fill=g_box, step=1cm](2, 6) rectangle(1, 5);
		\node at (2.5, 5.5) {1};
		\node at (3.5, 4.5) {1};
		\draw[color=g_box, fill=g_box, step=1cm](3, 3) rectangle(4, 4);
		\node at (4.5, 4.5) {1};
		\draw[color=g_box, fill=g_box, step=1cm](4, 2) rectangle(5, 3);
		\node at (5.5,  0.5) {1};
		\node at (6.5,  -0.5) {1};

		\node at (7.5,  -1.5) {1};

		\draw[step=1cm, color=black] (0, -2) grid (8, 9);
		\node at (1.5, 8.5) {$\eta_1$};
		\node at (2.5, 8.5) {$\eta_2$};
		\node at (3.5, 8.5) {$\eta_3$};
		\node at (4.5, 8.5) {$\eta_4$};
		\node at (5.5, 8.5) {$x_2$};
		\node at (6.5, 8.5) {$x_3$};
		\node at (7.5, 8.5) {$x_4$};
		\node at (0.5, 7.5) {$y_1$};
		\node at (0.5, 6.5) {$y_2$};
		\node at (0.5, 5.5) {$y_3$};
		\node at (0.5, 4.5) {$y_4$};
		\node at (0.5, 3.5) {$y_5$};
		\node at (0.5, 2.5) {$y_6$};
		\node at (0.5, 1.5) {$y_7$};
		\node at (0.5, 0.5) {$x_2$};
		\node at (0.5, -0.5) {$x_3$};
		\node at (0.5, -1.5) {$x_4$};

	\end{scope}
\end{tikzpicture} 
		&
		\begin{tikzpicture}[font=\sffamily\huge]
	\begin{scope}[scale=0.5, transform shape]
		\draw[color=black, fill=gray_box, step=1cm] (1, -2) rectangle(11, 8);
		\draw[color=g_box, fill=g_box, step=1cm](2, 8) rectangle(1, 7);
		\draw[color=g_box, fill=g_box, step=1cm](3, 7) rectangle(2, 6);
		\draw[color=g_box, fill=g_box, step=1cm](4, 6) rectangle(3, 5);
		\draw[color=g_box, fill=g_box, step=1cm](5, 5) rectangle(4, 4);
		\draw[color=g_box, fill=g_box, step=1cm](6, 4) rectangle(5, 3);
		\draw[color=g_box, fill=g_box, step=1cm](7, 3) rectangle(6, 2);
		\draw[color=g_box, fill=g_box, step=1cm](8, 2) rectangle(7, 1);

		\draw[color=g_box, fill=g_box, step=1cm](7, 4) rectangle(6, 3);
		\draw[color=g_box, fill=g_box, step=1cm](6, 3) rectangle(5, 2);
		\draw[step=1cm, color=black] (0, -2) grid (11,9);
		\node at (1.5, 8.5) {$y_1$};
		\node at (2.5, 8.5) {$y_2$};
		\node at (3.5, 8.5) {$y_3$};
		\node at (4.5, 8.5) {$y_4$};
		\node at (5.5, 8.5) {$y_5$};
		\node at (6.5, 8.5) {$y_6$};
		\node at (7.5, 8.5) {$y_7$};
		\node at (8.5, 8.5) {$x_2$};
		\node at (9.5, 8.5) {$x_3$};
		\node at (10.5, 8.5) {$x_4$};
		\node at (0.5, 7.5) {$y_1$};
		\node at (0.5, 6.5) {$y_2$};
		\node at (0.5, 5.5) {$y_3$};
		\node at (0.5, 4.5) {$y_4$};
		\node at (0.5, 3.5) {$y_5$};
		\node at (0.5, 2.5) {$y_6$};
		\node at (0.5, 1.5) {$y_7$};
		\node at (0.5, 0.5) {$x_2$};
		\node at (0.5, -0.5) {$x_3$};
		\node at (0.5, -1.5) {$x_4$};
	\end{scope} 
\end{tikzpicture}
\end{tabular}
	\begin{tabular}{rlrlrlrl}
	\lapbox[\width]{0.5em}{\raisebox{-0.3\height}{\begin{tikzpicture}[font=\sffamily\huge]
	\begin{scope}[scale=0.5, transform shape]
			\draw[color=black, fill=gray_box, step=1cm] (0, 0) rectangle(1, 1);
	\end{scope}
	\end{tikzpicture}}} & -- zero entry;
	&
	\lapbox[\width]{0.5em}{\raisebox{-0.3\height}{\begin{tikzpicture}[font=\sffamily\huge]
	\begin{scope}[scale=0.5, transform shape]
			\draw[color=black, fill=gray_box, step=1cm] (0, 0) rectangle(1, 1);
			\node at (0.5, 0.5) {1};
	\end{scope}
	\end{tikzpicture}}} & -- fixed $1.0$ entry;
\\

\\
\lapbox[\width]{0.5em}{\raisebox{-0.3\height}{\begin{tikzpicture}[font=\sffamily\huge]
	\begin{scope}[scale=0.5, transform shape]
			\draw[color=black, fill=gray_box, step=1cm] (0, 0) rectangle(1, 1);
			\draw[color=g_box, fill=g_box, step=1cm](0.05, 0.05) rectangle(0.95, 0.95);
	\end{scope}
	\end{tikzpicture}}} & -- parameter;
	&
	\lapbox[\width]{0.5em}{\raisebox{-0.3\height}{\begin{tikzpicture}[font=\sffamily\huge]
	\begin{scope}[scale=0.5, transform shape]
			\draw[color=black, fill=gray_box, step=1cm] (0, 0) rectangle(1, 1);
			\draw[color=y_box, fill=y_box, step=1cm](0.05, 0.05) rectangle(0.95, 0.95);
	\end{scope}
	\end{tikzpicture}}} & -- sample covariance;

\end{tabular}
	\caption{Matrices parameterization as implemented in \class{ModelMeans} class. Parameterized entries in symmetric matrices ($\Theta, \Psi$) have identical parameters in their lower and upper triangular parts. $\mathbbm{1}$ vector of ones are used for intercepts. }
\end{figure}
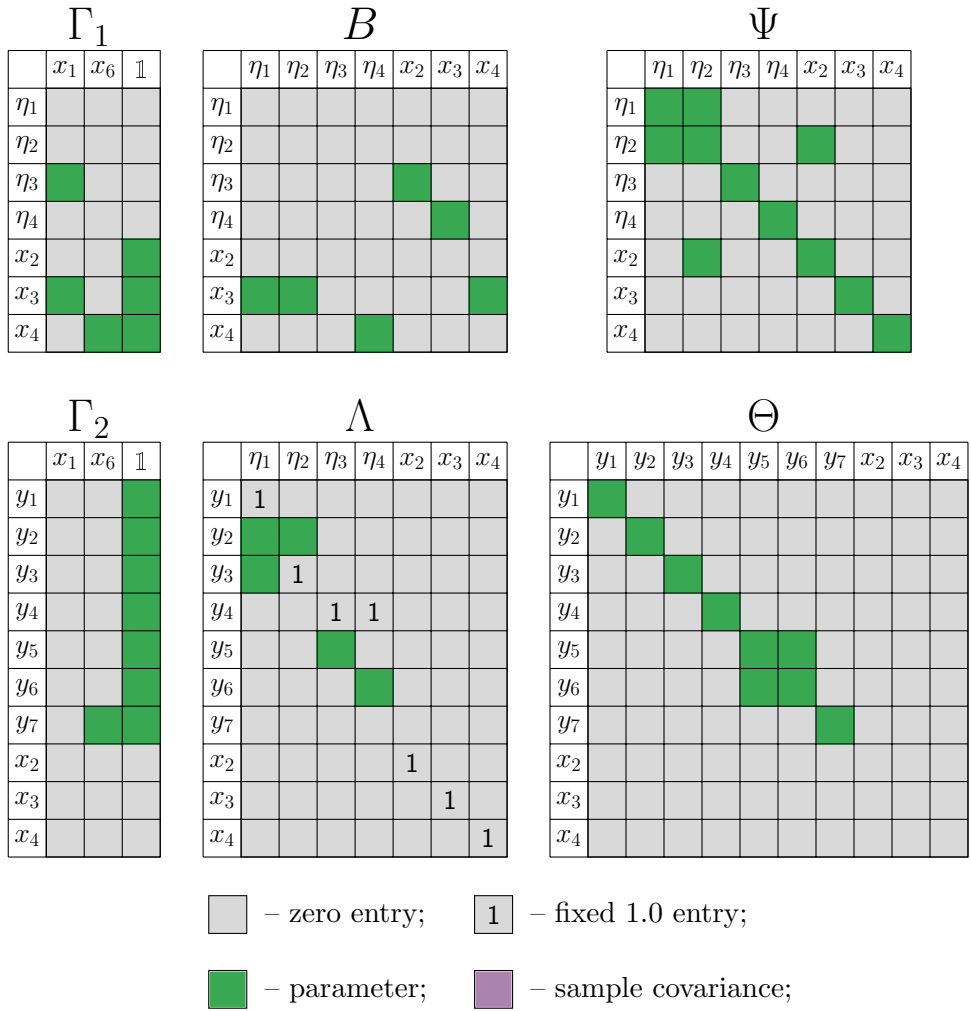

Previously, for Equation~\ref{eq:model}, $\EX[z] = 0$, but for Equation~\ref{eq:modelmeans}: $$\EX[z] = \EX[\Gamma_2 x^{(2)} + \Lambda \omega + \delta] = \EX[\Gamma_2 x^{(2)} + \Lambda C \Gamma_1 x^{(2)} + \Lambda C \epsilon] = \left(\Gamma_2 + \Lambda C \Gamma_1\right) x^{(2)}$$

As for the covariance matrix $\Sigma = \EX[(z - \EX[Z]) (z- \EX[Z])^T]$, it attains the same expression as previously in Equation~\ref{eq:sigma}. 

\subsubsection{Estimation methods}

Although generalization of ULS, GLS and WLS methods to \class{ModelMeans} is straightforward, efficient optimization of those functions is not. At the moment, \class{ModelMeans} provides only methods that assume a normal distribution of endogenous variables and developing an efficient least-squares approach for it is a subject for further research.

Before we proceed to methods, let's denote mean matrix $M$ as 
\begin{equation}\label{eq:mean_matrix}
	M(\theta) = \left(\Gamma_2(\theta) + \Lambda(\theta) C(\theta) \Gamma_1(\theta)\right) X^{(2)},
\end{equation}
where $X^{(2)}$ is a sample matrix of observed exogenous variables. Notice that if $Z$ is a matrix of observations, then $Z - M(\theta)$ is a "centered" (in a sense that is corrected for exogenous effects) matrix of observations. We need $M$ to write down objective functions in a more compact, readable way.
\begin{itemize}
	\item \textbf{Full information maximum likelihood (FIML)/Maximum likelihood (ML)}

	Similar to FIML for \class{Model}. When no missing data is present, FIML reduces to ML:
	
	\begin{equation}\label{eq:ml_modelmeans}
		F(\theta|Z) = tr\left\{(Z - M(\theta))^T \Sigma^{-1}(\theta) (Z - M(\theta)) \right\} + n \ln |\Sigma|		
	\end{equation}
	To estimate model parameters with FIML, supply \code{'FIML'} to argument \code{method} of \code{Model.fit}: \code{model.fit(..., method='FIML')}. It is the default option.
	\item \textbf{Restricted maximum likelihood (REML)}

	REML was proposed by \cite{Thompson:1962} for variance component estimation in linear models. Here, we adapt it to our SEM model.
	
	Let's infer $z$ from \ref{eq:modelmeans}:
	\begin{equation*}
		z = \left(\Gamma_2 + \Lambda C \Gamma_1\right) x^{(2)} + \Lambda C \epsilon + \delta
	\end{equation*}
	For each of $n$ observations holds
	\begin{equation*}
		z_{(i)} = \left(\Gamma_2 + \Lambda C \Gamma_1\right) x_{(i)}^{(2)} + \Lambda C \epsilon_{(i)} + \delta_{(i)},~i=1..n
	\end{equation*}
	We want to get rid of the fixed part $\left(\Gamma_2 + \Lambda C \Gamma_1\right) x_{(i)}^{(2)}$ to lessen the computational burden. That is possible if we find a matrix $P$ such that $X^{(2)} P = 0$ (that's the same as $x_{(i)} P = 0$ for all $i = 1..n$), $P \neq 0$ and the rank of $P$ is as big as possible. Good candidate for $P$ is
	$$P_0 = (I_{n} - {X^{(2)}}^T ({X^{(2)}} {X^{(2)}}^T)^{-1} {X^{(2)}})$$ 
	
	However, for reasons that will be more evident in Section~\ref{sect:modeleffects}, we shall use different $P$. First, we do an eigendecomposition on $P_0$:
	$$P_0 = Q D Q^T,$$
	where $Q$ is an orthogonal matrix of eigenvectors and $D$ is a diagonal matrix of eigenvalues. Let number of non-zero eigenvalues be $r$, (at least $n_{x^{(2)}}$ of eigenvalues are guaranteed to be zero), then a following low-rank decomposition is possible:
	$$D = \sqrt{D} \sqrt{D} = \underbrace{D_1}_{n \times r} \underbrace{D_2}_{r \times n},$$
	
	As $D$ is diagonal, $D_1$ and $D_2$ are just $\sqrt{D}$ with last $r$ columns and rows deleted respectively. Moreover, we can notice that $P_x = {X^{(2)}}^T({X^{(2)}} {X^{(2)}}^T)^{-1}{X^{(2)}}$ is a projection operator as $$P_x^2 = {X^{(2)}}^T({X^{(2)}} {X^{(2)}}^T)^{-1}{X^{(2)}} {X^{(2)}}^T({X^{(2)}} {X^{(2)}}^T)^{-1}{X^{(2)}} = P_x$$ It is known that all eigenvalues of projection operators are either $0$ or $1$. Hence, eigenvalues of $P_0 = I_{n} - P_x = Q Q^T - Q D Q^T = Q (I - D) Q^T$ are either $0$ or $1$ too. Therefore, the only non-zero elements of $D_1$ and $D_2$ are ones. Then, we can chose $P_1$ as $P$:
	$${X^{(2)}} P_0 = 0 \iff G Q D_1 D_2 Q^T * Q = 0 * Q \iff {X^{(2)}} \underbrace{Q D_1}_{P_1} D_2 = 0 \Rightarrow  {X^{(2)}} P_1 = 0$$
	
	\begin{equation}\label{eq:reml_p}
		P_1 = Q D_1
	\end{equation}
	
	Next, we transform data by $P_1 = [p_1, p_2, \dots, p_{n}]^T$:
	$$
	z_{(i)} p_{(i)} = \hat{z}_{(i)} = \Lambda C \epsilon p_{(i)} + \delta_{(i)} p_{(i)}, i=1..r
	$$
	Notice that $cov[\hat{z}] = cov[z] = \Sigma$, as
	$$cov[\hat{z}_{(i)}] = \EX\left[(\Lambda C \epsilon_{(i)} + \delta_{(i)}) \underbrace{p_1 p_1^T}_{1} (\Lambda_C \epsilon_{(i)} + \delta_{(i)})^T\right] = \Sigma$$
	However, a total number of observations "decreases" to $r$. It can be interpreted as taking degrees of freedom that are wasted on mean component into an account.
	
	So, at first we estimate variance components for transformed data as given my a REML likelihood:
	\begin{equation}\label{eq:reml}
		F(\theta|Z P_1) = tr\{P_1^T Z^T \Sigma^{-1}(\theta) Z P_1\} + r \ln |\Sigma| \propto tr\{\widehat{S} \Sigma^{-1}(\theta)\} + \ln |\Sigma|,
	\end{equation}
	where $\widehat{S}$ is a biased sample covariance matrix for transformed data $Z P_1$.
	
	After estimating $\hat{\Sigma}$ from Equation~\ref{eq:reml}, we can estimate mean components in $M(\theta)$ by fitting a likelihood of untransformed data:
	\begin{equation}\label{eq:reml2}
		F(\theta|Z, \hat{\Sigma}) = tr\{M(\theta)^T \hat{\Sigma}^{-1} M(\theta) \}
	\end{equation}
	
	REML has some advantages over ML:
	\begin{enumerate}
		\item It separates one big optimization problem into 2 smaller independent problems, hence practically decreasing computational burden;
		\item ML produces biased estimates of variance components, whereas REML provides unbiased estimates.
	\end{enumerate}
	The latter, however, does not necessarily mean that estimates will have lower mean squared error (MSE). In our tests (see Section~\ref{sect:experiments}), ML outperformed REML in terms of accuracy for exogenous loadings in $\Gamma_1, \Gamma_2$, yet was significantly behind in performance and a bit behind in estimates accuracy for the rest of the parameters.
	
	An interested reader can check literature review on REML given by \cite{Harville:1977}. An extensive coverage of REML is also present in "\textit{Variance components}" book \citep{Searle:1992}.

	To estimate model parameters with REML, supply \code{'REML'} to argument \code{method} of \code{ModelMeans.fit}: \code{model.fit(..., method='REML')}. Note that \code{fit} then returns a pair of optimization results -- one for Equation~\ref{eq:reml} and the other for Equation~\ref{eq:reml2}.
\end{itemize}

\subsubsection{Usage example}

	\consolein
\begin{lstlisting}[style=codeinput]
import semopy
ex = semopy.examples.example_article
desc, data = ex.get_model(), ex.get_data()
m = semopy.ModelMeans(desc)
m.fit(data)
print(m.inspect())
	\end{lstlisting}
	\consoleout
\begin{lstlisting}[style=codeoutput]
lval  op  rval  Estimate        Std. Err    z-value   p-value
0   eta3   ~    x2 -1.146671        0.065539 -17.495993       0.0
1   eta4   ~    x3  1.222544         0.03795  32.214424       0.0
2     x3   ~  eta1  1.831707   566870.787859   0.000003  0.999997
3     x3   ~  eta2 -0.345268        0.143255  -2.410154  0.015946
4     x4   ~  eta4  1.454411        0.040911  35.550814       0.0
5   eta3   ~    x1 -1.344433         0.07717 -17.421604       0.0
6     x3   ~    x1  1.291013        0.075713  17.051312       0.0
7     x4   ~    x6  0.839896        0.068167  12.321151       0.0
8     y7   ~    x6 -0.624860        0.057998 -10.773906       0.0
9     y1   ~  eta1  1.000000               -          -         -
10    y2   ~  eta1  1.212918  1237359.017737   0.000001  0.999999
11    y2   ~  eta2 -0.753647        0.830757   -0.90718  0.364311
12    y3   ~  eta1 -2.416658   1641829.10264  -0.000001  0.999999
13    y3   ~  eta2  1.000000               -          -         -
14    y4   ~  eta3  1.000000               -          -         -
15    y4   ~  eta4  1.000000               -          -         -
16    y5   ~  eta3  1.224545        0.049702  24.637799       0.0
17    y6   ~  eta4  1.443583        0.040787  35.393578       0.0
18    y7   ~    x4 -0.387550        0.014441 -26.837254       0.0
19    x3   ~     1  0.457809        0.101276   4.520432  0.000006
20    x4   ~     1  1.519828        0.097107  15.651084       0.0
21    x2   ~     1 -1.007762        0.062283 -16.180429       0.0
22    y7   ~     1  0.490313         0.06906   7.099783       0.0
23    y1   ~     1 -0.994713        0.080798 -12.311191       0.0
24    y2   ~     1 -0.646721        0.071053  -9.101999       0.0
25    y3   ~     1 -0.507852        0.086607  -5.863842       0.0
26    y4   ~     1 -1.045236        0.113045  -9.246199       0.0
27    y5   ~     1  1.038446        0.113512   9.148353       0.0
28    y6   ~     1  1.001274         0.10067   9.946154       0.0
29    x2  ~~  eta2 -0.091548        0.087398  -1.047491  0.294873
30  eta3  ~~  eta3  0.869757        0.111259   7.817419       0.0
31    x3  ~~    x3  1.131809        0.548416   2.063777  0.039039
32    x4  ~~    x4  1.009375        0.136543    7.39234       0.0
33  eta4  ~~  eta4  0.803528        0.090594   8.869577       0.0
34  eta2  ~~  eta2  5.369147  7739465.375667   0.000001  0.999999
35  eta2  ~~  eta1  2.356964       2097152.0   0.000001  0.999999
36  eta1  ~~  eta1  1.277327        0.451611   2.828376  0.004678
37    y6  ~~    y5 -0.105989        0.101863  -1.040511  0.298103
38    y6  ~~    y6  1.251810         0.15184   8.244278       0.0
39    y7  ~~    y7  1.096517         0.08953  12.247449       0.0
40    y1  ~~    y1  0.681144        0.429614    1.58548  0.112857
41    y3  ~~    y3  0.813174        0.941338   0.863849  0.387671
42    y4  ~~    y4  0.654815        0.111022   5.898087       0.0
43    y2  ~~    y2  0.894850        0.685349   1.305685   0.19166
44    y5  ~~    y5  0.823550        0.144001   5.719065       0.0		
	\end{lstlisting}

Notice new regression of $\mathbbm{1}$ onto observed variables - that's intercepts.

We also check again that the obtained estimates, including intercepts, are sensible:

	\consolein
\begin{lstlisting}[style=codeinput]
import numpy as np
params = ex.get_params()
mape = np.mean(semopy.utils.compare_results(m, params))
print('MAPE: {:.2f}%'.format(mape * 100))
	\end{lstlisting}
	\consoleout
\begin{lstlisting}[style=codeoutput]
MAPE: 20.24%
	\end{lstlisting}

\subsection[ModelEffects]{SEM with random effects: \class{ModelEffects}}\label{sect:modeleffects}

In \class{Model} and \class{ModelMeans} it is assumed that all observations are obtained independently. This assumption, however, is not reasonable in some cases, and some dependence structure between data samples is present. One way to model this dependence is to assume an extra random term $U$ (a \textit{random effect})with some covariance matrix $K$ of shape $n \times n$. $K$ matrix is not estimable in a general case, as the number of parameters is $\frac{n(n + 1)}{2} \ge n$, therefore some kind of a prior assumption on the $K$ structure should be provided. The best-case scenario is when we already know $K$ matrix up to some scalar multiplier -- for instance, this scenario often arises in bioinformatics and genomic studies, where $K$ plays a role of genomic relatedness matrix and is computed from genotypes \citep{Vanraden:2008}. That's the case \class{ModelEffects} is concerned with.

\begin{equation}\label{eq:modeleffects}
	\begin{cases}
		\begin{bmatrix}
			H \\ X^{(1)}
		\end{bmatrix} = W = \Gamma_1 X^{(2)} + B W + E, ~~~~~~~~E \sim \mathcal{MN}(0, \Psi, I_n) \\
		\begin{bmatrix}
			Y \\ X^{(1)}
		\end{bmatrix} = Z = \Gamma_2 X^{(2)} + \Lambda W + \Delta + U, ~~~\Delta \sim \mathcal{MN}(0, \Theta, I_n), U \sim \mathcal{MN}(0, D, K) 
	\end{cases}
\end{equation}

Previously, we have employed vectors to describe our models, but introduction of dependence across observations makes matrix notation much more convenient. In Equation~\ref{eq:modeleffects}, $H$ is a $n_\eta \times n$ matrix of latent variables, $X^{(1)}$ and $X^{(2)}$ are matrices of endogenous and exogenous observable variables of shapes $n_{x^{(1)}} \times n$ and $n_{x^{(2)}} \times n$ respectively, $D$ is a covariance matrix of random effect components $U$ that incorporates scale with respect to each of the observable variables and $K$ is a fixed covariance across-observations matrix. $\mathcal{MN}$ is a notation for matrix-variate normal distribution \citep{Gupta:2000}. Actually, it is just a synonym for a form of multivariate normal distribution: 
\begin{equation}\label{eq:matnormvec}
	\Omega \sim \mathcal{MN}(M, U, V) \iff vec(\Omega) \sim \mathcal{N}(vec(M), V \otimes U),
\end{equation}
where $vec$ is a vectorization operator and $\otimes$ is a Kronecker product symbol. It follows that
\begin{enumerate}
	\item $\EX[\Omega] ~~~~= M$;
	\item $cov[\Omega] ~~= \EX\left[(\Omega - \EX[\Omega])(X - \EX[X])^T \right] = tr\{V\} U$;
	\item $cov[\Omega^T] = \EX\left[(\Omega - \EX[\Omega])^T(X - \EX[X]) \right] = tr\{U\} V$;
	\item $U$ and $V$ are not estimable as $V \otimes U = c V \otimes \frac{1}{c}U$, where $c$ is an arbitrary scalar.
\end{enumerate}
The latter, however, is not bad news, as it will be shown next in Section~\ref{sect:modeleffects_estimation}. Using the first 3 properties we infer mean and covariances of $Z$:

\begin{equation}\label{eq:modeleffects_m}
	M = \EX[Z] = \left(\Lambda C \Gamma_1 + \Gamma_2 \right) X^{(2)}
\end{equation}
Covariance across rows (i.e. covariance between observable variables):
\begin{equation}\label{eq:modeleffects_L}
	L = cov[Z] = n \Sigma + tr\{K\} D
\end{equation}

Covariance across columns (i.e. covariance between observations):
\begin{equation}\label{eq:modeleffects_T}
	T = cov[Z^T] = tr\{\Sigma\}I_n + tr\{D\} K
\end{equation}

\begin{leftbar}
	There are multiple options to parameterize $D$ matrix in \textbf{semopy}:
	\begin{enumerate}
		\item Full parameterization: pass \code{d_mode='full'} argument to the constructor of \class{ModelMeans};
		\item Only diagonal parameterization: pass \code{d_mode='diag'} argument to the constructor of \class{ModelMeans} (the default);
		\item Single scale parameter $D = \sigma^2 I_{n_z}$: pass \code{d_mode='scale'} argument to the constructor of \class{ModelMeans}.
		\item Custom parameterization: use any of the above parameterizations as a base and then introduce custom parameters using \textbf{semopy} syntax (see Section~\ref{sect:syntax}).
	\end{enumerate}
\end{leftbar}

\subsubsection{Estimation methods}\label{sect:modeleffects_estimation}

\begin{itemize}
	\item \textbf{Matrix-variate maximum likelihood/ML}
	
	Negative loglikelihood for variable $Z \sim \mathcal{MN}(M(\theta), \widehat{L}(\theta), \widehat{T}(\theta))$:
	\begin{equation}\label{eq:mvn_density}
		f(\theta|Z) = tr\left\{\widehat{T}^{-1} (Z - M(\theta))^T \widehat{L}^{-1} (Z - M(\theta))\right\} + n \ln |\widehat{L}| + n_{z} \ln |\widehat{T}|
	\end{equation}
	Note that if $Z$ from Equation~\ref{eq:modeleffects} follows matrix-variate normal distribution $\mathcal{MN}(M, \widehat{L}, \widehat{T})$, it doesn't hold that $L = \widehat{L}$ or $T = \widehat{T}$. Hence, we cant' just substitute $L$ (\ref{eq:modeleffects_L}) and $T$ (\ref{eq:modeleffects_T}) to Equation~\ref{eq:mvn_density} and minimize it. As follows from properties 2 and 3 of $\mathcal{MN}$ above, 
	\begin{equation}\label{eq:lt}
		\begin{split}
			L = tr\{\widehat{T}\} \widehat{L} \iff \widehat{L} = \frac{1}{tr\{\widehat{T}\}} L\\
			T = tr\{\widehat{L}\} \widehat{T} \iff \widehat{T} = \frac{1}{tr\{\widehat{L}\}} T
		\end{split}
	\end{equation}
	From expression for $L$ we can infer $tr\{\widehat{L}\}$:
	$$tr\{\widehat{L}\} = \frac{tr\{L\}}{tr\{\widehat{T}\}}$$
	Then:
	\begin{equation}\label{eq:mvn_lt}
		\begin{split}
			\widehat{L} = \frac{1}{tr\{\widehat{T}\}} L\\
			\widehat{T} =\frac{tr\{\widehat{T}\}}{tr\{L\}} T
		\end{split}
	\end{equation}
	Next, we substitute Equation~\ref{eq:mvn_lt} to Equation~\ref{eq:mvn_density}:
	\begin{equation*}
		\begin{split}
			\underbrace{tr\left\{\left(\frac{tr\{\widehat{T}\}}{tr\{L\}} T\right)^{-1} (Z - M)^T \left(\frac{1}{tr\{\widehat{T}\}} L\right)^{-1} (Z - M)\right\}}_{A_1} + \underbrace{n \ln \left|{\frac{1}{tr\{\widehat{T}\}} L}\right| + n_{z} \ln \left|\frac{tr\{\widehat{T}\}}{tr\{L\}} T\right|}_{A_2}
		\end{split}
	\end{equation*}
	There is only one thing that we can't compute: $tr\{\widehat{T}\}$. Fortunately, it cancels out in both terms $A_1$ and $A_2$:
	$$A_1 = tr\{L\} tr\left\{T^{-1} (Z - M)^T L^{-1} (Z - M)\right\}$$
	In $A_2$, as $\left|\frac{1}{tr\{T\}}L\right| = \left(\frac{1}{tr\{T\}} \right)^{n_z} |L|$ and $\left|\frac{tr\{\widehat{T}\}}{tr\{L\}} T\right| = \left(\frac{tr\{\widehat{T}\}}{tr\{L\}}\right)^n |T|$:
	\begin{equation*} 
		\begin{split}
			A_2 = n \ln \left[\left(\frac{1}{tr\{T\}} \right)^{n_z} |L| \right] + n_z \ln \left[\left(\frac{tr\{\widehat{T}\}}{tr\{L\}}\right)^n |T| \right] = \\ = n ln |L| + n_z ln |T| - n n_z \ln tr\{L\}
		\end{split}
	\end{equation*}
	Therefore, the negative likelihood to minimize attains the form:
	\begin{equation}\label{eq:mvn_modelmeans}
		\begin{split}
			F(\theta|Z) = tr\{L(\theta)\} tr\left\{T^{-1}(\theta) (Z - M(\theta))^T L^{-1}(\theta) (Z - M(\theta))\right\} +\\+ n ln |L(\theta)| + n_z ln |T(\theta)| - n n_z \ln tr\{L(\theta)\}
		\end{split}
	\end{equation}
	It may appear that there is an asymmetry in Equation~\ref{eq:mvn_modelmeans} that arises from $tr\{L(\theta)\}$. Indeed, we could have inferred from Equation~\ref{eq:lt} $tr\{\widehat{T}(\theta)\}$ instead and then we would get seemingly different likelihood from of the Equation~\ref{eq:mvn_modelmeans}. However, they are actually the same as $tr\{L\} = tr\{T\}$. We've chosen to write down the likelihood function in terms of $tr\{L\}$ because it is faster to compute.
	
	This is the default method. If that changes, one can supply \code{'ML'} to argument \code{method} of \\ \code{ModelEffects.fit}: \code{model.fit(..., method='ML')}.
	
	\item \textbf{Restricted maximum likelihood (REML)}
	
	REML for \class{ModelEffects} follows the same logic as for \class{ModelMeans}. First, we transform data by $P_1$ (\ref{eq:reml_p}):
	$$\widehat{Z} = Z P_1 = \Lambda C \widehat{E} + \widehat{\Delta} + \widehat{U},$$ 
	where $\widehat{E} \sim \mathcal{MN}(0, \Psi, I_r), \widehat{\Delta} \sim \mathcal{MN}(0, \Theta, I_r), \widehat{U} \sim \mathcal{MN}(0, D, \underbrace{P_1^T K P_1}_{\widehat{K}})$.
	Second, we show that
	$$L_{R} = cov[\widehat{Z}] = r \Sigma + tr\{\widehat{K}\} D,$$
	and
	$$T_{R} = cov[\widehat{Z}^T] = tr\{\Sigma\}I_r + tr\{D\}\widehat{K}$$
	Third, we minimize the REML function:
	\begin{equation}\label{eq:modeleffects_reml}
		\begin{split}
			F_1(\theta|\widehat{Z}) = tr\{L_R(\theta)\} tr\left\{T_R^{-1}(\theta) \widehat{Z}^T L_R^{-1}(\theta) \widehat{Z}\right\} +\\+ r ln |L_R(\theta)| + n_z ln |T_R(\theta)| - r n_z \ln tr\{L(\theta)\}
		\end{split}
	\end{equation}	
	Fourth, after obtaining parameter estimates in covariance structures, we use them to obtain estimates for non-transformed $L$ and $T$ -- $\widetilde{L}$ and $\widetilde{T}$ respectively. Finally, we solve a kind of least squares problems to estimate parameters in $M(\theta)$:
	$$F_2(\theta|Z, \widetilde{L}, \widetilde{T}) = tr\{\widetilde{T}^{-1}(Z - M(\theta)^T )\widetilde{L}^{-1} (Z-M(\theta))\}$$
	Here, we can significantly decrease the cost of evaluating this function by transforming it to a conventional GLS problem as it was in Equation~\ref{eq:reml2}. To do it, we using Cholesky decomposition on $\widetilde{T}^{-1} = R^{-1} R^{-T}$:
	\begin{equation*}
		\begin{split}
			F_2(\theta|Z, \widetilde{L}, \widetilde{T}) = tr\{\widetilde{T}^{-1}(Z - M(\theta))^T L^{-1} (Z-M(\theta))\} = \\ = tr\{R^{-1} R^{-T}(Z - M(\theta))^T \widehat{L}^{-1} (Z-M(\theta))\}  =\\=
			tr\{(Z R^{-1} - M(\theta)R^{-1})^T \widehat{L}^{-1} (Z R^{-1}-M(\theta)R^{-1})\} 
		\end{split}
	\end{equation*}
	We replace $Z R^{-1}$ with $\widetilde{Z}$ and $M(\theta)R^{-1} = (\Lambda C \Gamma_1 + \Gamma_2)X^{(2)} X^{(2)} R^{-1}$ with $\widetilde{M}(\theta)$ and get the GLS equation: 
	\begin{equation}\label{eq:modeleffects_reml2}
		F_2(\theta|\widetilde{Z}, \widetilde{L}) - tr\{(\widetilde{Z} - \widetilde{M}(\theta))^T \widetilde{L}^{-1} (\widetilde{Z} - \widetilde{M}(\theta))\}
	\end{equation}
	
\end{itemize}

\subsubsection{Diagonalization trick}\label{sect:diag}
Both ML and REML objective functions require to compute inverse and determinant of $T(\theta)$ matrix at each evaluation. The complexity of those operations operation is $O(n^3)$, which is troublesome in cases when a number of observations $n$ is large or when high performance is demanded (in GWAS, for instance). We, similarly to \textbf{FaST-LMM} \citep{Lippert:2011}, use the fact that matrices matrices $I_n$ and $K$ from Equation~\ref{eq:modeleffects_T} are simultaneously diagonalizable, i.e.
\begin{equation*}
	\begin{split}
		T = tr\{\Sigma\}I_n + tr\{D\} K = tr\{\Sigma\}I_n + tr\{D\} Q S Q^T = Q(tr\{\Sigma\}Q^T Q + tr\{D\} S)Q^T =\\=  Q(tr\{\Sigma\}I_n + tr\{D\} S)Q^T,
	\end{split}
\end{equation*}
where $S$ is a diagonal matrix of eigenvalues. If we rotate data $Z$ by $Q^T$, covariance across rows remains unchanged
$$cov[Z Q^T] = \EX\left[(Z - \EX[Z])\underbrace{Q Q^T}_{I_n}(Z - \EX[Z])^T\right] = cov[Z] = L,$$
and covariance across columns transforms to
$$cov[Q Z^T] = Q^T Q(tr\{\Sigma\}I_n + tr\{D\} S)Q^T Q = tr\{\Sigma\}I_n + tr\{D\} S,$$
which is a diagonal matrix, hence the complexity of inversion and computing determinant is $O(n)$.

\subsubsection{Groups}
In \class{ModelEfects}, user doesn't have to provide an $n \times n$ $K$ matrix. Instead, one can pass a $p \times p$ covariance between groups matrix $V$ , provided that data is clustered into $p$ distinct groups and labeled accordingly. Then, $K$ is obtained as 
\begin{equation}\label{eq:zkz}
	K = Z^T V Z,
\end{equation}

where $Z$ is a $n \times p$ design matrix that assigns each of the $n$ observations to some of the $p$ groups.

\begin{leftbar}
	Furthermore, no $V$ can be provided at all. In that case, \textbf{semopy} assumes $V = I_p$, which should be roughly equivalent to centering data by groups. The latter can be done automatically in \class{Model} or \class{ModelMeans} by passing \code{groups} argument to the \code{fit} method
\end{leftbar}

\subsubsection{Usage example}
Like previously, we start by loading data and model from the \code{examples} submodule, but this time we pass \code{random_effects=1} argument to the \code{get_data} function to return the same data as before, but "spoiled" with unknown random effects, alongside with covariance matrix for those random effects:

	\consolein
\begin{lstlisting}[style=codeinput]
import semopy
ex = semopy.examples.example_article
desc = ex.get_model()
data, k = ex.get_data(random_effects=1)
print(data.shape, '\n', data.head())
	\end{lstlisting}
	\consoleout
\begin{lstlisting}[style=codeoutput]
(300, 13) 
y1        y2        y3        y4     ...      x6         group
0  0.570959 -0.014071  0.467279 -0.815086  ... -0.025494      0
1 -2.292133 -0.907395 -1.894549  0.939781  ... -0.217159      1
2  0.129772 -0.591307  0.112034 -0.157425  ...  0.058370      2
3 -1.010035 -0.863170 -1.924978 -0.921666  ... -0.603859      3
4  0.489841 -2.039194  0.264125  0.563737  ... -0.343206      4

[5 rows x 13 columns]
	\end{lstlisting}
	\consolein
\begin{lstlisting}[style=codeinput]
print(k)
	\end{lstlisting}
	\consoleout
\begin{lstlisting}[style=codeoutput]
group       0         1         2    ...       297       298       299
group                                ...                              
0      0.597172  0.193626  0.054028  ...  0.509318 -0.056519  0.073060
1      0.193626  1.558588 -0.246282  ...  0.590645  0.318278 -0.012598
2      0.054028 -0.246282  0.561246  ... -0.147389  0.015261 -0.206373
3      0.656677  0.143660 -0.529918  ...  0.409124  0.001428 -0.295317
4     -0.718798 -0.211845  0.048397  ... -0.498914  0.326037  0.183095
...       ...       ...  ...       ...       ...       ...
295   -0.015002  0.151484  0.082362  ... -0.028149  0.040969 -0.046783
296   -0.142605 -0.285580  0.190506  ... -0.552173 -0.126101 -0.523155
297    0.509318  0.590645 -0.147389  ...  0.770882 -0.078281  0.364578
298   -0.056519  0.318278  0.015261  ... -0.078281  1.369580 -0.319364
299    0.073060 -0.012598 -0.206373  ...  0.364578 -0.319364  0.689287

[300 rows x 300 columns]
	\end{lstlisting}

An extra column \code{'group'} is provided with the data frame. In this case, it is just equal to the data index as a number of "groups" is equal to a number of observations. Each unique group entry must have a corresponding row and column in \code{'k'} data frame.

Then, we fit \class{ModelEffects} instance to data:

	\consolein
\begin{lstlisting}[style=codeinput]
m = semopy.ModelEffects(desc,)
m.fit(data, group='group', k=k,)
	\end{lstlisting}

where \code{group} argument is a name of a column in dataframe that assigns individuals to groups.
\newpage
Finally, we verify the results  on true parameter values:

	\consolein
\begin{lstlisting}[style=codeinput]
import numpy as np
params = ex.get_params()
mape = np.mean(semopy.utils.compare_results(m, params))
print('MAPE: {:.2f}%'.format(mape * 100))
	\end{lstlisting}
	\consoleout
\begin{lstlisting}[style=codeoutput]
MAPE: 20.69%
	\end{lstlisting}

Note that fitting \class{ModelMeans} (or {Model}) to this data will provide worse results:

	\consolein
\begin{lstlisting}[style=codeinput]
m = semopy.ModelMeans(desc)
m.fit(data)
mape = np.mean(semopy.utils.compare_results(m, params))
print('MAPE: {:.2f}%'.format(mape * 100))
	\end{lstlisting}
	\consoleout
\begin{lstlisting}[style=codeoutput]
MAPE: 57.85%
	\end{lstlisting}

It will be demonstrated in Section~\ref{sect:experiments} that not taking random effects into an account can lead to arbitrary worse results.

\subsection[ModelGeneralizedEffects]{SEM with multiple generalized random effects: \class{ModelGeneralizedEffects}}\label{sect:modelgeneralizedeffects}

\class{ModelEffects} lets us account for only one random effect, which is sufficient in many cases. However, some studies might benefit from the inclusion of multiple effects. For example, in GWAS, one might include geographical information as a random effect alongside genetic relatedness. 
Furthermore, many population structure models don't assume a known $K$ matrix, but merely assume a structure of it; the exact form is supposed to be discovered. Moving average of first order (MA(1)) model is an example:
$$u_t = \alpha e_{t - 1} + e_t, ~~\EX[e_t] = 0, ~\EX[e_t e_{t-1}] = 0$$
Here, structure of covariance matrix $K$ is determined by an autocorrelation function of $u_t$:
$$K_{ij} = cov(u_{t_{i}}, u_{t_{j}}) = \begin{cases}
	1, ~~~~~~t_{i} - t_{j} = 0 \\
	\frac{\alpha}{1 + \alpha^2}, ~|t_{i} - t_{j}| = 1 \\ 0, ~~~~~~|t_i - t_j| > 1 \end{cases}, $$
$\alpha$ coefficient, however, is unknown.  

Hence, we propose \class{ModelGeneralizedEffects}, as defined by Equation~\ref{eq:modelgeneralizedeffects}:

\begin{equation}\label{eq:modelgeneralizedeffects}
	\begin{cases}
		W = \Gamma_1 X^{(2)} + B W + E, ~~~~~~~~~~~~~~~E \sim \mathcal{MN}(0, \Psi, I_n) \\
		Z = \Gamma_2 X^{(2)} + \Lambda W + \Delta + \sum_{i=1}^p U_{(i)}, \Delta \sim \mathcal{MN}(0, \Theta, I_n), U_{(i)} \sim \mathcal{MN}(0, D_{(i)}, K_{(i)}),
	\end{cases}
\end{equation}
where $p$ is a number of random effects and $\{K_{(i)}\}_{i=1..p}$ matrices are parameterized.

Similarly to \class{ModelEffects}, we infer expectation of model-implied data matrix $M$, covariance across rows (between variables) $L$ and covariance across columns (between observations) $T$:

\begin{flalign}
	M = &\EX[Z] =  \left(\Lambda C \Gamma_1 + \Gamma_2 \right) X^{(2)}\\
	L = & cov[Z] =  n \Sigma + \sum_{i=1}^p tr\{K_{(i)}\} D_{(i)}\\
	T = & cov[Z^T] =  tr\{\Sigma\} I_n + \sum_{i=1}^p tr\{D_{(i)}\} K_{(i)}
\end{flalign}

\subsubsection{Estimation methods}
At the moment, only one optimization method is implemented for \\ \code{ModelGeneralizedEffects}:
\begin{itemize}
	\item \textbf{Matrix-variate normal maximum likelihood (ML)}
\end{itemize}

REML is likely to appear in future versions.

\subsubsection{Random effects models}

\class{ModelGeneralizedEffects} constructor accepts a list of different effect model instances, each element in the list describes an $U_{(i)}$ from the Equation~\ref{eq:modelgeneralizedeffects}. All available effect models require a \code{columns} argument passed to the constructor that contains a list (or just a string if applicable) of column names with information on a population structure. For example, it can be group labels or times when a sample was observed. Next, we describe effect models that are available in the \textbf{semopy} as of version \textit{2.2.2}.

\begin{itemize}
	\item \textbf{Fixed effects model: \class{EffectStatic}}
	
	\class{EffectStatic} assumes a fixed covariance matrix $K$ just like in \class{ModelEffects}.
	
	\textit{Model-specific arguments}:
	\begin{itemize}
		\item \code{k}: $K$ matrix; indices and columns must have group labels.
	\end{itemize}

	\item \textbf{Fully parameterized model: \class{EffectFree}}
	
	With \class{EffectFree}, entire $V$ from Equation~\ref{eq:zkz} is estimated. This problem might be feasible if $p <<< n$.

	\textit{Model-specific arguments}:
	\begin{itemize}
		\item \code{diagonal}: if \code{True}, then $V$ is constrained to be a diagonal matrix. The default is \code{False}.
		\item \code{correlation}: if \code{True}, then $V$ is constrained to be a correlation matrix. The default is \code{False}.
	\end{itemize}
	
	\item \textbf{Moving average model: \class{EffectMA}}
	
	Moving average is a classical model in time series analysis, where the variable at present time is explained by a weighted sum of random independent errors from past times. 
	
	The first-order moving average model, denoted by MA(1), is:
	\begin{equation*}
		u_t =  \alpha \epsilon_{t-1} + \epsilon_t,
	\end{equation*}
	where $\{\epsilon_t\}_{t=1..T}$ are independent and centered. Autocorrelation function for MA(1) model:
	\begin{equation*}
		cor(u_{t_{i}}, u_{t_{j}}) = \begin{cases}
			1, ~~~~~~|t_{i} - t_{j}| = 0 \\
			\frac{\alpha}{1 + \alpha^2}, ~|t_{i} - t_{j}| = 1 \\ 0, ~~~~~~|t_i - t_j| > 1 \end{cases},
	\end{equation*}
	
	The second-order moving average model, denoted by MA(2), is:
	
	\begin{equation*}
		u_t =  \alpha_2 \epsilon_{t-2} + \alpha_1 \epsilon_{t-1} + \epsilon_t,
	\end{equation*}
	where $\epsilon_t, \epsilon_{t-1}, \epsilon_{t-2}$ are independent and centered. Autocorrelation function for MA(2) model:
	\begin{equation*}
		cor(u_{t_{i}}, u_{t_{j}}) = \begin{cases}
			1, ~~~~~~~~~~|t_{i} - t_{j}| = 0 \\
			\frac{\alpha_1 + \alpha_1 \alpha_2}{1 + \alpha_1^2 + \alpha_2^2},~|t_i-t_j| = 1\\
			\frac{\alpha_2}{1 + \alpha_1^2 + \alpha_2^2}, ~|t_{i} - t_{j}| = 2 \\ 0, ~~~~~~~~~~|t_i - t_j| > 2 \end{cases},
	\end{equation*}
	And general case of p-th order moving average model, denote by MA(p):
	\begin{equation*}
		u_t =  \epsilon_t + \sum_{i=1}^p \alpha_i \epsilon_{t - i} = \sum_{i=0}^p \alpha_i \epsilon_{t-i}
	\end{equation*}
	where $\{\epsilon_i\}_{i=t..t-p}$ are independent and centered, $\alpha_0 = 1$. The autocorrelation function for MA(p) model:
	\begin{equation*}
		cor(u_{t_{i}}, u_{t_{j}}) = \begin{cases}
			\frac{\sum_{i=0}^{p - k}\alpha_i \alpha_{i + p}}{\sum_{i=0}^p \alpha_i^2},~~|t_i - t_j| = k, k \leq p \\ 0, ~~~~~~~~~~~~~~~~|t_i - t_j| > p \end{cases},
	\end{equation*}
	
	\textit{Model-specific arguments}:
	\begin{itemize}
		\item \code{order}: Order of moving-average model. The default is 1.
		\item \code{dt_bounds}: Optional list of tuples that contain bounding boxes for different lags of autocorrelation function, i.e. if it is \\ \code{[(0, 1), (1, 3)]}, then all observations with time difference in interval $[0, 1)$ are assumed to have lag $0$ (similarly, if the difference is in $[1, 3)$, then the lag is $1$). In other words, if the time difference is in $k$-th interval, then the lag is $k$. It provides a mapping from a continuous time domain to a discrete domain of lags. The default is \code{[(0, 1), (1, 2), ... (p, p + 1)]}.
		\item \code{params}: Optional list of $\{\alpha_i\}_{i=1..p}$ parameters. If it is not provided, then parameters are estimated (that's the default approach).
	\end{itemize}
	
	\item \textbf{Autoregressive model: \class{EffectAR}}

	Autoregressive model  of the first order AR(1) is defined as
	$$u_t = \alpha u_{t-1} + \epsilon_t$$
	Its autocorrelation function is
	\begin{equation}\label{eq:ar1}
		cov(u_{t_i}, u_{t_j}) = \alpha^{k}, k = |t_{i} - t_{j}|
	\end{equation}
	Although higher-order AR(q) models are possible, at the moment \textbf{semopy} supports only first-order AR models.

	\textit{Model-specific arguments}:
	\begin{itemize}
		\item \code{dt}: $k$ from Equation~\ref{eq:ar1} is calculated as $\left[\frac{|t_{i} - t_{j}|}{dt}\right]$. The default is 1.
		\item \code{param}: Same as \code{params} in \class{EffectMA}.
	\end{itemize}
	
	\item \textbf{Spatial model: \class{EffectMatern}}

	For the task of modeling spatial dependencies, Matern autocovariance function has been proposed by \cite{Stein:1999}:
	\begin{equation*}
		cov(u_i, u_j)= C_\nu(d_{i j}) \propto \frac{2^{1-\nu}}{\Gamma(\nu)} \left(\sqrt{2\nu} \frac{d_{ij}}{\rho}\right)^{\nu} K_{\nu}\left(\sqrt{2\nu} \frac{d_{ij}}{\rho}\right),
	\end{equation*}
	where $d_{ij}$ is a Minkowski distance between $i$-th and $j$-th individual, $\Gamma$ is the gamma function, $K_\nu$ is the modified Bessel function of the second kind, $\rho$ and $\nu$ are positive hyperparameters (though $\rho$ can be turned into an active parameter).
	The special case of Matern covariance is when $\nu \rightarrow \infty$:
	\begin{equation*}
		C_{\infty} \propto exp\left({-\frac{d_{ij}^2}{2\rho^2}}\right),
	\end{equation*}
	which is Gaussian radial-basis function. 
	
	The other special case is absolute exponential kernel that is achieved when $\nu = \frac{1}{2}$:
	\begin{equation*}
		C_{\frac{1}{2}} \propto exp\left({-\frac{d_{ij}}{\rho}}\right)
	\end{equation*}
	
	\textit{Model-specific arguments}:
	\begin{itemize}
		\item \code{nu}: The $\nu$ hyperparameter. The default is $\infty$.
		\item \code{rho}: The $\rho$ parameter. The default is $1$.
		\item \code{active}: If \code{True}, then $\rho$ is an active parameter that is optimized. Otherwise, it is fixed. The default is \code{False}.
	\end{itemize}
	
	\begin{leftbar}
		By default (i.e. if \code{active=False}), \class{EffectMatern} has no active parameters, hence, it can be seen as a special case of \class{EffectStatic}. One might wish to enjoy the higher performance of \class{ModelEffects} if only one random effect is needed. If that's the case, a user can just call the \code{calc_zkz} method after instantiating and loading data into the effect, and then pass it as a $K$ matrix to \class{ModelEffects}.
	\end{leftbar}
	\item \textbf{Custom kernel model: \class{EffectKernel}}
	\textbf{semopy} provides a general effect model that is compatible with kernels of \textbf{sklearn} \citep{Pedregosa:2011} from the \code{gaussian_process.kernels} submodule. In fact, \class{EffectMatern} is a convenient specialization of \class{EffectKernel} for the Matern kernel.
	
	\textit{Model-specific arguments}:
	\begin{itemize}
		\item \code{kernel}: Any \code{sklearn.gaussian_process.kernels} compatible class.
		\item \code{params}: Dictionary of kernel parameters.
		\item \code{active}: If \code{True}, then parameters are active and estimated. The default is \code{False}.
	\end{itemize}
	
	\item \textbf{Custom random effects models}

	Do the models above cover all possible dependence structures? Probably not. The user is invited to build his own custom effect models when necessary. All effect classes in \textbf{semopy} are derived from the base class \class{EffectBase}. A developer has to implement the following methods:
	\begin{itemize}
		\item \code{load}
		
		A method that loads data and builds model-specific parameters vector of length $p_m$.
		
		\item \code{calc_k}
		
		Method that returns $n \times n$ $K$ matrix.
		
		\item \code{grad_k}
		
		A method that returns a list of length $p_m$ where $i$-th element is a derivative of $K$ matrix with respect to the $i$-th parameter.
		
		\item (Optional) \code{get_bounds}
		
		A method that returns a list of length $p_m$ where $i$-th element is a bounding box of the $i$-th parameter. This method is actually already implemented in the base class and if not overridden, then models parameters are unconstrained.
	\end{itemize}
	We advise a prospective developer to first take a look at source code of the demo class \class{EffectBlank} (an effect with $K = const = I_n$ and $0$ parameters) before proceeding to any of the "useful" effect models.
\end{itemize}

\subsubsection{Usage example}

	We are still working with the same model, but this time we load data that is "spoiled" with a mix of 2 random effects with different known $K$ matrices and an MA(2) process:
	\consolein
\begin{lstlisting}[style=codeinput]
import semopy
ex = semopy.examples.example_article
desc = ex.get_model()
data, (k1, k2) = ex.get_data(random_effects=2, moving_average=True)
	\end{lstlisting}
	Then, we build a list of \textit{effect models}:
	\consolein
\begin{lstlisting}[style=codeinput]
from semopy.effects import EffectStatic, EffectMA
ef = [EffectStatic('group', k1), EffectStatic('group', k2),
      EffectMA('time', 2)]
	\end{lstlisting}
	Finally, we build an instance of \class{ModelGeneralizedEffects} using \code{desc} and list of effect models \code{ef}:
	\consolein
\begin{lstlisting}[style=codeinput]
m = semopy.ModelGeneralizedEffects(desc, ef)
r = m.fit(data)
	\end{lstlisting}
	Let's see if parameter estimates are any close to the true ones:
	\consolein
\begin{lstlisting}[style=codeinput]
import numpy as np
params = ex.get_params()
mape = np.mean(semopy.utils.compare_results(m, params))
print('MAPE: {:.2f}%'.format(mape * 100))
	\end{lstlisting}
	\consoleout
\begin{lstlisting}[style=codeoutput]
MAPE: 34.63%
	\end{lstlisting}
	A reader can verify that exclusion of any of the random effect models from \code{ef} will produce estimates with significantly higher MAPE.

\subsection{Comparison and limitations}

\begin{table}[!t]
	\centering
	\begin{adjustbox}{max width=\linewidth}
		\begin{tabular}{p{2cm}|p{2cm}|p{2.1cm}|p{3cm}|p{3cm}}
			Model  name& \class{Model} & \class{ModelMeans} & \class{ModelEffects} & \class{ModelGeneralizedEffects}\\
			\hline
			Performance ranking & 1st & 2nd & 3rd & 4th \\
			\hline
			Non-normal continuous and ordinal variables & ULS, GLS, WLS, DWLS; Polychoric and polyserial correlations & Only exogenous variables, but any method&Only exogenous variables, but any method & Only exogenous variables, but any method
			\\
			\hline
			Intercepts & No; Possible to estimate intercepts after fitting model& Yes &  Yes & Yes 
			\\
			\hline
			Missing data & Pairwise deletions when computing $S$; FIML & FIML & No; Only group labels for random effects & No; Only group labels for random effects 
			\\
			\hline
			Population structure & Limited: group-wise means & Limited: group-wise means & Random effects & Random effects
			\\
			\hline
			Maximal number of random effects& 0 & 0 & 1 & $\infty$ \\
			\hline
			Generalized random effects (time series, spatial analysis, etc) & No & No & No & Yes
		\end{tabular}
	\end{adjustbox}
	\caption{Comparison of models.}
	\label{table:models}
\end{table}
\class{ModelGeneralizedEffects} can be seen as the most general model and all other classes can be interpreted as subsets of it. One cay say that \class{Model} $\subset$ \class{ModelMeans} $\subset$ \class{ModelEffects} $\subset$ \class{ModelGeneralizedEffects}. However, generality comes at price of lower performance, and at the moment, some features present for \class{Model} are lacking for other classes. A short review of \textbf{semopy} models can be seen at Table~\ref{table:models}.

Also, there are often identifiability issues of some variance components in \class{ModelEffects} and \\ \class{ModelGeneralizedEffects}. \cite{Wang:2013} showed that the exact conditions for variance components being non-identifiable in univariate LMMs depend on the structure of $K$ matrix, however, no research has been conducted in this area in the case of SEM and parameterized $K$ matrices. Yet, it appears that non-identifiable parameters always reside in $\Theta$ and $\{D_{(i)}\}_{i=1..p}$ matrices, the sum of those variances is always the same, and all other parameters are estimated correctly (in a sense that there is always a local minimum). Therefore, we don't consider the identifiability of random effects variance components to be an issue.

\section{Syntax}\label{sect:syntax}
To specify SEM models, \textbf{semopy} uses syntax that borrows from \textit{formula} syntax known to \texttt{R} userbase for describing regression models. \textbf{lavaan} package, for instance, enjoys the similar approach. SEM models are specified as a multiline string, where each line either constitutes a relationship between variables with \textit{operators}, or explains a properties of variables or parameters by means of \textit{operation} commands.

\subsection{Operators}

Three core operators are supported across all of the models:

\begin{itemize}
	\item "\code{~}": regression operator.
	
	For example, assume the following linear equation:
	
	\begin{equation*}
		y = a_1 x_1 + a_2 x_2 + a_3 x_3 + \epsilon,
	\end{equation*}
	where $y$ and $x_1, x_2, x_3$ are observable variables, and $a_1, a_2, a_3$ are regression coefficients. In \textbf{semopy} syntax it can be rewritten as:
	
\begin{lstlisting}[style=codeoutput]
y ~ x1 + x2 + x3
		\end{lstlisting}
	
	Here, regression coefficients are implicitly added to the model as active parameters.
	\item '\code{~~}': covariance operator.
	
	It is used to parameterize covariance between variables. For example, in model from Figure~\ref{fig:modelmeans}, covariances between $\eta_2$, $x_2$ and between $y_5, y_6$ are defined with those lines:
	
\begin{lstlisting}[style=codeoutput]
eta2 ~~ x2
y5 ~~ y6
		\end{lstlisting}

	\item "\code{=~}": measurement operator.
	
	This operator is used to define latent variables and regressions onto their indicators. For example, if a latent factor is explained by a set of equations
	\begin{equation*}
		\begin{cases}
			y_1 = 1.0 * \eta + \epsilon_1\\
			y_2 = a_1 * \eta + \epsilon_2\\
			y_3 = a_2 * \eta + \epsilon_3\\
			\eta \sim \mathcal{N}(0, \sigma^2_{\Psi}), \epsilon_k \sim \mathcal{N}(0, \sigma^2_{\epsilon})
		\end{cases}
	\end{equation*}
	where $\eta$ is a latent factor,n \textbf{semopy} it can be written down as:
	
\begin{lstlisting}[style=codeoutput]
eta =~ y1 + y2 + y3
		\end{lstlisting}
	
	In other words, "\code{=~}" operator does this:
	\begin{enumerate}
		\item Defines l-value as a latent variable;
		\item Regresses l-values onto r-values;
		\item Sets the first regression coefficient as appears in the formula to $1.0$ to make regression parameters estimable. This happens unless a user manually set any other regression in right part to some fixed value.
	\end{enumerate}
	It can be thought of as a syntax sugar. Later, in Section~\ref{sect:operations}, we will show how it can be rewritten in terms of "\code{~}" and \code{DEFINE} \textit{operation}.
\end{itemize}

There are also covariance operators that are specific to \class{ModelEffects} and \class{ModelGeneralizedEffects}:

\begin{itemize}
	\item (\class{ModelEffects}) "\code{~RF~}" covariance operator.
	
	Used to define covariances in $D$ matrix.
	
	\item (\class{ModelGeneralizedEffects}) "\code{~RFk~}" covariance operator.
	
	Here, "\code{k}" stands for the order in which random effects are specified. It is the same as "\code{~RF~}", but for $D_k$ matrix only. 
\end{itemize}

When specifying relationships between variables by means of operators, we always implicitly introduce new parameters. This can be done explicitly, however, by typing the name of the parameter in front of an r-value separated by an asterisk/multiplication "\code{*}" symbol. It can be used for further constraint specification, or to reuse the same parameter in other equations. For example,

\begin{lstlisting}[style=codeoutput]
y ~ a1 * x1 + a1 * x2 + x3
	\end{lstlisting}

Here, the same regression coefficient is forced for \code{x1} and \code{x2}. The same approach can be used to fix parameters, for example:

\begin{lstlisting}[style=codeoutput]
y ~ 2.0*x1 + x2 + x3
	\end{lstlisting}

Also, for brevity, you can specify multiple r-values separated by comma symbol "\code{,}" -- it will be separated into multiple formulas with the same r-values. For example,

\begin{lstlisting}[style=codeoutput]
y1 ~ x1 + x2 + 3 * x3
y2 ~ x1 + x2 + 3 * x3
y3 ~ x1 + x2 + 3 * x3
	\end{lstlisting}
	can be rewritten as
	
\begin{lstlisting}[style=codeoutput]
y1, y2, y3 ~ x1 + x2 + 3 * x3
	\end{lstlisting}

Let's see how the model that is depicted on Figure~\ref{fig:modelmeans} is described in \textbf{semopy} syntax:

	\consolein
\begin{lstlisting}[style=codeinput]
import semopy
desc = semopy.examples.example_article.get_model()
print(desc)
	\end{lstlisting}
	\consoleout
\begin{lstlisting}[style=codeoutput]
# Measurement part
eta1 =~ y1 + y2 + y3
eta2 =~ y3 + y2
eta3 =~ y4 + y5
eta4 =~ y4 + y6
# Structural part
eta3 ~ x2 + x1
eta4 ~ x3
x3 ~ eta1 + eta2 + x1
x4 ~ eta4 + x6
y7 ~ x4 + x6
# Additional covariances
y6 ~~ y5
x2 ~~ eta2
	\end{lstlisting}

\subsection{Operation commands}\label{sect:operations}

\begin{itemize}
	\item \code{DEFINE(latent)}
	
	Defines a list of variables that follow the command as latent variables. For example,

\begin{lstlisting}[style=codeoutput]
y1 ~ 1.0 * eta1
y2 ~ eta1 + 1.0 * eta2
y3 ~ eta1 + eta2
y4 ~ eta2 
DEFINE(latent) eta1 eta2
		\end{lstlisting}
		Notice that that's the same as 
\begin{lstlisting}[style=codeoutput]
eta1 =~ y1 + y2 + y3
eta2 =~ y2 + y3 + y4
		\end{lstlisting}
	
	\item \code{DEFINE(ordinal)}
	
	This operation has effect only for \class{Model}. When variables are defined as ordinal, polychoric and polyserial will be used for ordinal variables instead. An example:
	
\begin{lstlisting}[style=codeoutput]
y ~ x1 + cat1 + cat2
cat3 ~ x1
cat2 ~~ cat3
DEFINE(ordinal) cat1 cat2 cat3
		\end{lstlisting}
	
	\item \code{START(v)}

	Sets float value \code{v} as a starting value for a list of parameters that goes after the command. Parameters must be given a name beforehand. An example:

\begin{lstlisting}[style=codeoutput]
y ~ a1 * x1 + a2 * x2 + a3 * x3
START(0.0) a1 a2
START(10) a3
		\end{lstlisting}

	\item \code{BOUND(l, r)}
	
	\code{BOUND} operator restricts parameter values to an interval [\code{l}, \code{r}]. All variance parameters are restricted to an open interval [0, $\infty$] by default. An example:

\begin{lstlisting}[style=codeoutput]
y ~ a1 * x1 + a2 * x2 + a3 * x3
y ~~ v*y
BOUND(-1, 1) a2 a2 a3
BOUND(0, 10) v
		\end{lstlisting}

	\item \code{CONSTRAINT(constr)}
	
	Imposes a non-linear constraint \code{constr} of type equality or inequality on the model parameters. \code{constr} can be any \textbf{sympy}-compatible \citep{Meurer:2017} string. All parameters in the constraint must be named beforehand. An executable example:
	
		First, we load univariate regression model:
		\consolein
\begin{lstlisting}[style=codeinput]
import semopy
ex = semopy.examples.univariate_regression_many
desc, data = ex.get_model(), ex.get_data()
print(desc)
		\end{lstlisting}
		\consoleout
\begin{lstlisting}[style=codeoutput]
y ~ x1 + x2 + x3
		\end{lstlisting}
		Instead of using this \code{desc}, we name some of the parameters, apply constraints and then fit the model to the data:\newpage
		\consolein
\begin{lstlisting}[style=codeinput]
desc = '''y ~ a*x1 + b * x2 + c
# We explicitly define variance parameter:
y ~~ v * y
CONSTRAINT(exp(a) + exp(b) = 10)
CONSTRAINT(v < cos(a)^2 + sin(b)^2)'''
m = semopy.Model(desc)
r = m.fit(data)
m.inspect()
		\end{lstlisting}
		\consoleout
\begin{lstlisting}[style=codeoutput]
lval  op rval  Estimate  Std. Err    z-value       p-value
0    y   ~   x1  1.987709  0.090504  21.962690  0.000000e+00
1    y   ~   x2  0.993699  0.097202  10.223016  0.000000e+00
2    y   ~   x3  1.214214  0.085897  14.135694  0.000000e+00
3    y  ~~    y  0.866303  0.122514   7.071068  1.537437e-12
		\end{lstlisting}
		It can be seen that parameter estimates satisfy the imposed constraints.

\end{itemize}

\section{Prediction}
SEM can be interpreted as a data generating process, and the problem of predicting certain variables from given data is of interest. Specifically, given model parameters estimate, we want to:
\begin{enumerate}
	\item Predict missing entries in a dataset from, at least, some of the observed data;
	\item Predict factor scores/provide estimates of latent variables.
\end{enumerate} 
\subsection{SEM Regression/Imputation}
All \textbf{semopy} models can do regression by means of a conditional expectation of multivariate normal variable. Let $z$ be a vector of random variables distributed as $\mathcal{\mu, \Sigma}$. We partition $z$ into two subvectors $z_1$ and $z_2$, and, consequently, do the same with mean vector $\mu$ and the covariance matrix $\Sigma$: 
$$z = \begin{pmatrix}z_1\\ z_2 \end{pmatrix}, \mu = \begin{pmatrix} \mu_1 \\ \mu_2\end{pmatrix}, \Sigma = \begin{pmatrix}\Sigma_{11} & \Sigma_{12} \\ \Sigma_{21} & \Sigma_{22} \end{pmatrix}$$

It can be shown \citep{Eaton:1983} that the expectation of $z_1$ conditional on $z_2$ is
\begin{equation}\label{eq:condex}
	\EX[z_1|z_2] = \mu_1 + \Sigma_{12} \Sigma_{22}^{-1}(z_2 - \mu_2)
\end{equation}

If we interpret $z_2$ as observed variables that are present in the dataset and $z_1$ as missing variables, then Equation~\ref{eq:condex} provides a regression formula. It can also be thought of as an imputation method if parameter estimation was done with either \class{Model}s pairwise deletions or the FIML method on the incomplete data.

In \textbf{semopy}, it is done via \code{predict} method. See an example:

	First, we load the \code{political_democracy} dataset and then we randomly fill 10 data entries with NaNs:
	\consolein
\begin{lstlisting}[style=codeinput]
import semopy
import numpy as np
import random
random.seed(123)
ex = semopy.examples.political_democracy
desc, data = ex.get_model(), ex.get_data()
inds = list(np.ndindex(data.shape))
inds = tuple(zip(*random.sample(inds, 10)))
true_vals = data.values[inds]
data.values[inds] = np.nan
	\end{lstlisting}
	Then, we fit \class{ModelMeans} to the data with FIML, predict/impute the missing values and compare them to true values in terms of MAPE and mean-squared error (MSE):\newpage
	\consolein
\begin{lstlisting}[style=codeinput]
m = semopy.ModelMeans(desc)
r=m.fit(data, )
pred = m.predict(data,)[data.columns].values[inds]
mape = np.mean(abs((pred-true_vals) / true_vals)) * 100
mse = np.mean((pred - true_vals) ** 2)
print('MAPE: {:.2f}%, MSE: {:.2f}'.format(mape, mse))	
	\end{lstlisting}
	\consoleout
\begin{lstlisting}[style=codeoutput]
MAPE: 14.59%, MSE: 0.41
	\end{lstlisting}
	In practice, usually, predictions can be only as good as a variance of predicted variables.

\begin{leftbar}
	There is a limitation of \code{predict} method for \class{ModelMeans}, \class{ModelEffects} and\\ \class{ModelGeneralizedEffects}, however: it can't be used to predict exogenous observed variables that reside in $x^{(2)}$. Hence, if a user needs to predict $x^{(2)}$, he/she should either move the predicted variables from $x^{(2)}$ to $x^{(1)}$, or resort to \class{Model}.
\end{leftbar}

\subsection{Factor scores}
Factor scores, in context of our generalized SEM models, are the most likely values that latent variables would attain if we had observed them given data and parameter estimates. Numerous approaches have been proposed for different models, starting with exploratory factor analysis (EFA) \citep{Bartlett:1937} to generalizations of those approaches to CFA \citep{Bollen:1989} and SEM \citep{Yung:2013}. Yet, none of them can tackle \class{ModelMeans}, let alone to take random effects into an account. Hence, we propose our own generalized factor prediction scheme that is based on maximum a posteriori (MAP) approach. Namely, we assume a joint distribution function of data $Z$ and latent factors $H$:
\begin{equation}\label{eq:factorprob}
	f(Z, H|\theta) = f(Z|H, \theta)f(H|\theta),
\end{equation}
where $f(Z|H, \theta)$ is a conditional distribution of $Z$ given $H, \theta$ and $f(H|theta)$ is a distribution of $H$ given $\theta$, $\theta$ is a parameters vector. Then, we maximize Equation~\ref{eq:facestprob} with respect to $H$ and, finally, obtain MAP estimate of latent factors. In case of \class{ModelEffects} and \class{ModelGeneralizedEffects}, solving the maximization problem requires solving Sylvester equation of the form $A_1 H - H A_2 = A_3$, which is done by means of Bartels-Stewart algorithm \citep{Bartels:1972}. In case of \class{Model} and \class{ModelMeans}, the problem reduces to solving a system of linear equations. 

The derivation of our approach and an exact analytical formula is, unfortunately, too tedious to be included here, but an interested reader can refer to Appendix~\ref{app:factor_prediction}.

In \textbf{semopy}, method \code{predict_factors} is used for factor scores estimation. See an example:
	\consolein
\begin{lstlisting}[style=codeinput]
import semopy		
ex = semopy.examples.holzinger39
data, desc = ex.get_data(), ex.get_model()
m = semopy.Model(desc)
r = m.fit(data)
m.predict_factors(data).head()
	\end{lstlisting}
	\consoleout
\begin{lstlisting}[style=codeoutput]
speed   textual    visual
0  0.061510 -0.137550 -0.817675
1  0.625501 -1.012725  0.049531
2 -0.840548 -1.872284 -0.761421
3 -0.271325  0.018486  0.419333
4  0.194325 -0.122253 -0.415898
	\end{lstlisting}

\section{Solvers}
All the estimation methods require solving some sort of the maximization/minimization problem. \textbf{semopy} relies on \textbf{scipy} \citep{Virtanen:2020} optimization methods; any of the methods available to \textbf{scipy} can be chosen via setting the \code{solver} argument of the \code{fit} method. However, not all \textbf{scipy} solvers support constraints. The default solver in \textbf{semopy} is SLSQP \citep{Kraft:1988} as previously \citep{Igolkina:2020} it has demonstrated the best results. However, sometimes other solvers can provide better results. 

The one solver that deserves special attention is the differential evolution solver that is due to \cite{Storn:1997}; it is a global solver and might be useful in big and complex models with non-optimal local minima. However, it has two pitfalls that make it less attractive:
\begin{enumerate}
	\item A user might need to provide bound constraints for parameters, as differential evolution solver looks for optima inside of a box in a parameters space. If no bounding boxes are provided, then the package automatically assumes that they lie in the interval $[-10, 10]$, which is not always reasonable, or too large and will take a toll on performance;
	\item Computation times are huge and not suitable for applications where high performance is demanded.
\end{enumerate}

To use differential evolution solver, one should pass \code{"de"} as the \code{solver} argument to the \code{fit} method. Also, the default $[-10, 10]$ bound can be redefined as $[-b, b]$ by passing an extra \code{b_max=b} argument.

\section{Parametric bootstrap resampling for bias reduction}

Maximum likelihood estimators (MLE) are linchpins of statistical models due to their properties, however, it is known that MLE produces unbiased variance estimates up to the order $\frac{1}{\sqrt n}$. The bias becomes more noticeable in cases when the number of data samples $n$ is small. Fortunately, it is possible to reduce the bias to the order $\frac{1}{n^2}$. One way to do it is to generate pseudo-random samples from distribution to estimate bias and then subtract it from the previously obtained estimates. This approach, called parametric bootstrap resampling (PBE), is due to \cite{Efron:1982}.

Let $z \sim \mathcal{N}(0, \Sigma(\theta))$, where $\theta$ is a parameter vector obtained from MLE. Next, we sample data from $\mathcal{N}(0, \Sigma(\theta))$ $k$ times and for each of sampled datasets we calculate another ML estimate $\theta^{(i)}$. Then, we can estimate bias $b$ of $\theta$ as 
\begin{equation*}
	b = \frac{1}{k} \sum_{i=1}^k \left( \theta - \theta^{(i)} \right)
\end{equation*}
Finally, we can obtain ML estimates corrected for bias as 
$$\theta_{PBE} = \theta - b = 2\theta - \frac{1}{k}\sum_{i=1}^k \theta^{(i)}$$

To correct parameter estimates for bias, one should call function \code{bias_correction}:
\consolein
\begin{lstlisting}[style=codeinput]
semopy.bias_correction(model, n=1000)
\end{lstlisting}
where \code{n} is a number of bootstrap iterations (the default is 100). After that, parameters in \code{model} should be corrected for bias with the above PBE method. 

\begin{leftbar}
	By default, \textbf{semopy} uses biased sample covariance matrix similarly to \textbf{lavaan}. The bias might become big enough when number of samples is very small, therefore when incorporating the above PBE approach in scenarios with tiny datasets, it is advised to pass unbiased sample covariance matrix to the \code{fit} method via \code{cov} argument and/or using FIML.
\end{leftbar}

\section{Testing framework: model and data generation}\label{sect:modgen}
For numerical experiments with \textbf{semopy}, we implemented a versatile system that can:
\begin{enumerate}
	\item Given a configuration (i.e. the desired number of endogenous and exogenous variables, number of latent factors, etc), generate a random model description in \textbf{semopy} syntax;
	\item Given a model description, generate a random set of parameters for the model;
	\item Given a model, sample a random dataset that could be generated by the model; 
\end{enumerate}

Let's see how it works on an example:

	First, we generate a random model description:
	\consolein
\begin{lstlisting}[style=codeinput]
import random
import semopy
random.seed(12345)
modgen = semopy.model_generation
desc = modgen.generate_desc(n_endo=3, n_exo=2, n_lat=2,
                            n_inds=3, n_cycles=1, p_join=0.1)
print(desc)
	\end{lstlisting}
	\consoleout
\begin{lstlisting}[style=codeoutput]
# Measurement part:
eta1 =~ y1 + y2 + y3
eta2 =~ y4 + y5 + y6 + y2
# Structural part:
eta2 ~ eta1 + x3
x3 ~ g1
x1 ~ g1 + x3 + x2
x2 ~ g2
eta1 ~ g2 + x3 + x2 + x1
g1 ~ x3
	\end{lstlisting}
	Here, parameters have the following meaning: \code{n_endo} is a number of endogenous observable variables, \code{n_exo} is a number of exogenous observable variables, \code{n_lat} is a number of latent variables, \code{n_inds} is a number (or tuple, then for each factor number of indicators is chosen randomly from an interval) of indicators per random latent variable, \code{n_cycles} is a number of cycles in a model, \code{p_join} is a probability that an indicator will be shared with another latent variable (here, \code{y2} is an example). 
	
	Then, we generate a set of parameters for the given model:
	\consolein
\begin{lstlisting}[style=codeinput]
params, tmp = modgen.generate_parameters(desc)
	\end{lstlisting}
	\code{params} is a dataframe that is similar to ones returned by the \code{inspect} method, \code{tmp} is an auxiliary variable (to be more precise, it is a \class{ModelMeans} instance with parameters loaded from the \code{params} table) that is useful only for the next step:
	\consolein
\begin{lstlisting}[style=codeinput]
data = modgen.generate_data(tmp, n=200)
print(data.head())
	\end{lstlisting}
	\consoleout
\begin{lstlisting}[style=codeoutput]
y1        y2         y3  ...        x2        x3        g2
0  3.276384  0.525599   4.127978  ...  4.763574  2.031664 -2.128789
1  0.083196  1.600804   1.080848  ... -0.512638 -1.273207  1.008115
2 -4.294538 -1.885474  -1.581973  ... -1.025223  2.078174 -0.550860
3  7.844931  5.640587  11.752653  ...  4.871314 -3.475223 -1.385154
4 -3.326591 -2.099521  -3.087801  ...  0.575090  3.693363 -0.408098

[5 rows x 11 columns]
	\end{lstlisting}
	where \code{n} is a number of samples in a generated dataset.
	
	So, now we have a triplet of model description \code{desc}, dataset \code{data} and "true" parameter values \code{params}. Let's verify that \textbf{semopy} provides reasonable estimates for the generated model:
	\consolein
\begin{lstlisting}[style=codeinput]
m = semopy.Model(desc)
r = m.fit(data)
mape = np.mean(semopy.utils.compare_results(m, params)) * 100
print('MAPE: {:.2f}%'.format(mape))
	\end{lstlisting}
	\consoleout
\begin{lstlisting}[style=codeoutput]
MAPE: 15.45%
	\end{lstlisting}

\begin{leftbar}
	Note that \code{generate_data} can be applied to any \code{Model}, \code{ModelMeans} or even \code{ModelEffects}  and \code{ModelGeneralizedEffects} instance. It can be used to simulate different datasets that could be spewed out by a data generating process that is described by a SEM model.
\end{leftbar}

\section{Significance testing}

\subsection{Standard errors and p-values}

The \textbf{semopy} utilizes the Z-test to calculate p-values for parameters' estimates under the assumption that parameters are normally distributed; $H_0$: the value of a parameter is equal to zero. This approach considers z-score:
\[ Z(\hat{\theta}) = \frac{\hat{\theta}}{SE(\hat{\theta}) }, \]
where $\hat{\theta}$ is a vector of parameter estimates, $SE(\hat{\theta})$ is the standard error of estimates, which is proportional to variance: $SE(\hat{\theta}) = var(\hat{\theta})/\sqrt{n}$. Based on the Cramér–Rao bound, 
$$var(\hat{\theta}) \ge \mathrm{FIM}(\theta)^{-1}, $$
where $\mathrm{FIM}(\theta)$ is a Fisher information matrix (FIM). The $\mathrm{FIM}(\theta)$ matrix can be defined as an observed or expected FIM. The observed FIM is the Hessian of a likelihood function at $\hat{\theta}$, $H(\hat{\theta})$. The expected FIM is:

\[FIM(\hat{\theta}) = \left\{-E\left[\frac{\partial^2}{\partial \theta_i \partial \theta_j} \ln f(X, \hat{\theta})\right]\right\}_{i, j}^n, \] 
where $f(x, \hat{\theta})$ is a likelihood function. For instance, if $x \sim \mathcal{N}(\mu(\theta), \Sigma(\theta))$, the FIM is calculated as follows \citep{Mardia:1984}:
\begin{equation}\label{eq:fimmv}
	\begin{split}
		FIM(\theta) = \left\{E\left[\frac{\partial}{\partial \theta_i \partial \theta_j}f(\theta)\right]\right\}_{i, j}^n =  \left\{\frac{\mu^T}{\partial\theta_i}(\theta) \Sigma^{-1}(\theta) \frac{\mu}{\partial\theta_{j}}(\theta)\right\}_{i, j}^n + \\ + \left\{-\mathrm{tr}\left[\frac{\partial}{\partial \theta_i}\Sigma(\hat{\theta}) \Sigma(\hat{\theta})^{-1}\frac{\partial}{\partial \theta_j}\Sigma(\hat{\theta}) \Sigma(\hat{\theta})^{-1}\right]\right\}_{i, j}^n
	\end{split}
\end{equation}

\begin{leftbar}
	Applying property in the Equation~\ref{eq:matnormvec} to $x$, it is trivial to generalize the Equation~\ref{eq:fimmv} to a case of matrix-variate normal distribution, hence making it applicable to \class{ModelEffects} and \class{ModelGeneralizedEffects}. Unfortunately, this straightforward approach requires $O(n^3 m^3)$ operations for each of the FIM components, where $n$ is a number of observations and $m$ is a number of output observable variables. Therefore, we have inferred an alternative form of the FIM for the case of matrix-variate normal distributions that are computable in $O(n^3 + m^3)$ operations (see derivation in Appendix~\ref{app:fisher_information}):
	
	\begin{equation*}
		\begin{split}
			FIM(\theta)_{i,j} = tr\{T\} tr\left\{\frac{\partial M^T }{\partial \theta_i} L^{-1} \frac{\partial M }{\partial \theta_i} T^{-1} \right\} + 
			\frac{1}{2}(
			n tr\{A_i A_k\} +
			m tr\{B_i B_k\} +\\
			+
			tr\{A_i\} tr\{B_k\} +
			tr\{A_k\} tr\{B_i\} +
			n m \alpha_i \alpha_k 
			-
			n \alpha_k tr\{A_i\} - \\
			-
			m \alpha_k tr\{B_i\} -
			n \alpha_i tr\{A_k\} -
			m \alpha_i tr\{B_k\}
			),
		\end{split}
	\end{equation*}
	where $A_i = T^{-1} \frac{\partial T}{\partial \theta_i}$,$A_k = T^{-1} \frac{\partial T}{\partial \theta_k}$, $B_i =  L^{-1} \frac{\partial L}{\partial \theta_i}$,  $B_k =  L^{-1} \frac{\partial L}{\partial \theta_k}$, $\alpha_i = \frac{tr\{\frac{\partial T}{\partial \theta_i}\}}{tr\{T\}}$ and  $\alpha_k = \frac{tr\{\frac{\partial T}{\partial \theta_k}\}}{tr\{T\}}$.
\end{leftbar}

The p-values for parameter estimates are based on the Z-test assumption, that  $Z(\theta)$ follows a multivariate normal distribution. This assumption follows from properties of ML estimators, namely, that parameter estimates $\hat{\theta}$ of a parameter vector $\theta$ converge in distribution to $\mathcal{N}(\theta, {FIM}^{-1}(\theta))$:
$$\sqrt{n}(\widehat{\theta_{i}}) \overset{d}{\rightarrow} \mathcal{M}(0, {FIM}^{-1}(\hat{\theta}))$$

In \textbf{seompy}, standard errors, alongside p-values, are calculated automatically when \code{inspect} method is called. By default, expected FIM is used, but a user may want to estimate standard errors with the observed FIM instead by setting the \code{information} argument to \code{"observed"}:

\consolein
\begin{lstlisting}[style=codeinput]
model.inspect(information="observed")
\end{lstlisting}

\subsection{Robust standard errors}

When the assumption of data normality is violated, standard errors should not be trusted. The problem can be mitigated by using the so-called "Huber sandwich estimator" of an asymptotic covariance matrix proposed by \cite{Huber:1967}. Standard errors, obtained from the sandwich estimator, are often referred to as "robust standard errors". The intuition behind this method and discussion on whether it is reasonable is provided by \cite{Freedman:2006}. In \textbf{semopy}, user can substitute conventional standard errors with robust ones by passing \code{robust=True} argument to the \code{inspect} method:
\consolein
\begin{lstlisting}[style=codeinput]
model.inspect(robust=True)
\end{lstlisting}
\section{Regularization}

\begin{table}[!t]
	\centering
	\begin{tabular}{r|p{12cm}}
		
		Name  & Description \\
		\hline
		\texttt{"l1-naive"} & Naive implementation of L1 regularization: the gradient does not exist at $0$.\\
		\texttt{"l1-smooth"} & Instead of L1 penalty, a smooth approximation is used \citep{Schmidt:2007}: $\alpha\left( ln(1 + exp(-x/\alpha)) + ln(1 + exp(x/\alpha))\right)$.   \\
		\texttt{"l1-thresh"} & l1 regularization, but the gradient is a soft-thresholding operator \citep{Daubechies:2004} with hyperparemter $\alpha$.\\
		\texttt{"l2-naive"} & L2 regularization, the gradient is not differentiable at 0.\\
		\texttt{"l2-square"} & L2-squared regularization, the gradient exists everywhere. 
	\end{tabular}
	\caption{The list of available penalty functions. In all cases, the less $\alpha$ is, the better is approximation, but too small values lead to numerical instability.$\alpha$ is a hyperparemter that can be submitted via \code{alpha} argument. The default value for the \code{alpha} is $10^{-6}$.}
	\label{table:regularizations}
\end{table}

Regularization, in the context of SEM, is not well studied, neither have we did any substantial research that could contribute to the field. So far, its applications are limited, but not non-existent. For example, penalized SEM has been applied for variable (and model in a broader sense) selection by \cite{Jacobucci:2019}. Therefore, we added support of regularization to \textbf{semopy} as a promising feature for an ingenious user and as a basis for possible further research. We consider a regularized problem of the form

$$G(\theta) = F(\theta) + c R(\theta_{\aleph}) \rightarrow min,$$
where $F$ is an objective function such as log-likelihood or a least-squares loss, $R$ is a regularization constant, $R$ is a penalty function and $\theta_{\aleph}$ is a subvector of $\theta$ that is penalized.

To apply regularization in \textbf{semopy}, a user should first call the \code{create_regularization} function to instantiate an auxiliary structure that will be used by the \code{fit} method of a model. This function has the following parameters of interest:

\begin{itemize}
	\item \code{model} -- a model instance;
	\item \code{regularization} -- a name of penalty function (see Table~\ref{table:regularizations});
	\item \code{c} -- a regularization constant (the default is $1.0$);
	\item \code{alpha} -- an $\alpha$ parameter that is used in \code{"l1_smooth"} and \code{"l1_thresh"} (the default is $10^{-6}$);
	\item \code{param_names} -- an optional list of parameter names (as provided by a model syntax) that are penalized;
	\item \code{mx_names} -- an optional list of matrix names whose parameters are penalized.  
\end{itemize} 
At least one of \code{param_names} or \code{mx_names} arguments must be provided.

The list of supported penalty functions is provided in Table~\ref{table:regularizations}. Note that as \code{regularization} accepts a list of regularization instances, one can combine different penalties on the same subset of parameters. It can be used to obtain the elastic net penalty, for instance. See an example:

	\consolein
\begin{lstlisting}[style=codeinput]
import semopy
ex = semopy.examples.political_democracy
desc, data = ex.get_model(), ex.get_data()
m = semopy.Model(desc)
reg1 = semopy.create_regularization(m, 'l1-thresh', mx_names=['Beta'])
reg2 = semopy.create_regularization(m, 'l2-square', c=0.5,
mx_names=['Beta'])
regs = [reg1, reg2]
r = m.fit(data, regularization=regs)
	\end{lstlisting}

\section{Sparse exploratory factor analysis}

Although an exact definition for Exploratory Factor Analysis (EFA) is arguable, it is fair to generalize and say that EFA is a set of multivariate statistical techniques used to extract an underlying latent structure from observed variables. EFA is often used as the first step in SEM model specification in cases when the researcher can't provide a reasonable prior latent structure. Numerous implementations of EFA exist, the most popular is probably a two-step procedure of firstly determining the number of latent factors via a kind of scree test and secondly performing a factor analysis (FA) to discover factors' indicators. That's easily achievable with great number of statistical packages,such as \textbf{statsmodels} \citep{Seabold:2010} and \textbf{sklearn}. 

Though not necessarily a con, the above approach is prone to a researcher subjective reasoning: first, a researcher has to select a number of latent factors, second, he has to decide what loading values obtained from FA are significant. The latter can be dealt with using a sparse variation of EFA, for example, sparse PCA (SPCA) \citep{Zou:2006}. In \textbf{semopy}, we provide an enhanced heuristic approach that not only manages to uncover latent structures with a greater degree of success than plain SPCA but also provides a way to automatically detect a number of latent factors hidden in data.

First, we compute correlations $\{\rho_{ij}\}_{i,j=1}^m$ between phenotypes. Then we construct a distance matrix $\mathbb{D}$, where $\mathbb{D}_{ij} = 1 - |\rho_{ij}|$. Then, we pass $\mathbb{D}$ to \textbf{OPTICS} \citep{Ankerst:1999} clustering algorithm with minimal cluster size set to 2. We chose \textbf{OPTICS} as a cluster due to its ability to identify clusters of uneven sizes and densities: it was observed that first 3 eigenvectors of covariance matrix might explain up to 90\% of the variance while the number of latent factors in an actual model is significantly higher in our synthetic datasets, thus making it impossible to obtain even a correct number of latent factors using conventional EFA approaches and clustering algorithms. A number of clusters identified by \textbf{OPTICS} are considered to be a number of latent factors (and variables' labels can be interpreted as loadings onto a respective latent factor; in fact, this representation is often already good enough, but in this approach, we don't use it further).

Second, we run SPCA to determine loadings of latent factors onto observed variables. At this point, we've obtained a CFA model, but it is usually quite overidentified in the sense that there are some abundant loadings.

Finally, we refine the CFA model by evaluating it through \class{Model} and dropping loadings with high p-values.

The goal of the proposed algorithm is to provide a convenient EFA technique within a \textbf{semopy} framework, rather than to contribute to the field of sparse learning or EFA, as the algorithm is very heuristic and not theoretically sound, yet performed well in our tests (see Section~\ref{sect:exp_efa}). To use it, user should call the \code{explore_cfa_model} function of the \code{efa} submodule. See an example:

	Before all else, we generate a random model that consists only of CFA part and a 100 exogenous variables that load onto latent factors. The exogenous variables here play a role as variance contributors to latent factors, we omit them shortly from the analysis for a cleaner example:
	\consolein
\begin{lstlisting}[style=codeinput]
import semopy 
import numpy as np
np.random.seed(123)
modgen = semopy.model_generation
desc = modgen.generate_desc(0, 100, 4, 3, 0, 0.05)
params, t = modgen.generate_parameters(desc)
data = modgen.generate_data(t, 200, )
print(desc.split('# ')[1])
	\end{lstlisting}
	\consoleout
\begin{lstlisting}[style=codeoutput]
Measurement part:
eta1 =~ y1 + y2 + y3 + y9
eta2 =~ y4 + y5 + y6
eta3 =~ y7 + y8 + y9 + y5
eta4 =~ y10 + y11 + y12 + y2
	\end{lstlisting}
	\consolein
\begin{lstlisting}[style=codeinput]
cols = [c for c in data.columns if not c.startswith('g')]
cols = sorted(cols, key=lambda x: int(x.split('y')[-1]))
data = data[cols]
	\end{lstlisting}
	Next, to make the task of discovering the underlying latent structure a bit harder, we introduce 40 fake variables into a dataset.
	\consolein
\begin{lstlisting}[style=codeinput]
for i in range(40):
	data[f'fake{i+1}'] = np.random.normal(size=len(data))
	\end{lstlisting}
	Finally, we run EFA procedure on \code{data}:
	\consolein
\begin{lstlisting}[style=codeinput]
desc = semopy.efa.explore_cfa_model(data, mode='spca')
print('Predicted CFA model:\n', desc)
	\end{lstlisting}
	\consoleout
\begin{lstlisting}[style=codeoutput]
Predicted CFA model:
eta1 =~ y7 + y8 + y5 + y9
eta2 =~ y3 + y1 + y2 + y9
eta3 =~ y10 + y11 + y12 + y2
eta4 =~ y6 + y4 + y5
	\end{lstlisting}
	It is the same model as that was generated previously up to indicator permutations.


\section{Visualization}

\begin{figure}[t!]
	\centering
	\begin{adjustbox}{max width=\textwidth}
	\includegraphics{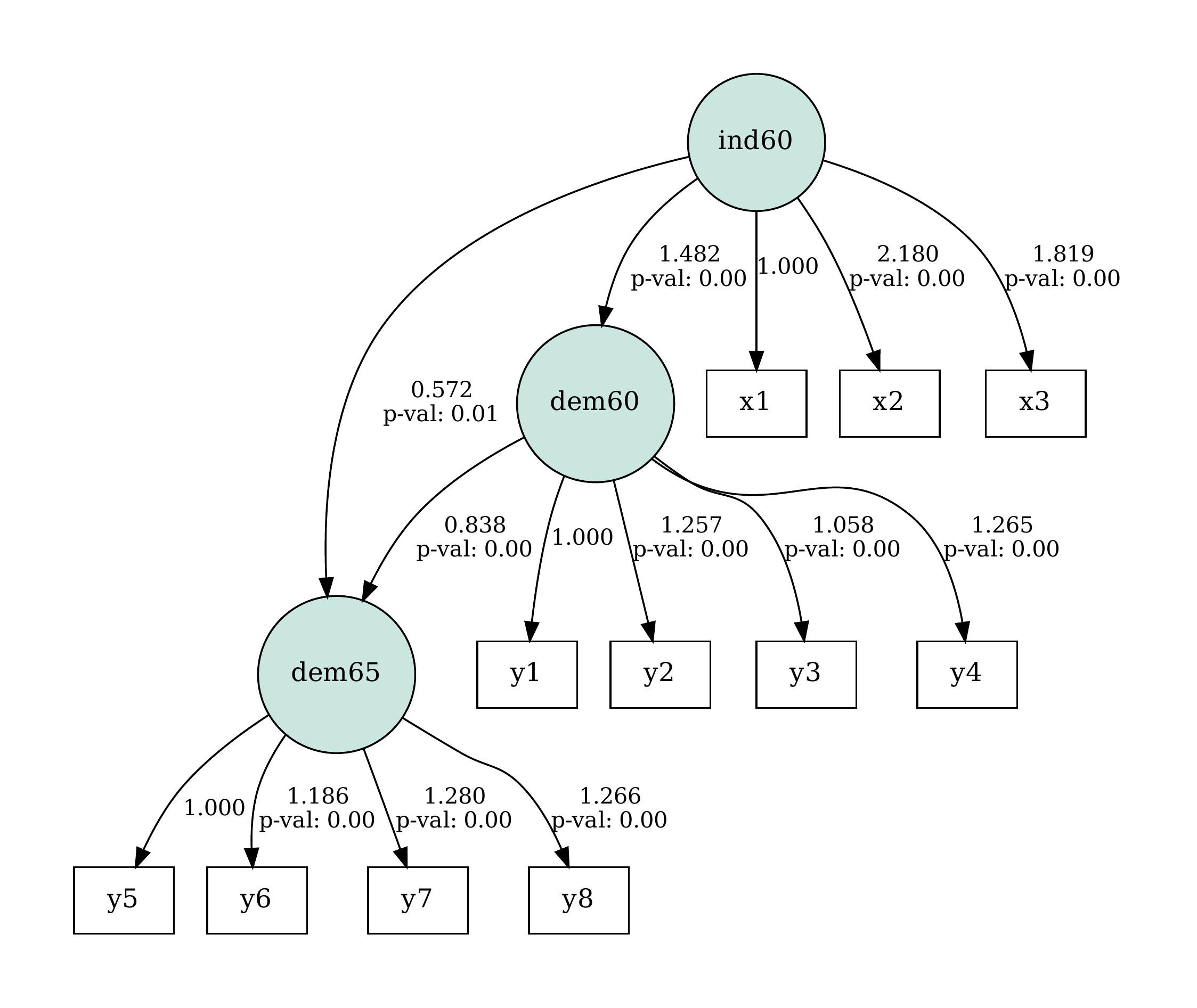}
	\end{adjustbox}
	\caption{Visualization of estimated Political Democracy model.}
	\label{fig:vis}
\end{figure}

\begin{figure}[t!]
	\centering
	\begin{adjustbox}{max height=0.4\textheight}
		\includegraphics{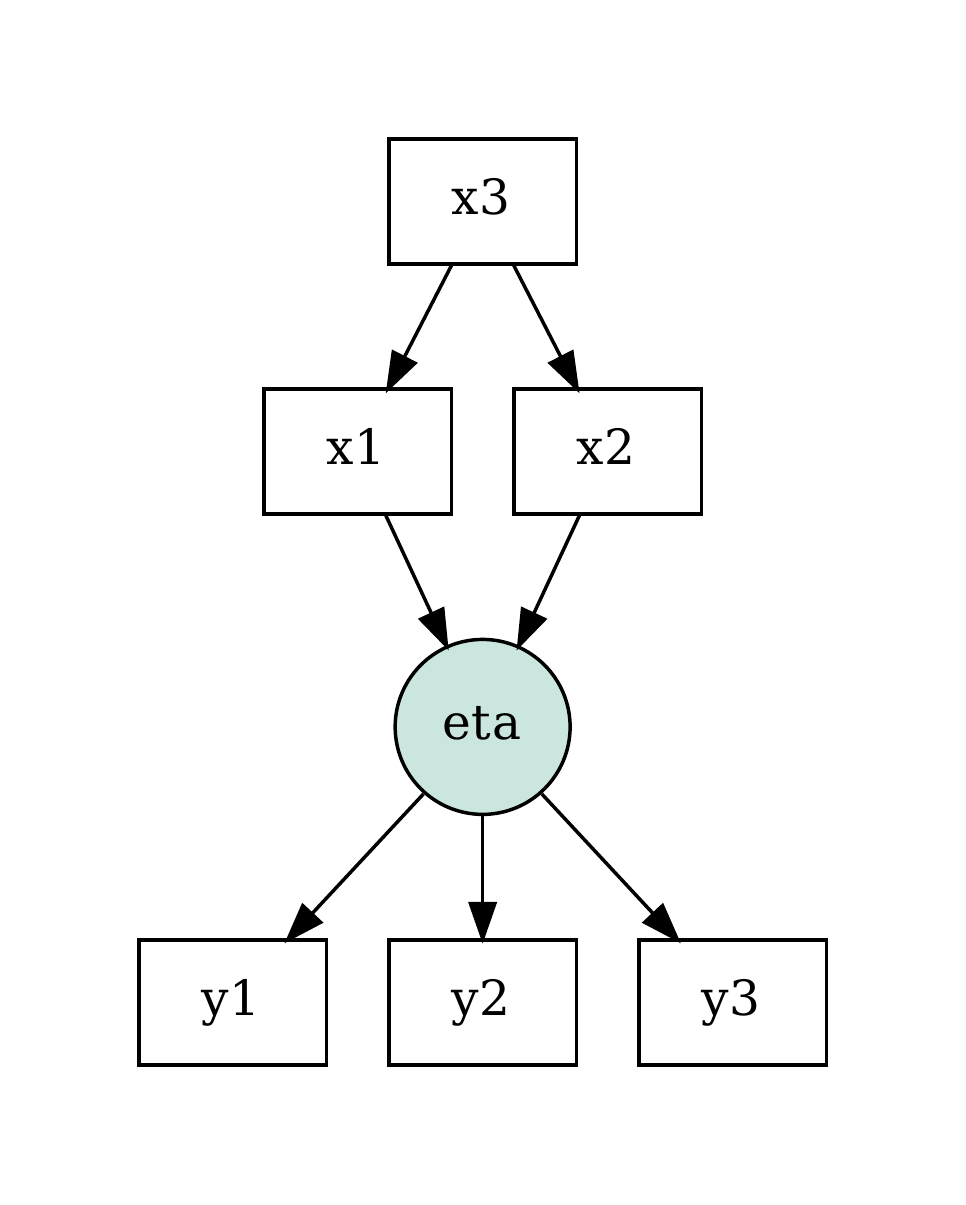}
	\end{adjustbox}
	\caption{Visualization of a model given only text description.}
	\label{fig:vis2}
\end{figure}

It is often tiresome to analyze SEM results in a form of plain text, hence \textbf{semopy} provides a simpler wrapper around \textbf{graphviz} \citep{Ellson:2003}. See an example:

	First, we load and fit the model:
	\consolein
\begin{lstlisting}[style=codeinput]
import semopy
ex = semopy.examples.political_democracy
desc, data = ex.get_model(), ex.get_data()
m = semopy.Model(desc)
r = m.fit(data)
	\end{lstlisting}
	Then, we plot it and save to \textit{"out.pdf"} file:
	\consolein
\begin{lstlisting}[style=codeinput]
g = semopy.semplot(m, 'out.pdf')
	\end{lstlisting}
	The result is shown on Figure~\ref{fig:vis}. Also, in \code{g} resides graph in dot-format \citep{Gansner:2015}, that a user is free to adjust for a more fancy look.
	
	Alternatively, one can just pass a model description to \textbf{semplot} instead of a \class{Model} instance. For example:\newpage
	\consolein
\begin{lstlisting}[style=codeinput]
desc = '''
eta =~ y1 + y2 + y3
eta ~ x1 + x2
x1, x2 ~ x3
'''
g = semopy.semplot(desc, 'out.pdf')
	\end{lstlisting}
	The result is shown on Figure~\ref{fig:vis2}

\section{Reporting results}

Although most of information of interest can be extracted pragmatically via \code{inspect} method and fit indices can be computed via \code{calc_stats} function, \textbf{semopy} provides a convenient report generation feature:

\consolein
\begin{lstlisting}[style=codeinput]
semopy.report(model, "Title or model name")	
\end{lstlisting}

For example, let's fit \class{ModelMeans} to the \code{political_democracy} dataset and then generate a report:
\newpage
\consolein
\begin{lstlisting}[style=codeinput]
from semopy import examples, ModelMeans, report
ex = examples.political_democracy
model = ModelMeans(ex.get_desc())
r = model.fit(ex.get_data())
report(model, "Political Democracy")
\end{lstlisting}

\begin{figure}[t!]
	\centering
	\begin{adjustbox}{max height=0.8\textheight, max width=\textwidth}
	\includegraphics{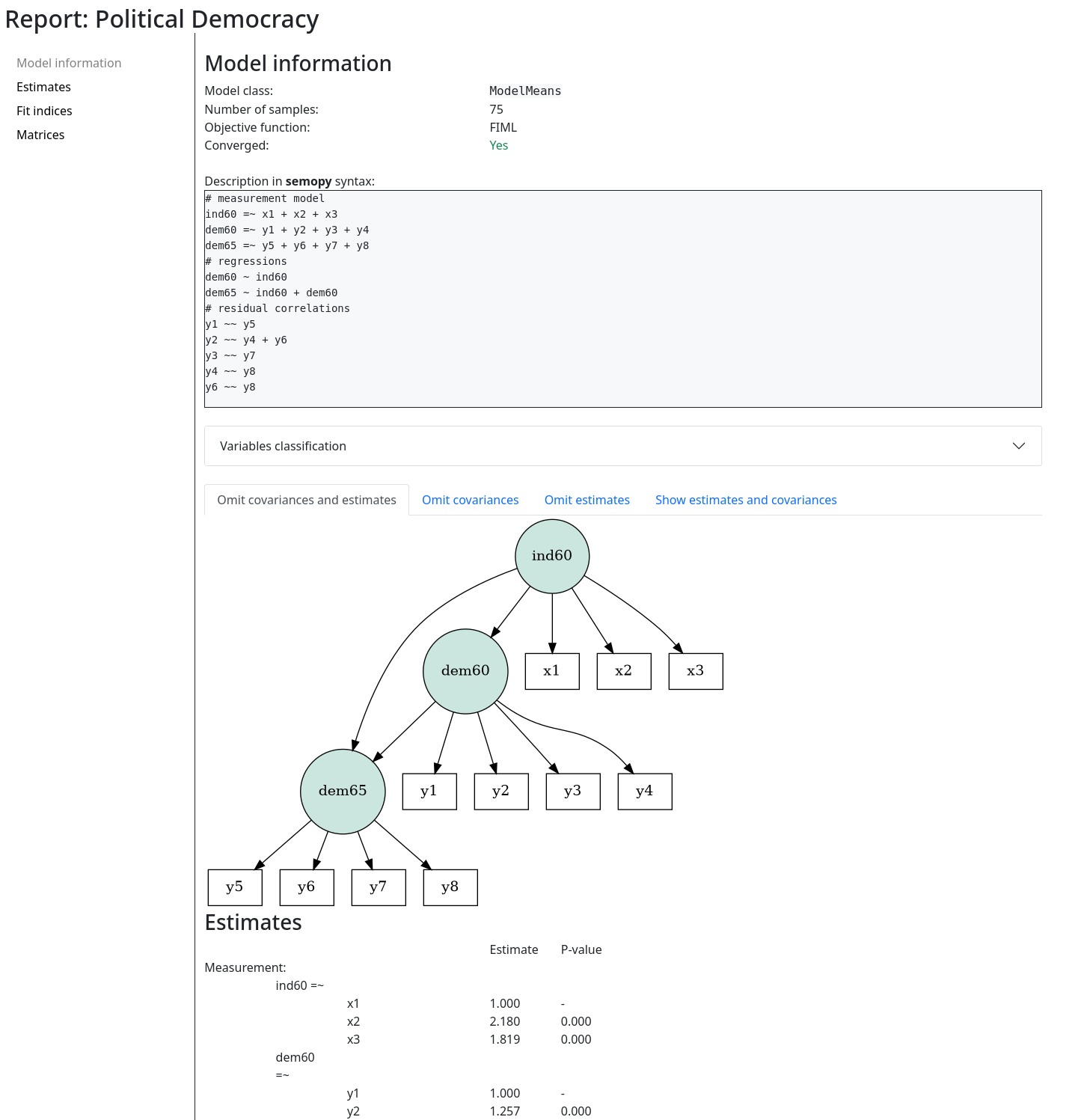}
	\end{adjustbox}
	\caption{A portion of the HTML report for the Political Democracy dataset.}
	\label{fig:report}
\end{figure}

Afterwards, a report with parameter estimates, visualization and fit indices is generated in HTML format. The resulting HTML file can be found at \url{https://semopy.com/report_example/report.html} or in supplementary files. A small fraction of the report can be seen in Figure~\ref{fig:report}.

\section{Numerical experiments}\label{sect:experiments}

\subsection{Estimation accuracy}
\begin{table}[t!]
	\centering
	\begin{tabular}{llllllll}
		\toprule
		{}&\ \# &  \texttt{n} & \texttt{n\_exo} & \texttt{n\_endo} &\texttt{n\_lat} &  \texttt{n\_cycles} & Random effects  \\
		\midrule
		\multirow{1}{*}{\Large A}&1 & 100 & 4 & 3 & 3 & 0 & No\\
		\hline
		\multirow{6}{*}{\Large B}& 2 & 100 & \textbf{2} & 3 & 3 & 0 & No\\
		& 3 & 100 & \textbf{3} & 3 & 3 & 0 & No\\
		& 4 & 100 & \textbf{4} & 3 & 3 & 0 & No\\
		& 5 & 100 & \textbf{5} & 3 & 3 & 0 & No\\
		& 6 & 100 & \textbf{6} & 3 & 3 & 0 & No\\
		& 7 & 100 & \textbf{7} & 3 & 3 & 0 & No\\
		& 8 & 100 & \textbf{8} & 3 & 3 & 0 & No\\
		\hline
		\multirow{6}{*}{\Large C} & 9 & 100 & 4 & \textbf{2} & 3 & 0 & No\\
		& 10 & 100 & 4 & \textbf{3} & 3 & 0 & No\\
		& 11 & 100 & 4 & \textbf{4} & 3 & 0 & No\\
		& 12 & 100 & 4 & \textbf{5} & 3 & 0 & No\\
		& 13 & 100 & 4 & \textbf{6} & 3 & 0 & No\\
		& 14 & 100 & 4 & \textbf{7} & 3 & 0 & No\\
		\hline
		\multirow{6}{*}{\Large D}& 15 & 100 & 4 & 3 & \textbf{1} & 0 & No\\
		& 16 & 100 & 4 & 3 & \textbf{2} & 0 & No\\
		& 17 & 100 & 4 & 3 & \textbf{3} & 0 & No\\
		& 18 & 100 & 4 & 3 & \textbf{4} & 0 & No\\
		& 19 & 100 & 4 & 3 & \textbf{5} & 0 & No\\
		& 20 & 100 & 4 & 3 & \textbf{6} & 0 & No\\
		\hline
		\multirow{4}{*}{\Large E} & 21 & 100 & 4 & 3 & 3 & \textbf{1} & No\\
		& 22 & 100 & 4 & 3 & 3 & \textbf{2} & No\\
		& 23 & 100 & 4 & 3 & 3 & \textbf{3} & No\\
		& 24 & 100 & 4 & 3 & 3 & \textbf{4} & No\\
		\hline
		\multirow{8}{*}{\Large F}& 25 & \textbf{50} & 4 & 3 & 3 & 0 & No\\
		& 26 & \textbf{100} & 4 & 3 & 3 & 0 & No\\
		& 27 & \textbf{150} & 4 & 3 & 3 & 0 & No\\
		& 28 & \textbf{200} & 4 & 3 & 3 & 0 & No\\
		& 29 & \textbf{250} & 4 & 3 & 3 & 0 & No\\
		& 30 & \textbf{300} & 4 & 3 & 3 & 0 & No\\
		& 31 & \textbf{350} & 4 & 3 & 3 & 0 & No\\
		& 32 & \textbf{400} & 4 & 3 & 3 & 0 & No\\
		\hline
		\multirow{1}{*}{\Large G} & 33 & 100 & 4 & 3 & 3 & 0 & \textbf{Yes}\\
		
		\bottomrule
	\end{tabular}
	\caption{Configurations of generated models and their respective datasets. The generated sets are separated into subsets A, B, C, D, E, F, G with respect to a varying parameter. Parameters \code{n\_inds} and \code{p\_join} are fixed to $3$ and $0.05$ respectively.}
	\label{table:ne_models}
\end{table}

To test \textbf{semopy} performance, we evaluated different \textit{models} and methods on the series of synthetic datasets and SEM models generated by our testing framework (see Section~\ref{sect:modgen}). We generated 33 synthetic sets of varying configurations (i.e. parameters that were passed to \code{generate_description} and \code{generate_data}): each set consists of 40 unique SEM model descriptions and for each of the models 10 random pairs of parameters and data were generated (in total, that's 400 testing subjects per set). To measure the quality of estimates, we employed two metrics:
\begin{itemize}
	\item Distance between parameter estimates and the true parameter values.
	
	Throughout most of this section (with an exception of Table~\ref{table:ne_base}), we shall consider only the MAPE statistics as it is easy to interpret, but root-mean-square error (RMSE) also can be found in supplementary data.
	\item Number of non-convergent models.
	
	We say that an optimization process failed to converge to the true parameter estimates if any of the 3 conditions are met:
	\begin{enumerate}
		\item MAPE $>$ 40\%;
		\item Numerical optimizer returns an error;
		\item The resulting objective function value is NaN.
	\end{enumerate}
\end{itemize} 

Throughout this section, wherever MAPE or RMSE are provided, they were calculated only on the intersection of sets of converged models from all of the compared methods.

\subsubsection{Comparison to the previous version}
\begin{table}[t!]
	\centering
	\begin{tabular}{lllll}
		\toprule
		{} &            N &      MAPE &      RMSE &        Time, s \\
		\midrule
		\class{Model}[DWLS]      &           78 &          14.76 &           0.11 &           0.18 \\
		\class{Model}[FIML]      &  \textbf{18} &  \textbf{11.70} &  \textbf{0.09} &           0.35 \\
		\class{Model}[GLS]       &           63 &           13.30 &            0.10 &           0.17 \\
		semopy1[GLS]     &           68 &          15.07 &           0.12 &           0.22 \\
		\class{Model}[MLW]       &           18 &          11.82 &           0.09 &  \textbf{0.12} \\
		semopy1[MLW]     &           21 &          13.11 &            0.10 &           0.21 \\
		\class{ModelMeans}[ML]   &           19 &          11.82 &           0.09 &            0.60 \\
		\class{ModelMeans}[REML] &           35 &          15.76 &           0.11 &           0.15 \\
		\class{Model}[ULS]       &           93 &          15.96 &           0.14 &           0.28 \\
		semopy1[ULS]     &          112 &          18.09 &           0.15 &            0.30 \\
		\bottomrule
	\end{tabular}
	\caption{Comparison of the core \textbf{semopy} methods to the older 1.3.1 version.}
	\label{table:ne_base}
\end{table}

First, we examine both estimates accuracy and working times of the new \textbf{semopy} to the older version of \textit{1.3.1} on the testing subset \textit{A}: as evident from the Table~\ref{table:ne_base}, the newer version has an edge not only in terms of working times, but also in terms of estimation accuracy. Here, we also would like to remind a reader that previously \citep{Igolkina:2020} we have demonstrated that \textbf{semopy} outperforms the most popular SEM package \textbf{lavaan}, therefore the new results solidify the past result.

\subsubsection{Performance on testing sets}
For subsets \textit{B}-\textit{F}, we omit results for the older version of \textbf{semopy} and leave only some of the best-performing methods as reported by Table~\ref{table:ne_base}, and provide plots of number of non-covergent models $N$ in Figure~\ref{fig:performance}. Tables with exact $N$ and MAPE statistic can be found in Appendix~\ref{app:tables}.
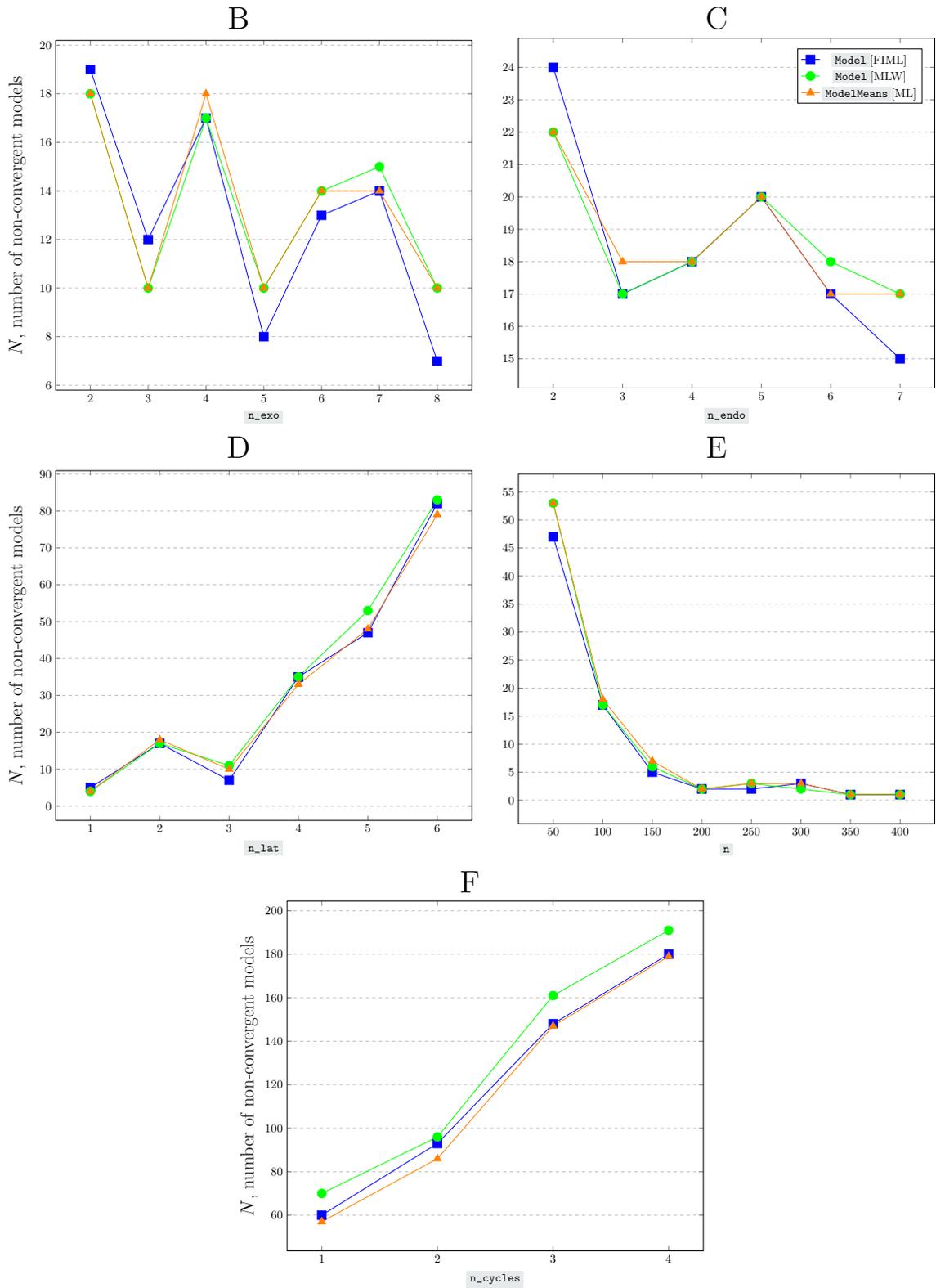
\begin{figure}
	\centering
	
	\begin{tabular}{cc}
	{\Large B} & {\Large C} \\
	\begin{tikzpicture}
		\begin{scope}[scale=0.5, transform shape]
			\begin{axis}[
				label style={font=\LARGE},
				height=0.5\textheight,
				xlabel={\code{n\_exo}},
				ylabel={$N$, number of non-convergent models},
				ymajorgrids=true,
				grid style=dashed,
				xtick={2,3,4,5,6,7,8}
				]
				\addplot[style1, ]
				coordinates {
					(2,19)(3,12)(4,17)(5,8)(6,13)(7,14)(8,7)
				};
				\addplot[style2, ]
				coordinates {
					(2,18)(3,10)(4,17)(5,10)(6,14)(7,15)(8,10)
				};
				\addplot[style3,]
				coordinates {
					(2,18)(3,10)(4,18)(5,10)(6,14)(7,14)(8,10)
				};
				
			\end{axis}
		\end{scope}
	\end{tikzpicture}

	&
	
	\begin{tikzpicture}
		\begin{scope}[scale=0.5, transform shape]
			\pgfkeys{{/pgfplots/legend entry/.code=\addlegendentry{#1}}}
			\begin{axis}[
				label style={font=\LARGE},
				height=0.5\textheight,
				xlabel={\code{n\_endo}},
				legend pos=north east,
				ymajorgrids=true,
				grid style=dashed,
				xtick={2,3,4,5,6,7}
				]
				\addplot[style1, legend entry={\code{Model}[FIML]}]
				coordinates {
					(2,24)(3,17)(4,18)(5,20)(6,17)(7,15)
				};
				\addplot[style2, legend entry={\code{Model}[MLW]}]
				coordinates {
					(2,22)(3,17)(4,18)(5,20)(6,18)(7,17)
				};
				\addplot[style3, legend entry={\code{ModelMeans}[ML]}]
				coordinates {
					(2,22)(3,18)(4,18)(5,20)(6,17)(7,17)
				};
			\end{axis}
		\end{scope}
	\end{tikzpicture}
	\\
	
	{\Large D} & {\Large E} \\
	\begin{tikzpicture}
		\begin{scope}[scale=0.5, transform shape]
			\begin{axis}[
				label style={font=\LARGE},
				height=0.5\textheight,
				xlabel={\code{n\_lat}},
				ylabel={$N$, number of non-convergent models},
				legend pos=north east,
				ymajorgrids=true,
				grid style=dashed,
				xtick={1,2,3,4,5,6}
				]
				\addplot[style1, ]
				coordinates {
					(1,5)(2,17)(3,7)(4,35)(5,47)(6,82)
				};
				\addplot[style2, ]
				coordinates {
					(1,4)(2,17)(3,11)(4,35)(5,53)(6,83)
				};
				\addplot[style3,]
				coordinates {
					(1,4)(2,18)(3,10)(4,33)(5,48)(6,79)
				};
			\end{axis}
		\end{scope}
	\end{tikzpicture}
	&
	
	\begin{tikzpicture}
		\begin{scope}[scale=0.5, transform shape]
			\begin{axis}[
				label style={font=\LARGE},
				height=0.5\textheight,
				xlabel={\code{n}},
				legend pos=north east,
				ymajorgrids=true,
				grid style=dashed,
				xtick={50,100,150,200,250,300,350,400}
				]
				\addplot[style1, ]
				coordinates {
					(50,47)(100,17)(150,5)(200,2)(250,2)(300,3)(350,1)(400,1)
				};
				\addplot[style2, ]
				coordinates {
					(50,53)(100,17)(150,6)(200,2)(250,3)(300,2)(350,1)(400,1)
				};
				\addplot[style3, ]
				coordinates {
					(50,53)(100,18)(150,7)(200,2)(250,3)(300,3)(350,1)(400,1)
				};
			\end{axis}
		\end{scope}
	\end{tikzpicture}
	
\end{tabular}
\begin{tabular}{c}
	{\Large F}\\
	
	\begin{tikzpicture}
		\begin{scope}[scale=0.5, transform shape]
			\begin{axis}[
				label style={font=\LARGE},
				height=0.5\textheight,
				xlabel={\code{n\_cycles}},
				ylabel={$N$, number of non-convergent models},
				legend pos=north east,
				ymajorgrids=true,
				grid style=dashed,
				xtick={1,2,3,4}
				]
				\addplot[style1,]
				coordinates {
					(1,60)(2,93)(3,148)(4,180)
				};
				\addplot[style2,]
				coordinates {
					(1,70)(2,96)(3,161)(4,191)
				};
				\addplot[style3, ]
				coordinates {
					(1,57)(2,86)(3,147)(4,179)
				};
			\end{axis}
		\end{scope}
	\end{tikzpicture}

\end{tabular}

	\caption{Number of non-convergent models for sets B-F.}
	\label{fig:performance}
\end{figure}
\subsubsection{Performance on the testing set G with random effects}

\begin{table}[H]
	\centering
	
	\begin{tabular}{llll}
		\toprule
		{} & N & MAPE & RMSE\\
		\midrule
		\class{Model} & 195&26.67 & 0.17\\
		\class{ModelEffects} & 59&15.36 & 0.12\\
		\bottomrule
	\end{tabular}
	\caption{Models performances on data with random effects.}
	\label{table:rf}
\end{table}

Here, we draw a comparison of \class{ModelEffects} and \class{Model} evaluations on datasets "contaminated" with random effects. The purpose of this comparison is to showcase a drastic negative effect of not taking random effects into an account: see Table~\ref{table:rf}. As expected, \class{Model} demonstrates very poor performance in this setting, whereas \class{ModelEffects} tackles the situation considerably better.

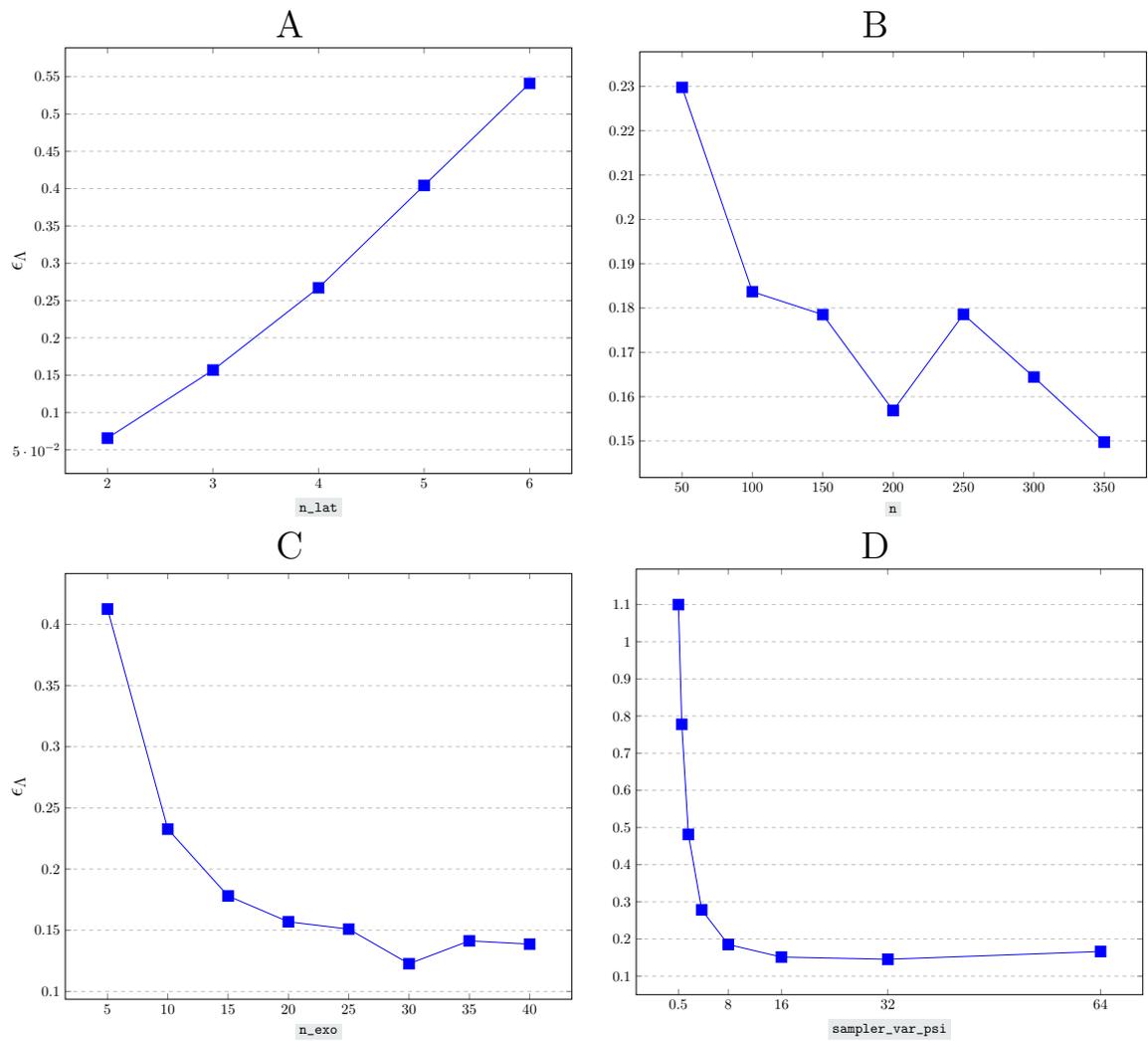
\begin{figure}
	\centering
	\begin{tabular}{cc}
	{\Large A} & {\Large B} \\
	
	\begin{tikzpicture}
		\begin{scope}[scale=0.5, transform shape]
			\begin{axis}[
				label style={font=\LARGE},
				height=0.5\textheight,
				xlabel={\code{n\_lat}},
				ylabel={$\epsilon_\Lambda$},
				ymajorgrids=true,
				grid style=dashed,
				xtick={2,3,4,5,6}
				]
				\addplot[style1]
				coordinates {
					(2,0.06572420634920635)(3,0.15688973063973066)(4,0.267127385142091)(5,0.40430904237470283)(6,0.5410405531518232)
				};
			\end{axis}
		\end{scope}
	\end{tikzpicture}
	&
	
	\begin{tikzpicture}
		\begin{scope}[scale=0.5, transform shape]
			\begin{axis}[
				label style={font=\LARGE},
				height=0.5\textheight,
				xlabel={\code{n}},
				ymajorgrids=true,
				grid style=dashed,
				xtick={50,100,150,200,250,300,350}
				]
				\addplot[style1,]
				coordinates {
					(50,0.22977456185789522)(100,0.18367542950876287)(150,0.17851031684365018)(200,0.15688973063973066)(250,0.17856330397997064)(300,0.1644194077527411)(350,0.1497511439178106)
				};
			\end{axis}
		\end{scope}
	\end{tikzpicture}
	\\
	{\Large C} & {\Large D} \\
	
	\begin{tikzpicture}
		\begin{scope}[scale=0.5, transform shape]
			\begin{axis}[
				label style={font=\LARGE},
				height=0.5\textheight,
				xlabel={\code{n\_exo}},
				ylabel={$\epsilon_\Lambda$},
				ymajorgrids=true,
				grid style=dashed,
				xtick={5,10,15,20,25,30,35,40}
				]
				\addplot[style1,]
				coordinates {
					(5,0.412540145040145)(10,0.2326834585167918)(15,0.17798692048692052)(20,0.15688973063973066)(25,0.15091740050073382)(30,0.12264061555728224)(35,0.14132349132349134)(40,0.138665285331952)
				};
			\end{axis}
		\end{scope}
	\end{tikzpicture}
	&
	
	\begin{tikzpicture}
		\begin{scope}[scale=0.5, transform shape]
			\begin{axis}[
				label style={font=\LARGE},
				height=0.5\textheight,
				xlabel={\code{sampler\_var\_psi}},
				ymajorgrids=true,
				grid style=dashed,
				xtick={0.5,8.0,16.0,32.0,64.0}
				]
				\addplot[style1,]
				coordinates {
					(0.5,1.0999919062419063)(1.0,0.7777052577052577)(2.0,0.4812857204523871)(4.0,0.27824095657428993)(8.0,0.18520514978848313)(16.0,0.15141673141673143)(32.0,0.1454008029008029)(64.0,0.16636493136493138)
				};
			\end{axis}
		\end{scope}
	\end{tikzpicture}

\end{tabular}
	\caption{EFA procedure performance on different datasets.}
	\label{fig:efa_results}
\end{figure}

\subsection{EFA}\label{sect:exp_efa}

\begin{table}
	\centering
	\begin{tabular}{llllll}
		\toprule
		{}&\ \#  & \texttt{n\_lat} & \texttt{n} & \texttt{n\_exo} & $\sigma_\eta$   \\
		\midrule
		\multirow{6}{*}{\Large A} & 1 & \textbf{1} & 200 & 20 & 1.0 \\
		& 2 & \textbf{2} & 200 & 20 & 1.0 \\
		& 3 & \textbf{3} & 200 & 20 & 1.0 \\
		& 4 & \textbf{4} & 200 & 20 & 1.0 \\
		& 5 & \textbf{5} & 200 & 20 & 1.0 \\
		& 6 & \textbf{6} & 200 & 20 & 1.0 \\
		\hline
		\multirow{7}{*}{\Large B} & 7 & 3 & \textbf{50} & 20 & 1.0 \\
		& 8 & 3 & \textbf{100} & 20 & 1.0 \\
		& 9 & 3 & \textbf{150} & 20 & 1.0 \\
		& 10 & 3 & \textbf{200} & 20 & 1.0 \\
		& 11 & 3 & \textbf{250} & 20 & 1.0 \\
		& 12 & 3 & \textbf{300} & 20 & 1.0 \\
		& 13 & 3 & \textbf{350} & 20 & 1.0 \\
		\hline
		\multirow{8}{*}{\Large C} & 14 & 3 & 200 & \textbf{5} & 1.0 \\
		& 15 & 3 & 200 & \textbf{10} & 1.0 \\
		& 16 & 3 & 200 & \textbf{15} & 1.0 \\
		& 17 & 3 & 200 & \textbf{20} & 1.0 \\
		& 18 & 3 & 200 & \textbf{25} & 1.0 \\
		& 19 & 3 & 200 & \textbf{30} & 1.0 \\
		& 20 & 3 & 200 & \textbf{35} & 1.0 \\
		& 21 & 3 & 200 & \textbf{40} & 1.0 \\
		\hline
		\multirow{9}{*}{\Large D} & 22 & 3 & 200 & 0 & \textbf{0.5} \\
		& 24 & 3 & 200 & 0 & \textbf{1.0} \\
		& 25 & 3 & 200 & 0 & \textbf{2.0} \\
		& 26 & 3 & 200 & 0 & \textbf{4.0} \\
		& 27 & 3 & 200 & 0 & \textbf{8.0} \\
		& 28 & 3 & 200 & 0 & \textbf{16.0} \\
		& 29 & 3 & 200 & 0 & \textbf{32.0} \\
		& 30 & 3 & 200 & 0 & \textbf{64.0} \\
		
		\bottomrule
	\end{tabular}
	\caption{Configurations of generated models and their respective datasets. The generated sets are separated into subsets A, B, C, D with respect to a varying parameter. Parameters \code{n\_inds} and \code{p\_join} are fixed to $3$ and $0.05$ respectively. $\sigma_\eta$ is a variance of latent variables: variance of latent variable is sampled from the distribution $\sigma_\eta * \mathcal{U}(0.7, 1.4)$.}
	\label{table:efa_models}
\end{table}

We also used our testing framework to generate $K$ datasets and ran our EFA procedure on each of them (see Table~\ref{table:efa_models}). Then, we compared the resulting CFA models to the measurement part of true models. To perform this comparison, we invented a simple metric $\epsilon_\Lambda = \frac{m}{n}$, where $m$ is a number of falsely identified/not-identified loadings and $n$ is a total number of loadings in the true model. Note that if the EFA procedure failed to correctly estimate the number of latent factors, $\epsilon_\Lambda$ can be greater than $1.0$. The results are depicted in Figure~\ref{fig:efa_results} that is built by points from tables in Appendix~\ref{app:tables}. 

\section{Concluding remarks}

Many options and features that are present in the package, didn't make it to this article, for the sake of not bloating it. \textbf{semopy} is better documented at the project's website (\url{https://semopy.com}), although without technical and mathematical details.

Despite that \textbf{semopy} may appear as a rather big project, we abstain from declaring it complete. We welcome prospective users and we expect any kind of feedback, as well as feature requests. 

\section*{Computational details}

All numerical results that are present in this article were obtained with \texttt{Python} \textit{3.9.4} and \textbf{semopy} version \textit{2.2.2}. \textbf{OpenBLAS} was used as an backend for \textbf{BLAS}/\textbf{LAPACK} APIs. Computations were done on a \textit{Intel(R) Core(TM) i5-3230M CPU @ 2.60GHz}

\section*{Data availability}

Randomly generated datasets and models that were used in Section~\ref{sect:experiments} are available at the file storage \url{https://drive.google.com/file/d/1MRZ1_HrDjsEQL2tgnABRafF02XWmlwpI/view?usp=sharing}, alongside with scripts that were used to generate and evaluate them. Alternatively, one can fetch scripts only from the repository \url{https://gitlab.com/georgy.m/semopy-tests}, re-generate datasets and then reproduce the results.

The package \textbf{semopy} can be retrieved either directly from the PyPi repository via \textit{pip}, i.e.
\begin{lstlisting}
pip install semopy
\end{lstlisting}
or via git repository \url{https://gitlab.com/georgy.m/semopy}.

To ensure that results are exactly reproducible, version \textit{2.2.2} should be installed:
\begin{lstlisting}
pip install semopy==2.2.2
\end{lstlisting}

\section*{Acknowledgments}

This work was supported by RFBR Grant No. 18‒29‒13033.

\pagestyle{fancy}
\fancyhf{}
\rhead{}
\lhead{\leftmark}
\rfoot{\thepage}
\bibliography{refs}
\pagestyle{fancy}
\fancyhf{}
\lhead{\leftmark}
\rfoot{\thepage}
\newpage

\begin{appendix}
	
	\section{Toy datasets} \label{app:toys}
	A reader can use results below to double-check output of \pkg{semopy} on their machine.
\subsection*{\class{Model}}

\consolein
\begin{lstlisting}[style=codeinput]
from semopy import Model, examples
\end{lstlisting}

\subsubsection*{\code{univariate\_regression}}

	\consolein
\begin{lstlisting}[style=codeinput]
ex = examples.univariate_regression
desc, data = ex.get_model(), ex.get_data()
m = Model(desc)
r = m.fit(data)
print(m.inspect())
	\end{lstlisting}
	\consoleout
\begin{lstlisting}[style=codeoutput]
  lval  op rval  Estimate  Std. Err    z-value       p-value
0    y   ~    x -1.221069  0.083165 -14.682538  0.000000e+00
1    y  ~~    y  0.670367  0.094804   7.071068  1.537437e-12
	\end{lstlisting}

\subsubsection*{\code{univariate\_regression\_many}}

	\consolein
\begin{lstlisting}[style=codeinput]
ex = examples.univariate_regression_many
desc, data = ex.get_model(), ex.get_data()
m = Model(desc)
r = m.fit(data)
print(m.inspect())
	\end{lstlisting}
	\consoleout
\begin{lstlisting}[style=codeoutput]
  lval  op rval  Estimate  Std. Err    z-value       p-value
0    y   ~   x1  1.399551  0.091138  15.356385  0.000000e+00
1    y   ~   x2  0.450561  0.097883   4.603051  4.163465e-06
2    y   ~   x3  1.190470  0.086499  13.762839  0.000000e+00
3    y  ~~    y  0.878486  0.124237   7.071068  1.537437e-12
	\end{lstlisting}

\subsubsection*{\code{multivariate\_regression}}

	\consolein
\begin{lstlisting}[style=codeinput]
ex = examples.multivariate_regression
desc, data = ex.get_model(), ex.get_data()
m = Model(desc)
r = m.fit(data)
print(m.inspect())
	\end{lstlisting}
	\consoleout
\begin{lstlisting}[style=codeoutput]
   lval  op rval  Estimate  Std. Err    z-value       p-value
0    y1   ~   x1 -1.389754  0.073417 -18.929470  0.000000e+00
1    y1   ~   x2 -1.138405  0.087966 -12.941462  0.000000e+00
2    y1   ~   x3 -0.317893  0.072576  -4.380132  1.186073e-05
3    y2   ~   x1 -0.745837  0.097974  -7.612623  2.686740e-14
4    y2   ~   x2  1.074436  0.117388   9.152855  0.000000e+00
5    y2   ~   x3 -1.130890  0.096851 -11.676597  0.000000e+00
6    y3   ~   x1  0.702778  0.064270  10.934755  0.000000e+00
7    y3   ~   x2  1.235044  0.077006  16.038334  0.000000e+00
8    y3   ~   x3 -0.920469  0.063534 -14.487925  0.000000e+00
9    y1  ~~   y1  0.637755  0.090192   7.071068  1.537437e-12
10   y3  ~~   y3  0.488735  0.069118   7.071068  1.537437e-12
11   y2  ~~   y2  1.135729  0.160616   7.071068  1.537437e-12
	\end{lstlisting}

\subsubsection*{\code{political\_democracy}}

	\consolein
\begin{lstlisting}[style=codeinput]
ex = examples.political_democracy
desc, data = ex.get_model(), ex.get_data()
m = Model(desc)
r = m.fit(data)
print(m.inspect())
	\end{lstlisting}
	\consoleout
\begin{lstlisting}[style=codeoutput]
     lval  op   rval  Estimate  Std. Err    z-value   p-value
0   dem60   ~  ind60  1.482379  0.399024   3.715017  0.000203
1   dem65   ~  ind60  0.571912  0.221383   2.583364  0.009784
2   dem65   ~  dem60  0.837574  0.098446   8.507992       0.0
3      x1   ~  ind60  1.000000         -          -         -
4      x2   ~  ind60  2.180494  0.138565  15.736254       0.0
5      x3   ~  ind60  1.818546  0.151993   11.96465       0.0
6      y1   ~  dem60  1.000000         -          -         -
7      y2   ~  dem60  1.256819  0.182687   6.879647       0.0
8      y3   ~  dem60  1.058174  0.151521   6.983699       0.0
9      y4   ~  dem60  1.265186  0.145151   8.716344       0.0
10     y5   ~  dem65  1.000000         -          -         -
11     y6   ~  dem65  1.185743  0.168908   7.020032       0.0
12     y7   ~  dem65  1.279717  0.159996    7.99841       0.0
13     y8   ~  dem65  1.266084  0.158238   8.001141       0.0
14  dem60  ~~  dem60  3.950849  0.920451   4.292296  0.000018
15  dem65  ~~  dem65  0.172210  0.214861   0.801494  0.422846
16  ind60  ~~  ind60  0.448321  0.086677   5.172345       0.0
17     y1  ~~     y5  0.624423  0.358435   1.742083  0.081494
18     y1  ~~     y1  1.892743   0.44456   4.257565  0.000021
19     y2  ~~     y4  1.319589   0.70268   1.877937   0.06039
20     y2  ~~     y6  2.156164  0.734155   2.936934  0.003315
21     y2  ~~     y2  7.385292  1.375671   5.368501       0.0
22     y3  ~~     y7  0.793329  0.607642   1.305585  0.191694
23     y3  ~~     y3  5.066628  0.951722   5.323646       0.0
24     y4  ~~     y8  0.347222  0.442234   0.785154  0.432363
25     y4  ~~     y4  3.147911  0.738841   4.260605   0.00002
26     y6  ~~     y8  1.357037    0.5685   2.387047  0.016984
27     y6  ~~     y6  4.954364  0.914284   5.418843       0.0
28     x3  ~~     x3  0.466732  0.090168   5.176276       0.0
29     y8  ~~     y8  3.256389   0.69504   4.685182  0.000003
30     y7  ~~     y7  3.430032  0.712732   4.812512  0.000001
31     y5  ~~     y5  2.351910  0.480369   4.896044  0.000001
32     x2  ~~     x2  0.119894  0.069747   1.718973  0.085619
33     x1  ~~     x1  0.081573  0.019495   4.184317  0.000029
	\end{lstlisting}

\subsubsection*{\code{holzinger39}}

	\consolein
\begin{lstlisting}[style=codeinput]
ex = examples.holzinger39
desc, data = ex.get_model(), ex.get_data()
m = Model(desc)
r = m.fit(data)
print(m.inspect())
	\end{lstlisting}
	\consoleout
\begin{lstlisting}[style=codeoutput]
       lval  op     rval  Estimate  Std. Err    z-value   p-value
0        x1   ~   visual  1.000000         -          -         -
1        x2   ~   visual  0.554421  0.099727   5.559413       0.0
2        x3   ~   visual  0.730526   0.10918   6.691009       0.0
3        x4   ~  textual  1.000000         -          -         -
4        x5   ~  textual  1.113076  0.065392  17.021522       0.0
5        x6   ~  textual  0.926120  0.055425  16.709493       0.0
6        x7   ~    speed  1.000000         -          -         -
7        x8   ~    speed  1.179980  0.165045   7.149459       0.0
8        x9   ~    speed  1.082517  0.151354   7.152197       0.0
9   textual  ~~  textual  0.980034  0.112145   8.739002       0.0
10  textual  ~~    speed  0.173603  0.049316   3.520223  0.000431
11  textual  ~~   visual  0.408277  0.073527    5.55273       0.0
12   visual  ~~   visual  0.808310  0.145287   5.563548       0.0
13    speed  ~~    speed  0.383377  0.086171   4.449045  0.000009
14    speed  ~~   visual  0.262135  0.056252   4.659977  0.000003
15       x8  ~~       x8  0.487934  0.074167   6.578856       0.0
16       x3  ~~       x3  0.843731  0.090625    9.31016       0.0
17       x7  ~~       x7  0.799708  0.081387   9.825966       0.0
18       x9  ~~       x9  0.565804  0.070757   7.996483       0.0
19       x5  ~~       x5  0.446208  0.058387   7.642264       0.0
20       x6  ~~       x6  0.356171   0.04303   8.277334       0.0
21       x4  ~~       x4  0.371117  0.047712   7.778264       0.0
22       x2  ~~       x2  1.133391  0.101711  11.143202       0.0
23       x1  ~~       x1  0.550161  0.113439    4.84983  0.000001
	\end{lstlisting}

\subsection*{\class{ModelMeans}} 
\consolein
\begin{lstlisting}[style=codeinput]
	from semopy import ModelMeans, examples
\end{lstlisting}
\subsubsection*{\code{univariate\_regression}}

	\consolein
\begin{lstlisting}[style=codeinput]
ex = examples.univariate_regression
desc, data = ex.get_model(), ex.get_data()
m = ModelMeans(desc)
r = m.fit(data)
print(m.inspect())
	\end{lstlisting}
	\consoleout
\begin{lstlisting}[style=codeoutput]
  lval  op rval  Estimate  Std. Err    z-value       p-value
0    y   ~    x -1.221251  0.083167 -14.684298  0.000000e+00
1    y   ~    1 -1.421354  0.082150 -17.301864  0.000000e+00
2    y  ~~    y  0.670406  0.094810   7.071068  1.537437e-12
	\end{lstlisting}

\subsubsection*{\code{univariate\_regression\_many}}

	\consolein
\begin{lstlisting}[style=codeinput]
ex = examples.univariate_regression_many
desc, data = ex.get_model(), ex.get_data()
m = ModelMeans(desc)
r = m.fit(data)
print(m.inspect())
	\end{lstlisting}
	\consoleout
\begin{lstlisting}[style=codeoutput]
  lval  op rval  Estimate  Std. Err    z-value       p-value
0    y   ~   x1  1.399662  0.091104  15.363298  0.000000e+00
1    y   ~   x2  0.450742  0.097847   4.606606  4.092950e-06
2    y   ~   x3  1.190553  0.086467  13.768895  0.000000e+00
3    y   ~    1  1.120529  0.094777  11.822861  0.000000e+00
4    y  ~~    y  0.877835  0.124145   7.071068  1.537437e-12
	\end{lstlisting}

\subsubsection*{\code{multivariate\_regression}}

	\consolein
\begin{lstlisting}[style=codeinput]
ex = examples.multivariate_regression
desc, data = ex.get_model(), ex.get_data()
m = ModelMeans(desc)
r = m.fit(data)
print(m.inspect())
	\end{lstlisting}
	\consoleout
\begin{lstlisting}[style=codeoutput]
   lval  op rval  Estimate  Std. Err    z-value       p-value
0    y1   ~   x1 -1.389720  0.073417 -18.929022  0.000000e+00
1    y1   ~   x2 -1.138398  0.087966 -12.941384  0.000000e+00
2    y1   ~   x3 -0.317953  0.072576  -4.380965  1.181546e-05
3    y2   ~   x1 -0.745740  0.097969  -7.611965  2.708944e-14
4    y2   ~   x2  1.074526  0.117383   9.154020  0.000000e+00
5    y2   ~   x3 -1.130938  0.096847 -11.677611  0.000000e+00
6    y3   ~   x1  0.702778  0.064269  10.934870  0.000000e+00
7    y3   ~   x2  1.235047  0.077005  16.038535  0.000000e+00
8    y3   ~   x3 -0.920454  0.063533 -14.487837  0.000000e+00
9    y1   ~    1 -1.456621  0.080268 -18.147061  0.000000e+00
10   y2   ~    1  0.929294  0.107110   8.676035  0.000000e+00
11   y3   ~    1  0.992040  0.070266  14.118332  0.000000e+00
12   y1  ~~   y1  0.637755  0.090192   7.071068  1.537437e-12
13   y3  ~~   y3  0.488724  0.069116   7.071068  1.537437e-12
14   y2  ~~   y2  1.135630  0.160602   7.071068  1.537437e-12
	\end{lstlisting}

\subsubsection*{\code{political\_democracy}}

	\consolein
\begin{lstlisting}[style=codeinput]
ex = examples.political_democracy
desc, data = ex.get_model(), ex.get_data()
m = ModelMeans(desc)
r = m.fit(data)
print(m.inspect())
	\end{lstlisting}
	\consoleout
\begin{lstlisting}[style=codeoutput]
     lval  op   rval  Estimate  Std. Err    z-value   p-value
0   dem60   ~  ind60  1.482999  0.399149   3.715401  0.000203
1   dem65   ~  ind60  0.572322  0.221313   2.586027  0.009709
2   dem65   ~  dem60  0.837346  0.098351   8.513859       0.0
3      x1   ~  ind60  1.000000         -          -         -
4      x2   ~  ind60  2.180375   0.13851  15.741609       0.0
5      x3   ~  ind60  1.818522   0.15196  11.967144       0.0
6      y1   ~  dem60  1.000000         -          -         -
7      y2   ~  dem60  1.256753  0.182439   6.888611       0.0
8      y3   ~  dem60  1.057746  0.151385   6.987136       0.0
9      y4   ~  dem60  1.264790  0.145006   8.722314       0.0
10     y5   ~  dem65  1.000000         -          -         -
11     y6   ~  dem65  1.185687   0.16881   7.023806       0.0
12     y7   ~  dem65  1.279531  0.159903   8.001935       0.0
13     y8   ~  dem65  1.265935   0.15811    8.00667       0.0
14     x1   ~      1  5.054393  0.084062  60.126661       0.0
15     x2   ~      1  4.792218   0.17327  27.657545       0.0
16     x3   ~      1  3.557715  0.161233  22.065738       0.0
17     y1   ~      1  5.464665  0.301854   18.10368       0.0
18     y2   ~      1  4.256439   0.44987   9.461478       0.0
19     y3   ~      1  6.563146  0.375891  17.460248       0.0
20     y4   ~      1  4.452537   0.38391  11.597855       0.0
21     y5   ~      1  5.136254  0.300511  17.091727       0.0
22     y6   ~      1  2.978057  0.385931   7.716551       0.0
23     y7   ~      1  6.196289  0.377202  16.426972       0.0
24     y8   ~      1  4.043383  0.371319  10.889254       0.0
25  dem60  ~~  dem60  3.956039  0.921185   4.294508  0.000018
26  dem65  ~~  dem65  0.172487  0.214803   0.803001  0.421974
27  ind60  ~~  ind60  0.448436  0.086692   5.172765       0.0
28     y1  ~~     y5  0.623671  0.358319   1.740548  0.081763
29     y1  ~~     y1  1.891402  0.444423   4.255861  0.000021
30     y2  ~~     y4  1.313085   0.70198   1.870545  0.061408
31     y2  ~~     y6  2.152828  0.733771   2.933925  0.003347
32     y2  ~~     y2  7.372791  1.373882   5.366395       0.0
33     y3  ~~     y7  0.794961  0.607702   1.308143  0.190825
34     y3  ~~     y3  5.067487  0.951738   5.324453       0.0
35     y4  ~~     y8  0.348246  0.442238   0.787464   0.43101
36     y4  ~~     y4  3.147907  0.738783   4.260936   0.00002
37     y6  ~~     y8  1.356165  0.568281   2.386434  0.017013
38     y6  ~~     y6  4.953952  0.914241   5.418647       0.0
39     x3  ~~     x3  0.466708  0.090157   5.176583       0.0
40     y8  ~~     y8  3.254068  0.694596   4.684837  0.000003
41     y7  ~~     y7  3.431334  0.712843   4.813589  0.000001
42     y5  ~~     y5  2.350969  0.480238   4.895427  0.000001
43     x2  ~~     x2  0.119802  0.069721   1.718314  0.085739
44     x1  ~~     x1  0.081551   0.01949   4.184259  0.000029
	\end{lstlisting}

\subsubsection*{\code{holzinger39}}

	\consolein
\begin{lstlisting}[style=codeinput]
ex = examples.holzinger39
desc, data = ex.get_model(), ex.get_data()
m = ModelMeans(desc)
r = m.fit(data)
print(m.inspect())
	\end{lstlisting}
	\consoleout
\begin{lstlisting}[style=codeoutput]
       lval  op     rval  Estimate  Std. Err    z-value   p-value
0        x1   ~   visual  1.000000         -          -         -
1        x2   ~   visual  0.553493  0.099663    5.55366       0.0
2        x3   ~   visual  0.729357  0.109106   6.684861       0.0
3        x4   ~  textual  1.000000         -          -         -
4        x5   ~  textual  1.113076   0.06542  17.014187       0.0
5        x6   ~  textual  0.926147  0.055449  16.702625       0.0
6        x7   ~    speed  1.000000         -          -         -
7        x8   ~    speed  1.179973  0.164992   7.151679       0.0
8        x9   ~    speed  1.081572  0.151176   7.154404       0.0
9        x1   ~        1  4.935774  0.067178   73.47271       0.0
10       x2   ~        1  6.088037  0.067754  89.854497       0.0
11       x3   ~        1  2.250415   0.06508   34.57913       0.0
12       x4   ~        1  3.060910  0.066987  45.694203       0.0
13       x5   ~        1  4.340535  0.074258   58.45232       0.0
14       x6   ~        1  2.185574  0.063044  34.667251       0.0
15       x7   ~        1  4.185915  0.062695   66.76601       0.0
16       x8   ~        1  5.527082  0.058269  94.854886       0.0
17       x9   ~        1  5.374125   0.05807  92.546353       0.0
18  textual  ~~  textual  0.979483  0.112105   8.737173       0.0
19  textual  ~~    speed  0.173487  0.049313   3.518059  0.000435
20  textual  ~~   visual  0.408245  0.073525   5.552498       0.0
21   visual  ~~   visual  0.809338  0.145464   5.563848       0.0
22    speed  ~~    speed  0.383726  0.086207   4.451217  0.000009
23    speed  ~~   visual  0.262232  0.056277   4.659693  0.000003
24       x8  ~~       x8  0.487697  0.074193   6.573326       0.0
25       x3  ~~       x3  0.844326  0.090622   9.316999       0.0
26       x7  ~~       x7  0.799415  0.081382   9.822992       0.0
27       x9  ~~       x9  0.566112  0.070737   8.003081       0.0
28       x5  ~~       x5  0.446256  0.058393   7.642312       0.0
29       x6  ~~       x6  0.356202  0.043035   8.277039       0.0
30       x4  ~~       x4  0.371174  0.047718   7.778523       0.0
31       x2  ~~       x2  1.133843  0.101723  11.146333       0.0
32       x1  ~~       x1  0.549053  0.113601    4.83318  0.000001
	\end{lstlisting}

	\newpage

	\section{Gradient of matrix-variate normal loglikelihood} \label{app:matnormgrad}
		\begin{equation}\label{eq:loglikg}
		l(\theta) = tr\{L\} tr\{T^{-1} \widehat{Z}^T L^{-1} \widehat{Z}\} + m \ln|T| + n \ln|L| - n m \ln tr\{L\}
	\end{equation}
	We are seeking to find a derivative of $l(\theta)$ from Equation \ref{eq:loglikg} with respect to a parameter $\theta_i$ from the vector $\theta$. Next, we are using the product rule for the differentiation and the cyclic permutation property of the trace operator:
	
	\begin{equation*}
	\begin{split}
	\frac{\partial}{\partial \theta_i} l(\theta) = tr\{\frac{\partial L}{\partial \theta_i}\} tr\{T^{-1} \widehat{Z}^T L^{-1} \widehat{Z}\} + tr\{L\}tr\{T^{-1} \frac{\partial \widehat{Z}^T}{\partial \theta_i} L^{-1} \widehat{Z} + T^{-1} \widehat{Z}^{T} L^{-1} \frac{\partial \widehat{Z}}{\partial \theta_i}\} - \\ 
	- tr\{L\} tr\{\underbrace{T^{-1}\frac{\partial T}{\partial \theta_i}}_{A_i}T^{-1} \widehat{Z}^T L^{-1} \widehat{Z} + T^{-1} \widehat{Z}^T \underbrace{\frac{\partial L}{\partial \theta_i}L^{-1}}_{B_i} L^{-1} \widehat{Z}\} + \\ + m tr\{\underbrace{T^{-1} \frac{\partial T}{\partial \theta_i}}_{A_i}\} + n tr\{\underbrace{ \frac{\partial L}{\partial \theta_i}L^{-1}}_{B_i}\} - n m \frac{tr\{\frac{\partial L}{\partial \theta_i}\}}{tr\{L\}} 
	\end{split}
	\end{equation*}
	Notice that
	\begin{equation*}
		\begin{split}
		tr\{T^{-1} \frac{\partial \widehat{Z}^T}{\partial \theta_i} L^{-1} \widehat{Z} + T^{-1} \widehat{Z}^{T} L^{-1} \frac{\partial \widehat{Z}}{\partial \theta_i}\} = tr\{T^{-1} \frac{\partial \widehat{Z}^T}{\partial \theta_i} L^{-1} \widehat{Z}\} + tr\{T^{-1} \widehat{Z}^{T} L^{-1} \frac{\partial \widehat{Z}}{\partial \theta_i}\} = \\
		tr\{ \widehat{Z}^T L^{-1} \frac{\partial \widehat{Z}}{\partial \theta_i} T^{-1} \} + tr\{T^{-1} \widehat{Z}^{T} L^{-1} \frac{\partial \widehat{Z}}{\partial \theta_i}\} = 
			tr\{T^{-1} \widehat{Z}^T L^{-1} \frac{\partial \widehat{Z}}{\partial \theta_i} \} + tr\{T^{-1} \widehat{Z}^{T} L^{-1} \frac{\partial \widehat{Z}}{\partial \theta_i}\} = \\
			2 tr\{\underbrace{T^{-1} \widehat{Z}^{T} L^{-1}}_{C_0} \frac{\partial \widehat{Z}}{\partial \theta_i}\} = 2 tr\{C_0 \frac{\partial \widehat{Z}}{\partial \theta_i}\},
	\end{split}
\end{equation*}
 here, we have used the invariance of trace under transposition and cyclic permutations properties. We continue to simplify the derivative:
 \begin{equation*}
 	\begin{split}
 	\frac{\partial}{\partial \theta_i} l(\theta) = tr\{\frac{\partial L}{\partial \theta_i}\} tr\{\underbrace{C_0 \widehat{Z}}_{C_1}\} + 2 tr\{L\}tr\{C_0 \frac{\partial \widehat{Z}}{\partial \theta_i}\} - \\ 
 	- tr\{L\} tr\{A_i \underbrace{C_0 \widehat{Z}}_{C_1} +B_i \underbrace{\widehat{Z} C_0}_{C_2} \}  + m tr \{A_i\} + n tr\{B_i\} - n m \frac{tr\{\frac{\partial L}{\partial \theta_i}\}}{tr\{L\}} 
 \end{split}
\end{equation*}
	The final compact form of the gradient:
	\begin{equation}\label{eq:matnormgrad}
	\begin{split}
	\frac{\partial}{\partial \theta_i} l(\theta) = tr\{\frac{\partial L}{\partial \theta_i}\} tr\{C_1\} + 2 tr\{L\}tr\{C_0 \frac{\partial \widehat{Z}}{\partial \theta_i}\} 
	- tr\{L\} tr\{A_i C_1 +B_i C_2 \} + \\ + m tr \{A_i\} + n tr\{B_i\} - n m \frac{tr\{\frac{\partial L}{\partial \theta_i}\}}{tr\{L\}} 
	\end{split},
	\end{equation}
	where $C_0 = T^{-1} \widehat{Z}^T L^{-1}$, $C_1 = C_0 \widehat{Z}$, $C_2 = \widehat{Z} C_0$ are matrices that need to be calculated only once, and $A_i = T^{-1} \frac{\partial T}{\partial \theta_i}$, $B_i = \frac{\partial L}{\partial \theta_i} L^{-1}$ have to be calculated for each of elements of the parameter vector $\theta$.

	\newpage
	\section{Fisher information matrix} \label{app:fisher_information}
	
	For the task of deriving Fisher Information Matrix (FIM) for the matrix-variate distribution, we use the fact that $vec(Z) \sim \mathcal{MN}(M, \widehat{L}, \widehat{T})$ iff $Z \sim \mathcal{N}(vec(M), \widehat{T} \otimes \widehat{L})$, where $\otimes$ is a Kronecker product and $vec$ is a vectorization operator. The result for multivariate-normal FIM $I(\theta)$ is known. If $x \sim \mathcal{N}(\mu, \Sigma)$, then
	\begin{equation}\label{eq:mvnfim}
		I(\theta)_{i,j} = \frac{\partial \mu^T }{\partial \theta_i} \Sigma^{-1} \frac{\partial \mu }{\partial \theta_k} + \frac{1}{2}tr\left\{\Sigma^{-1} \frac{\partial{\Sigma}}{\partial \theta_i} \Sigma^{-1} \frac{\partial{\Sigma}}{\partial \theta_k}
		\right\}
	\end{equation}
	One could compute $I(\theta)$ for $Z$ just by using the Equation \ref{eq:mvnfim} by plugging $\mu = vec(M)$, $\Sigma = \widehat{T} \times \widehat{L}$, however, we don't have actual $\widehat{L}$ or $\widehat{T}$ available, therefore $\Sigma$ is not known. Instead, we are working with some model-implied matrices $L = cov(Z), T = cov(Z^T)$. This is a problem, as for matrix-variate distributions, $L= tr\{\widehat{T}\} \widehat{L}, T = tr\{\widehat{L}\} \widehat{T}$. Although we can't estimate $tr\{\widehat{T}\}, tr\{\widehat{L}\}$, it is not necessary: this is completely similar to the problem we've encountered when computing likelihood function. Firstly, we observe that $tr\{T\} = tr\{L\} tr\{\widehat{T}\} \leftrightarrow tr\{L\} = \frac{tr\{T\}}{tr\{\widehat{T}\}}$. Secondly, we substitute
	$$\widehat{L} = \frac{1}{tr\{\widehat{T}\}} L, \widehat{T} = \frac{tr\{\widehat{T}\}}{tr\{T\}} T$$
	So the $\Sigma$ will have the following form:
	\begin{equation}\label{eq:sigma_fim}
	\Sigma =  \widehat{T} \otimes \widehat{L} = \left(\frac{tr\{\widehat{T}\}}{tr\{T\}} T \right) \otimes \left(\frac{1}{tr\{\widehat{T}\}} L\right) = \frac{1}{tr\{T\}} T \otimes L
	\end{equation}
	Most importantly, this straightforward approach would be very inefficient: each component of $I$ is computed in $O(n^3 m^3)$.

	 Next, we'll show that FIM can be computed very efficiently. First, we analyze the trace term of the Equation \ref{eq:mvnfim}, and then tackle the first quadratic term. We shall use the following 5 properties of the Kronecker product and \textit{vec} operator:
	\begin{enumerate}
		\item Inverse of Kronecker product: $$(A \otimes B)^{-1} = A^{-1} \otimes B^{-1}$$
		\item Mixed product:
		$$(A \otimes B)(C \otimes D) = (A C) \otimes (B D)$$
		\item Trace of Kronecker product:
		$$tr\{A \otimes B\} = tr\{A\} tr\{B\}$$
		\item Link of Kronecker product to vectorization operator:
		$$vec(ABC) = (C^T \otimes A) vec(B)$$
		\item Link between vectorization operator and dot product:
		$$vec(A^T) vec(B) = tr\{A^T B\}$$
	\end{enumerate}
	\subsection*{Trace term}
	We start off by rewriting the trace term in terms of $\Sigma$ from Equation \ref{eq:sigma_fim}. First, we use the first property to show that
	$$\Sigma^{-1} = tr\{T\} T^{-1} \otimes L^{-1}$$
	Second, we shall use the product rule for differentiation to obtain an expression for $\frac{\partial \Sigma}{\partial \theta_i}$:
	$$\frac{\partial \Sigma}{\partial \theta_i} = \frac{1}{tr\{T\}}\left( \frac{\partial T}{\partial \theta_i} \otimes L +  T \otimes \frac{\partial L}{\partial \theta_i} -\frac{tr\{\frac{\partial T}{\partial \theta_i}\}}{tr\{T\}} T \otimes L\right)$$
	And similarly for $\frac{\partial \Sigma}{\partial \theta_k}$:
	$$\frac{\partial \Sigma}{\partial \theta_k} = \frac{1}{tr\{T\}}\left( \frac{\partial T}{\partial \theta_k} \otimes L +  T \otimes \frac{\partial L}{\partial \theta_k} -\frac{tr\{\frac{\partial T}{\partial \theta_k}\}}{tr\{T\}} T \otimes L\right)$$
	Now, for brevity, we rename some parts in the trace term as $C_1$ and $C_2$: 
	\begin{equation}\label{eq:traceterm}
		\frac{1}{2}tr\left\{\underbrace{\Sigma^{-1} \frac{\partial{\Sigma}}{\partial \theta_i}}_{C_1} \underbrace{\Sigma^{-1} \frac{\partial{\Sigma}}{\partial \theta_k}}_{C_2}\right\} = \frac{1}{2} tr\{C_1 C_2\}
	\end{equation}
	We expand $C_1$:
	\begin{equation*}
	\begin{split}
		 C_1 = T^{-1} \otimes L^{-1} \left( \frac{\partial T}{\partial \theta_i} \otimes L +  T \otimes \frac{\partial L}{\partial \theta_i} -\frac{tr\{\frac{\partial T}{\partial \theta_i}\}}{tr\{T\}} T \otimes L\right) = \\ 
		 = \underbrace{(T^{-1} \otimes L^{-1}) \left(\frac{\partial T}{\partial \theta_i} \otimes L\right)}_{C_{11}} +  \underbrace{(T^{-1} \otimes L^{-1}) \left(T \otimes \frac{\partial L}{\partial \theta_i}\right)}_{C_{12}} -  \underbrace{\frac{tr\{\frac{\partial T}{\partial \theta_i}\}}{tr\{T\}} (T^{-1} \otimes L^{-1}) (T \otimes L)}_{C_{13}}
	\end{split}
	\end{equation*}
	We will take advantage of the mixed product property 2 to tackle each of the terms individually:
	$$C_{11} = (T^{-1} \otimes L^{-1}) \left(\frac{\partial T}{\partial \theta_i} \otimes L\right) = \left(T^{-1 }\frac{\partial T}{\partial \theta_i} \right) \otimes \left( L^{-1} L \right) =  \left(T^{-1 }\frac{\partial T}{\partial \theta_i} \right) \otimes I_n$$
	
	$$C_{12} =(T^{-1} \otimes L^{-1}) \left(T \otimes \frac{\partial L}{\partial \theta_i}\right)= (T^{-1} T) \otimes \left( L^{-1} \frac{\partial L}{\partial \theta_i}\right) = I_m \otimes \left( L^{-1} \frac{\partial L}{\partial \theta_i}\right) $$
	\begin{equation*}
	\begin{split}
	C_{13} = \frac{tr\{\frac{\partial T}{\partial \theta_i}\}}{tr\{T\}} (T^{-1} \otimes L^{-1}) (T \otimes L) = \frac{tr\{\frac{\partial T}{\partial \theta_i}\}}{tr\{T\}} (T^{-1} T) \otimes (L^{-1}  L) = \\ = \frac{tr\{\frac{\partial T}{\partial \theta_i}\}}{tr\{T\}} I_m \otimes I_n = \frac{tr\{\frac{\partial T}{\partial \theta_i}\}}{tr\{T\}} I_{n m}
	\end{split}
	\end{equation*}
	In the latter case of $C_{13}$, we used the trivial fact that $I_m \otimes I_n = I_{n m}$.
	
	For an extra brevity, we also rename replace some terms with a singular matrices, namely $A_i = T^{-1} \frac{\partial T}{\partial \theta_i}$,$A_k = T^{-1} \frac{\partial T}{\partial \theta_k}$, $B_i =  L^{-1} \frac{\partial L}{\partial \theta_i}$ and  $B_k =  L^{-1} \frac{\partial L}{\partial \theta_k}$. We also denote scalars $\frac{tr\{\frac{\partial T}{\partial \theta_i}\}}{tr\{T\}}$ as $\alpha_i$ and $\frac{tr\{\frac{\partial T}{\partial \theta_k}\}}{tr\{T\}}$ as $\alpha_k$.
	
	So, after the above simplifications, $C_1$ takes the following form:
	
	$$C_1 = A_i \otimes I_n + I_m \otimes B_i - \alpha_i I_{n m}$$
	
	Likewise, we can infer $C_2$:
	
	$$C_2 = A_k \otimes I_n + I_m \otimes B_k - \alpha_k I_{n m}$$
	
	We substitute the obtained results for $C_1, C_2$ to the Equation \ref{eq:traceterm} and expand it:
	\begin{equation*}
	\begin{split}
		\frac{1}{2} tr\{C_1 C_2\} = \frac{1}{2} tr\{\left( A_i \otimes I_n + I_m \otimes B_i - \alpha_i I_{n m}\right) \left( A_k \otimes I_n + I_m \otimes B_k - \alpha_k I_{n m}\right) \} = \\
		= \frac{1}{2}tr\{
		( A_i \otimes I_n)( A_k \otimes I_n)
		+
		( A_i \otimes I_n ) ( I_m \otimes B_k)
		-
		 \alpha_k(A_i \otimes I_n)I_{n m}
		 +
		 \\
		 +
		 (I_m \otimes B_i)( A_k \otimes I_n)
		 +
		 (I_m \otimes B_i)( I_m \otimes B_k)
		 -
		 \alpha_k(I_m \otimes B_i) I_{n m} -
		 \\
		 -
		 \alpha_i I_{n m}( A_k \otimes I_n)
		 -
		 \alpha_i I_{n m}( I_m \otimes B_k)
		 +
		 \alpha_i \alpha_k I_{n m}
		\} = \frac{1}{2}tr\{ \\
		( A_i \otimes I_n)( A_k \otimes I_n)
		+
		( A_i \otimes I_n ) ( I_m \otimes B_k)
		-
		\alpha_k(A_i \otimes I_n)
		+
		\\
		+
		(I_m \otimes B_i)( A_k \otimes I_n)
		+
		(I_m \otimes B_i)( I_m \otimes B_k)
		-
		\alpha_k(I_m \otimes B_i)  -
		\\
		-
		\alpha_i( A_k \otimes I_n)
		-
		\alpha_i ( I_m \otimes B_k)
		+
		\alpha_i \alpha_k I_{n m}
		\} = \frac{1}{2}( \\
		\underbrace{tr\{( A_i \otimes I_n)( A_k \otimes I_n)\}}_{D_1} + \underbrace{tr\{( A_i \otimes I_n ) ( I_m \otimes B_k)\}}_{D_2} - \underbrace{\alpha_k tr\{A_i \otimes I_n\}}_{D_3} + \\
		+
		\underbrace{tr\{(I_m \otimes B_i)( A_k \otimes I_n)\}}_{D_4} 
		+
		\underbrace{tr\{(I_m \otimes B_i)( I_m \otimes B_k)\}}_{D_5}
		-
		\underbrace{\alpha_k tr\{(I_m \otimes B_i)\}}_{D_6}
		-
		\\
		\underbrace{\alpha_i tr\{ A_k \otimes I_n\}}_{D_7}
		-
		\underbrace{\alpha_i tr\{I_m \otimes B_k\}}_{D_8}
		+
		\underbrace{\alpha_i \alpha_k tr\{I_{n m}\}}_{D_9}
		) = \\ = \frac{1}{2}(D_1 + D_2 - D_3 + D_4 + D_5 - D_6 - D_7 - D_8 + D_9)
	\end{split}
	\end{equation*}
Now, we shall deal with each of the $\{D_k\}_{k=1..9}$ terms.

For $D_1$, $D_2$, $D_4$, $D_5$ we use the second property followed by the third property. As for terms $D_3$, $D_6$, $D_7$, $D_8$, just the third property suffices for them, and $D_9$ is trivial.

$$D_1 = tr\{( A_i \otimes I_n)( A_k \otimes I_n)\} = tr\{(A_i A_k) \otimes I_{n}\} = n tr\{A_i A_k\}$$
$$D_2  = tr\{( A_i \otimes I_n ) ( I_m \otimes B_k)\} = tr\{A_i \otimes B_k\} = tr\{A_i\} tr\{B_k\}$$
$$D_3 = \alpha_k tr\{A_i \otimes I_n\} = n \alpha_k tr\{A_i\}$$
$$D_4 = tr\{(I_m \otimes B_i)( A_k \otimes I_n)\} = tr\{A_k \otimes B_i\} = tr\{A_k\} tr\{B_i\}$$
$$D_5 = tr\{(I_m \otimes B_i)( I_m \otimes B_k)\} = m tr\{B_i B_k\}$$
$$D_6 = \alpha_k tr\{I_m \otimes B_i\} = m \alpha_k tr\{B_i\}$$
$$D_7 = \alpha_i tr\{A_k \otimes I_n\}= n \alpha_i tr\{A_k\}$$
$$D_8 = \alpha_i tr\{I_m \otimes B_k\} = m \alpha_i tr\{B_k\}$$
$$D_9 = \alpha_i \alpha_k tr\{I_{nm}\} = n m \alpha_i \alpha_k$$ 

So, the trace term takes the final form of

\begin{equation}\label{eq:tracetermfinal}
\begin{split}
\frac{1}{2}(
n tr\{A_i A_k\} +
m tr\{B_i B_k\} +
tr\{A_i\} tr\{B_k\} +
tr\{A_k\} tr\{B_i\} +\\
+
n m \alpha_i \alpha_k -
n \alpha_k tr\{A_i\} -
m \alpha_k tr\{B_i\} -
n \alpha_i tr\{A_k\} -
m \alpha_i tr\{B_k\}
)
\end{split}
\end{equation}
\subsection*{Quadratic term}
Let's rewrite the quadratic term in $L, T, M$ matrices:
\begin{equation*}
\frac{\partial \mu^T }{\partial \theta_i} \Sigma^{-1} \frac{\partial \mu }{\partial \theta_k} = tr\{T\}\frac{\partial vec(M)^T }{\partial \theta_i} \left(T^{-1} \otimes L^{-1} \right)\frac{\partial vec(M) }{\partial \theta_k}
\end{equation*}
First, we use the fourth property on $\left(T^{-1} \otimes L^{-1} \right)\frac{\partial vec(M) }{\partial \theta_k}$:
$$tr\{T\}\frac{\partial vec(M)^T }{\partial \theta_i} \left(T^{-1} \otimes L^{-1} \right)\frac{\partial vec(M) }{\partial \theta_k} = tr\{T\}\frac{\partial vec(M)^T }{\partial \theta_i}vec\left(L^{-1} \frac{\partial M }{\partial \theta_k} T^{-1}\right)$$
Second, we use the fifth property and obtain the final form for the quadratic term:
\begin{equation}\label{eq:quadraticterm}
	tr\{T\}\frac{\partial vec(M)^T }{\partial \theta_i}vec\left(L^{-1} \frac{\partial M }{\partial \theta_k} T^{-1}\right) = tr\{T\}tr\left\{\frac{\partial M^T }{\partial \theta_i} L^{-1} \frac{\partial M }{\partial \theta_k} T^{-1} \right\}
\end{equation}
\subsection*{Final result}
We combine Equations \ref{eq:tracetermfinal} and \ref{eq:quadraticterm} to obtain the final representation for Fisher Information Matrix:

\begin{equation}\label{eq:fim_final}
	\begin{split}
	I(\theta)_{i,j} = tr\{T\} tr\left\{\frac{\partial M^T }{\partial \theta_i} L^{-1} \frac{\partial M }{\partial \theta_i} T^{-1} \right\} + 
	\frac{1}{2}(
	n tr\{A_i A_k\} +
	m tr\{B_i B_k\} +\\
	+
	tr\{A_i\} tr\{B_k\} +
	tr\{A_k\} tr\{B_i\} +
	n m \alpha_i \alpha_k 
	-
	n \alpha_k tr\{A_i\} - \\
	-
	m \alpha_k tr\{B_i\} -
	n \alpha_i tr\{A_k\} -
	m \alpha_i tr\{B_k\}
	)
	\end{split},
\end{equation}
where $A_i = T^{-1} \frac{\partial T}{\partial \theta_i}$,$A_k = T^{-1} \frac{\partial T}{\partial \theta_k}$, $B_i =  L^{-1} \frac{\partial L}{\partial \theta_i}$,  $B_k =  L^{-1} \frac{\partial L}{\partial \theta_k}$, $\alpha_i = \frac{tr\{\frac{\partial T}{\partial \theta_i}\}}{tr\{T\}}$ and  $\alpha_k = \frac{tr\{\frac{\partial T}{\partial \theta_k}\}}{tr\{T\}}$.

The complexity of computing the Equation \ref{eq:fim_final} is $O(n^3 + m^3 + n m^2 + m n^2)$. The most practically troublesome term of complexity $n^3$ arises from the necessity of computing $B_i, B_k$. This is, however, a big improvement on the original complexity of $O(n^3 m^3)$.
	\newpage
	\section{Generalized latent factor prediction scheme} \label{app:factor_prediction}
	
\makeatletter
\newenvironment{sqcases}{%
	\matrix@check\sqcases\env@sqcases
}{%
	\endarray\right.%
}
\def\env@sqcases{%
	\let\@ifnextchar\new@ifnextchar
	\left\lbrack
	\def\arraystretch{1.2}%
	\array{@{}l@{\quad}l@{}}%
}
\makeatother

We propose a MAP-like (maximum a posteriori) approach for estimating factor scores. The core idea is to express the joint density $f(Z, H)$ of observed variables and latent variables through a product of conditional densitiy $f(Z|H)$ and an unconditional density $f(H)$ of latent variables $H$:
\begin{equation}\label{eq:joint}
	f(Z, H) = f(Z|H) f(H) \leftrightarrow \ln f(z, H) = \ln f(z|H) + \ln f(H),
\end{equation}
and then, maximizing it with respect to $H$. The task of deriving those densities is cumbersome, however. 
\begin{equation*}
	\begin{cases}
		\begin{bmatrix}H \\ X \end{bmatrix} = W = \Gamma_1 G +B W + E \\
		Z = \Gamma_2 G + \Lambda W + \Delta + \sum_{i=1}^p U_i
	\end{cases}
\end{equation*}
$$\Updownarrow$$
\begin{equation*}
	~~~~~~~~~~~~
	\begin{cases}
		H = F_1 C (\Gamma_1 G + E) \\
		X = F_2 C (\Gamma_1 G + E) \\
		Z = \Gamma_2 G + \Lambda_H H + \Lambda_X X + \Delta + \sum_{i=1}^p U_i
	\end{cases},
\end{equation*}

Here, 	$U_i \sim \mathbb{MN}(0, D_i, K_i)$, $F_1 = \begin{pmatrix} I_{n_\eta} & 0_{n_\eta, n_x}\end{pmatrix}$, $F_2 = \begin{pmatrix}0_{n_x, n_\eta} & I_{n_x}\end{pmatrix}$, and $0_{a,b}$ is a zero matrix of shape $a \times b$. 

\subsection*{Derivation of $ln f(z|H)$}

First, we infer $Z$ from the system of equations under the assumption of known $H$ matrix:

$$	\begin{cases}
	X = F_2 C (\Gamma_1 G + E) \\
	Z = \Gamma_2 G + \Lambda_H H + \Lambda_X X + \Delta + \sum_{i=1}^p U_i
\end{cases}$$
$$\Updownarrow$$
$$~~~~~~~~~~~~~~~~~~~~~~~~~~~~~~~~	\begin{cases}
	X = F_2 C (\Gamma_1 G + E) \\
	Z = (\Gamma_2  + \Lambda_X F_2 C \Gamma_1)G + \Lambda_H H + \Lambda_X F_2 C E + \Delta + \sum_{i=1}^p U_i
\end{cases}$$
Expectation of $Z$: $$\EX[Z|H] =(\Gamma_2  + \Lambda_X F_2 C \Gamma_1)G + \Lambda_H H $$ 
Covariance matrix across rows (across variables):
$$L_{Z|H} = n \Lambda_X F_2 C \Psi C^T F_2^T \Lambda_X^T + n \Theta + \sum_{i=1}^p tr\{K_i\} D_i$$
Covariance matrix across columns (across observations):
\begin{equation*}
	\begin{split}
		T_{Z|H}  = \EX[E^T \underbrace{C^T F_2^T \Lambda_X^T \Lambda_X F_2 C}_{A_0} E ] + tr\{\Theta \}I_n + \sum_{i=1}^{p} tr\{D_i\} K_i = \\ =tr\{\Psi A_0\} I_n + tr\{\Theta \}I_n + \sum_{i=1}^{p} tr\{D_i\} K_i
	\end{split}
\end{equation*}

The formula for $\ln f(z|H)$ (constants omitted):
\begin{equation}\label{eq:lnf}
\ln f(z|H) = tr\{L_{Z|H}\} tr\{T_{Z|H}^{-1} (Z - \EX[Z|H])^T L_{Z|H}^{-1}(Z - \EX[Z|H])\}
\end{equation}
We rearrange the term $Z - \EX[Z|H]$ this way:
$$Z - \EX[Z|H] = \underbrace{Z - (\Gamma_2  + \Lambda_X F_2 C \Gamma_1)G}_{M_H} - \Lambda_H H = M_H - \Lambda_H H$$
With this rearrangement in mind, we can now more easily expand the formula \ref{eq:lnf} (it will be helpful later):
\begin{equation}\label{eq:lnf2}
\begin{split}
\ln f(z|H) = tr\{L_{Z|H}\} tr\{T_{Z|H}^{-1} (M_H - \Lambda_H H)^T L_{Z|H}^{-1}(M_H - \Lambda_H H)\} =  \\ = tr\{L_{Z|H}\} tr\{(M_H^T - H^T \Lambda_H^T )^T (L_{Z|H}^{-1}M_H T_{Z|H}^{-1} - L_{Z|H}^{-1} \Lambda_H H T_{Z|H}^{-1}) \} =  \\
= tr\{L_{Z|H}\} tr\{ M_H^T L_{Z|H}^{-1}M_H T_{Z|H}^{-1} - M_H^T L_{Z|H}^{-1} \Lambda_H H T_{Z|H}^{-1} 
-\\- H^T \Lambda_H^T L_{Z|H}^{-1}M_H T_{Z|H}^{-1} + H^T \Lambda_H^T L_{Z|H}^{-1} \Lambda_H H T_{Z|H}^{-1}   \}
\end{split}
\end{equation}
To make Equation \ref{eq:lnf2} look less intimidating, we group some matrices together and substitute them with single matrices:
\begin{equation}\label{eq:lnfsubs}
	\begin{sqcases}
		A = tr\{L_{Z|H}\} \Lambda_H^T L_{Z|H}^{-1}M_H T_{Z|H}^{-1} \\
		A_0 = tr\{L_{Z|H}\}  \Lambda_H^T L_{Z|H}^{-1} \Lambda_H 
	\end{sqcases}
\end{equation}
Then, the Equation \ref{eq:lnf2} transforms to:
 
\begin{equation}\label{eq:lnff}
		\ln f(z|H)
		= tr\{ M_H^T L_{Z|H}^{-1}M_H T_{Z|H}^{-1} - A^T H - H^T A + H^T A_0 H T_{Z|H}^{-1} \}
\end{equation}

\paragraph{Derivation of $\ln f(H)$}
\begin{equation}\label{eq:H}
H = F_1 C \Gamma_1 G + F_1 C E
\end{equation}
From Equation \ref{eq:H}, we can infer expectation of $H$ and both of the covariance matrices. 

Expectation of $H$:
$$\EX[H] =  M = F_1 C \Gamma_1 G $$

Covariance matrix across rows (variables):

$$\widehat{L_{H}} = n F_1 C \Psi C^T F_1^T $$

Covariance matrix across columns (observations):

\begin{equation}\label{eq:TH}
T_{H} = \EX[E^T C^T F_1^T F_1 C E] = tr\{\Psi C^T F_1^T F_1 C\} I_n 
\end{equation}

Remember, that we did make independence assumption when talking about latent variables. Hence, as it can be seen from both model and the Equation \ref{eq:TH}, the matrix-variate distribution here degenerates straightforwardly to a multivariate normal with $\widehat{L_{H}}$ covariance matrix and a zero mean. Constants $n$ and $tr\{\Psi C^T F_1^T F_1 C\}$ will canceled out, therefore, for the sake of simplicity, we shall use normalized by $\frac{1}{n}$ matrix $L_{H}$  instead:
$$L_{H} = \frac{1}{n}\widehat{L_{H}} =  F_1 C \Psi C^T F_1^T $$
Then, the loglikelihood $ln f(H)$ is:
\begin{equation}\label{eq:lnfhpre}
	\begin{split}
	\ln f(H) = tr\{(H - M)^T L_{H}^{-1} (H - M)\} =  tr\{(H^T - M^T)  (L_{H}^{-1} H - L_{H}^{-1} M)\} = \\ = tr\{H^T L_{H}^{-1} H - H^T L_{H}^{-1} M - M^T L_{H}^{-1} H + M^T L_{H}^{-1} M\}
	\end{split}
\end{equation}
Here, we collect only one term into a single matrix, namely, $L_{H}^{-1} M = A_1$:
\begin{equation}\label{eq:lnfh}
		\ln f(H) =tr\{H^T L_{H}^{-1} H - H^T A_1 - A_1^T H + M^T A_1\}
\end{equation}

\subsection*{Estimation of latent factors}
We can hope to get analytical expression for factor scores estimates. For it to happen, solution to the problem of maximizaiton of the Equation \ref{eq:joint} should have closed form, or, equivalently, we should solve the system of equations:
\begin{equation}\label{eq:facestprob}
	\frac{\partial ln f(Z, H)}{\partial H} = \frac{\partial ln f(Z|H)}{\partial H} + \frac{\partial ln f(H)}{\partial H} = 0
\end{equation}

To compute derivatives $\frac{\partial ln f(Z|H)}{\partial H}, \frac{\partial ln f(H)}{\partial H}$, one can refer to matrix calculus handbooks, such Matrix Calculus by G. Golub, or simply to the Matrix Cookbook.

$$\frac{\partial ln f(Z|H)}{\partial H} =2 A_0 H T_{Z|H}^{-1} - 2 A$$

$$\frac{\partial ln f(H)}{\partial H} =2 L_{H}^{-1} H  - 2 A_1$$
So, the Equation \ref{eq:facestprob} becomes:

\begin{equation}\label{eq:factorsys}
	\begin{split}
		2 A_0 H T_{Z|H}^{-1} - 2A + 2 L_{H}^{-1} H  - 2A_1 = 0 \leftrightarrow \\ 
		 A_0 H T_{Z|H}^{-1} + L_{H}^{-1} H = A + A_1
	\end{split}
\end{equation}
Let's multiply left and right sides of the Equation \ref{eq:factorsys} by $A_0^{-1}$:
\begin{equation}\label{eq:factorfinal}
	H \underbrace{T_{Z|H}^{-1}}_{=-A_3} + \underbrace{A_0^{-1} L_{H}^{-1}}_{=A_2} H = A_0^{-1}\left(A + A_1\right) = \widehat{A}
\end{equation}
The Equation \ref{eq:factorfinal} is actually a form of the Sylvester equation:
\begin{equation}\label{eq:sylvester}
	 A_2 H - H A_3 = \widehat{A}
\end{equation}
Sylvester equations, like in \ref{eq:sylvester}, can be solved in $O(n^3 + m^3)$ with Bartels–Stewart algorithm.

If there are no random effects in the model, $T_{Z|H}^{-1}$ is effectively $\frac{1}{tr\{\Sigma\}}I_n$. With that in mind, we can solve Equation~\ref{eq:factorsys} just like we solve regular systems of linear equations:

\begin{equation}\label{eq:factornorf}
 H = \left(\frac{1}{tr\{\Sigma\}} A_0 + L_H^{-1}\right)^{-1} \left(A_1 + A\right)
\end{equation}

	\newpage
	\section{Multivariate BLUP} \label{app:blup}
	
	Assume that we have already estimated model parameters $\theta$ that reside in covariance matrices $L, T, \{D_i\}_{i=1..p}, \{K_i\}_{i=1..p}$ and mean components matrix $M$. We would like to estimate $i$-th random effect. We can deduce it in the same fashion as we did when predicting latent factors, i.e. by using the MAP-estimator.
	\begin{equation}\label{eq:mtblupmap}
		f(Z, U| \theta) = f(Z|U, \theta) f(U|\theta) \leftrightarrow 	\ln f(Z, U| \theta) = \ln f(Z|U, \theta) + \ln f(U|\theta) ,
	\end{equation}
	where 
	
	\begin{equation}\label{eq:mtbluplnf}
	\begin{split}
	\ln f(Z, U| \theta) \propto tr\{T^{-1}(\underbrace{Z - M}_{\widehat{Z}} - U)^T L^{-1}(\underbrace{Z - M}_{\widehat{Z}} - U_i)\} = \\ = tr\{T^{-1}(\widehat{Z} - U_i)^T L^{-1} (\widehat{Z} - U_i)\},
	\end{split}
	\end{equation}
	where $L$ and $T$ are computed without the $i$-th $U_i$.

	$$\ln f(U|\theta) \propto tr\{T_i^{-1}U^T L_i^{-1} U\}, T_i =  tr\{D_i\} K_i,L_i = tr\{K_i\} D_i$$ 
	
	Maximizing Equation \ref{eq:mtblupmap} with respect to $U$ is the same problem as inferring $U$ from $\frac{\partial }{\partial U}\ln f(Z, U|\theta) = 0$. Before computing the derivative, we start off by expanding the Equation \ref{eq:mtbluplnf}:
	\begin{equation*}
	\begin{split}
		\ln f(Z, U| \theta) \propto  tr\{T^{-1}(\widehat{Z} - U_i)^T L^{-1} (\widehat{Z} - U_i)\} = tr\{(\widehat{Z}^T - U_i^T)  (L^{-1} \widehat{Z} T^{-1} -\\- L^{-1} U_i T^{-1})\} = tr\{\widehat{Z}^T L^{-1} \widehat{Z} T^{-1} +  U_{i}^T L^{-1} U_i T^{-1}- 2 U_i^T L^{-1} \widehat{Z} T^{-1} \}
	\end{split}
	\end{equation*}
	Here, we used the trace invariance under transposition and cyclic permutation properties to to combine two negative terms into $2 \widehat{Z}^T L^{-1} U_i T^{-1}$.

	Next, we compute the derivative of the Equation \ref{eq:mtblupmap} and set it to zero:
	
	\begin{equation}\label{eq:mtblupder}
		\begin{split}
			\frac{\partial }{\partial U}\ln f(Z, U|\theta) = 2 L^{-1} U_{i} T^{-1} + 2 L_i^{-1} U_{i} T_i^{-1} - 2 L^{-1} \widehat{Z} T^{-1} = 0 \leftrightarrow \\
			L^{-1} U_i T^{-1} + L_i^{-1} U_i T_i^{-1} = L^{-1}\widehat{Z}T^{-1}
		\end{split}
	\end{equation}
	We multiply both sides of the Equation \ref{eq:mtblupder} by $T_i$ from the right and by $L$ from the left:
	
	\begin{equation}\label{eq:mtblup}
	U_i T^{-1} T_i^{-1} + L L_i^{-1} U_i = \widehat{Z} T^{-1} T_{i}
	\end{equation}
	Again, we've obtained a from of the Sylvester Equation that is solvable by Stewart-Bartels algorithm.
	
	Notice, that the Equation \ref{eq:mtblup} has been obtained for the case of \class{ModelGeneralizedEffects}. It is easy to restrict it to the case of \class{ModelEffects}: the $f(Z|U, \theta)$ will degenerate to multivariate-normal distribution (as $T$ becomes proportional to $I_n$), and the resulting BLUP estimator is
	
	$$	U T^{-1} T_u^{-1} + L L_u^{-1} U_u = \widehat{Z} T_{u}, L_u = tr\{K\}D, T_u = tr\{D\}K $$

	\newpage
	\section{Numerical experiments tables}\label{app:tables}

	\begin{table}[H]
		\centering
		\begin{adjustbox}{max width=\textwidth}
			
			\begin{tabular}{l|cc|cc|cc|cc|cc|cc|cc}
				\toprule
				\texttt{n\_exo} & \multicolumn{2}{c}{2} & \multicolumn{2}{c}{3} & \multicolumn{2}{c}{4} & \multicolumn{2}{c}{5} & \multicolumn{2}{c}{6} & \multicolumn{2}{c}{7} & \multicolumn{2}{c}{8}\\
				\midrule
				{} & N & MAPE & N & MAPE & N & MAPE & N & MAPE & N & MAPE & N & MAPE & N & MAPE\\
				\texttt{Model}[FIML] & 19&13.27 & 12&13.02 & 17&13.10 & 8&12.85 & 13&12.25 & 14&12.18 & 7&12.07\\
				\texttt{Model}[MLW] & 18&13.34 & 10&13.08 & 17&13.23 & 10&12.94 & 14&12.29 & 15&12.30 & 10&12.12\\
				\texttt{ModelMeans}[ML] & 18&13.34 & 10&13.08 & 18&13.23 & 10&12.94 & 14&12.29 & 14&12.30 & 10&12.12\\
				\texttt{ModelMeans}[REML] & 24&16.75 & 21&16.83 & 34&16.89 & 31&17.50 & 31&17.37 & 30&16.91 & 35&17.25\\
				\bottomrule
			\end{tabular}
		\end{adjustbox}
		\caption{N and MAPE for the subset B: varying number of exogenous variables.}
	\end{table}
	
	\begin{table}[H]
		\centering
		\begin{adjustbox}{max width=\textwidth}
			
			\begin{tabular}{l|cc|cc|cc|cc|cc|cc}
				\toprule
				\texttt{n\_endo} & \multicolumn{2}{c}{2} & \multicolumn{2}{c}{3} & \multicolumn{2}{c}{4} & \multicolumn{2}{c}{5} & \multicolumn{2}{c}{6} & \multicolumn{2}{c}{7}\\
				\midrule
				{} & N & MAPE & N & MAPE & N & MAPE & N & MAPE & N & MAPE & N & MAPE\\
				\texttt{Model}[FIML] & 24&13.05 & 17&13.28 & 18&12.87 & 20&12.56 & 17&12.79 & 15&13.28\\
				\texttt{Model}[MLW] & 22&13.14 & 17&13.43 & 18&12.99 & 20&12.68 & 18&12.87 & 17&13.34\\
				\texttt{ModelMeans}[ML] & 22&13.14 & 18&13.43 & 18&12.99 & 20&12.68 & 17&12.87 & 17&13.34\\
				\bottomrule
			\end{tabular}
		\end{adjustbox}
		\caption{N and MAPE for the subset C: varying number of endogenous variables.}
	\end{table}
	
	\begin{table}[H]
		\centering
		\begin{adjustbox}{max width=\textwidth}
			
			\begin{tabular}{l|cc|cc|cc|cc|cc|cc}
				\toprule
				\texttt{n\_lat} & \multicolumn{2}{c}{1} & \multicolumn{2}{c}{2} & \multicolumn{2}{c}{3} & \multicolumn{2}{c}{4} & \multicolumn{2}{c}{5} & \multicolumn{2}{c}{6}\\
				\midrule
				{} & N & MAPE & N & MAPE & N & MAPE & N & MAPE & N & MAPE & N & MAPE\\
				\texttt{Model}[FIML] & 5&12.45 & 17&13.28 & 7&14.20 & 35&14.22 & 47&14.49 & 82&15.45\\
				\texttt{Model}[MLW] & 4&12.58 & 17&13.43 & 11&14.27 & 35&14.27 & 53&14.40 & 83&15.51\\
				\texttt{ModelMeans}[ML] & 4&12.58 & 18&13.43 & 10&14.27 & 33&14.25 & 48&14.42 & 79&15.51\\
				\bottomrule
			\end{tabular}
		\end{adjustbox}
		\caption{N and MAPE for the subset D: varying number of latent variables.}
	\end{table}
	
	\begin{table}[H]
		\centering
		\begin{adjustbox}{max width=\textwidth}
			
			\begin{tabular}{l|cc|cc|cc|cc}
				\toprule
				\texttt{n\_cycles} & \multicolumn{2}{c}{1} & \multicolumn{2}{c}{2} & \multicolumn{2}{c}{3} & \multicolumn{2}{c}{4}\\
				\midrule
				{} & N & MAPE & N & MAPE & N & MAPE & N & MAPE\\
				\texttt{Model}[FIML] & 60&15.09 & 93&17.46 & 148&18.00 & 180&20.10\\
				\texttt{Model}[MLW] & 70&15.39 & 96&17.87 & 161&18.37 & 191&20.67\\
				\texttt{ModelMeans}[ML] & 57&15.30 & 86&17.65 & 147&17.95 & 179&20.49\\
				\bottomrule
			\end{tabular}
		\end{adjustbox}
		\caption{N and MAPE for the subset E: varying number of cycles.}
	\end{table}
	
	\begin{table}[H]
		\centering
		\begin{adjustbox}{max width=\textwidth}
			
			\begin{tabular}{l|cc|cc|cc|cc|cc|cc|cc|cc}
				\toprule
				\texttt{n} & \multicolumn{2}{c}{50} & \multicolumn{2}{c}{100} & \multicolumn{2}{c}{150} & \multicolumn{2}{c}{200} & \multicolumn{2}{c}{250} & \multicolumn{2}{c}{300} & \multicolumn{2}{c}{350} & \multicolumn{2}{c}{400}\\
				\midrule
				{} & N & MAPE & N & MAPE & N & MAPE & N & MAPE & N & MAPE & N & MAPE & N & MAPE & N & MAPE\\
				\texttt{Model}[FIML] & 47&19.15 & 17&13.28 & 5&11.08 & 2&9.25 & 2&8.26 & 3&7.24 & 1&6.59 & 1&6.43\\
				\texttt{Model}[MLW] & 53&19.49 & 17&13.43 & 6&11.10 & 2&9.20 & 3&8.19 & 2&7.18 & 1&6.62 & 1&6.39\\
				\texttt{ModelMeans}[ML] & 53&19.48 & 18&13.43 & 7&11.16 & 2&9.30 & 3&8.30 & 3&7.28 & 1&6.68 & 1&6.44\\
				\bottomrule
			\end{tabular}
		\end{adjustbox}
		\caption{N and MAPE for the subset F: varying number of data samples.}
	\end{table}
	
	\begin{table}[H]
		\centering
		\begin{adjustbox}{max width=\textwidth}
			\begin{tabular}{l|cc|cc|cc|cc|cc}
				\toprule
				\texttt{n\_lat} & \multicolumn{2}{c}{2} & \multicolumn{2}{c}{3} & \multicolumn{2}{c}{4} & \multicolumn{2}{c}{5} & \multicolumn{2}{c}{6}\\
				\midrule
				{} & $\epsilon_\Lambda$ & Time, s & $\epsilon_\Lambda$ & Time, s & $\epsilon_\Lambda$ & Time, s & $\epsilon_\Lambda$ & Time, s & $\epsilon_\Lambda$ & Time, s\\
				$\lambda=0.75$ & 0.08&0.49 & 0.17&1.14 & 0.27&3.02 & 0.43&6.21 & 0.60&14.46\\
				$\lambda=1$ & 0.07&0.41 & 0.17&0.92 & 0.27&2.50 & 0.42&4.96 & 0.57&12.10\\
				$\lambda=1.25$ & 0.07&0.35 & 0.16&0.83 & 0.27&2.09 & 0.40&4.18 & 0.54&10.63\\
				\bottomrule
			\end{tabular}
		\end{adjustbox}
		\caption{$\epsilon_\Lambda$ and working time for the subset A: varying number of latent factors.}
		\label{table:app_efa_lat}
	\end{table}
	
	\begin{table}[H]
		\centering
		\begin{adjustbox}{max width=\textwidth}
			\begin{tabular}{l|cc|cc|cc|cc|cc|cc|cc}
				\toprule
				\texttt{n} & \multicolumn{2}{c}{50} & \multicolumn{2}{c}{100} & \multicolumn{2}{c}{150} & \multicolumn{2}{c}{200} & \multicolumn{2}{c}{250} & \multicolumn{2}{c}{300} & \multicolumn{2}{c}{350}\\
				\midrule
				{} & $\epsilon_\Lambda$ & Time, s & $\epsilon_\Lambda$ & Time, s & $\epsilon_\Lambda$ & Time, s & $\epsilon_\Lambda$ & Time, s & $\epsilon_\Lambda$ & Time, s & $\epsilon_\Lambda$ & Time, s & $\epsilon_\Lambda$ & Time, s\\
				$\lambda=0.75$ & 0.26&0.85 & 0.21&0.99 & 0.19&1.08 & 0.17&1.14 & 0.18&1.37 & 0.18&1.46 & 0.16&1.44\\
				$\lambda=1$ & 0.24&0.74 & 0.20&0.82 & 0.18&0.91 & 0.17&0.96 & 0.18&1.08 & 0.16&1.19 & 0.15&1.18\\
				$\lambda=1.25$ & 0.23&0.57 & 0.18&0.70 & 0.18&0.74 & 0.16&0.78 & 0.18&0.88 & 0.16&0.96 & 0.15&1.01\\
				\bottomrule
			\end{tabular}
		\end{adjustbox}
		\caption{$\epsilon_\Lambda$ and working time for the subset B: varying number of data samples.}
	\end{table}
	
	\begin{table}[H]
		\centering
		\begin{adjustbox}{max width=\textwidth}
			\begin{tabular}{l|cc|cc|cc|cc|cc|cc|cc|cc}
				\toprule
				\texttt{n\_exo} & \multicolumn{2}{c}{5} & \multicolumn{2}{c}{10} & \multicolumn{2}{c}{15} & \multicolumn{2}{c}{20} & \multicolumn{2}{c}{25} & \multicolumn{2}{c}{30} & \multicolumn{2}{c}{35} & \multicolumn{2}{c}{40}\\
				\midrule
				{} & $\epsilon_\Lambda$ & Time, s & $\epsilon_\Lambda$ & Time, s & $\epsilon_\Lambda$ & Time, s & $\epsilon_\Lambda$ & Time, s & $\epsilon_\Lambda$ & Time, s & $\epsilon_\Lambda$ & Time, s & $\epsilon_\Lambda$ & Time, s & $\epsilon_\Lambda$ & Time, s\\
				$\lambda=0.75$ & 0.44&0.73 & 0.26&0.96 & 0.20&1.27 & 0.17&1.17 & 0.16&1.40 & 0.13&1.52 & 0.14&1.52 & 0.15&1.70\\
				$\lambda=1$ & 0.42&0.63 & 0.25&0.75 & 0.19&0.96 & 0.17&0.90 & 0.15&1.11 & 0.12&1.13 & 0.14&1.23 & 0.14&1.45\\
				$\lambda=1.25$ & 0.41&0.51 & 0.23&0.64 & 0.18&0.80 & 0.16&0.82 & 0.15&0.93 & 0.12&1.01 & 0.14&0.97 & 0.14&1.17\\
				\bottomrule
			\end{tabular}
		\end{adjustbox}
		\caption{$\epsilon_\Lambda$ and working time for the subset C: varying number of exogenous variables.}
	\end{table}
	
	\begin{table}[H]
		\centering
		\begin{adjustbox}{width=\textwidth}
			\begin{tabular}{l|cc|cc|cc|cc|cc|cc|cc|cc}
				\toprule
				$k$ & \multicolumn{2}{c}{0.5} & \multicolumn{2}{c}{1.0} & \multicolumn{2}{c}{2.0} & \multicolumn{2}{c}{4.0} & \multicolumn{2}{c}{8.0} & \multicolumn{2}{c}{16.0} & \multicolumn{2}{c}{32.0} & \multicolumn{2}{c}{64.0}\\
				\midrule
				{} & $\epsilon_\Lambda$ & Time, s & $\epsilon_\Lambda$ & Time, s & $\epsilon_\Lambda$ & Time, s & $\epsilon_\Lambda$ & Time, s & $\epsilon_\Lambda$ & Time, s & $\epsilon_\Lambda$ & Time, s & $\epsilon_\Lambda$ & Time, s & $\epsilon_\Lambda$ & Time, s\\
				$\lambda=1.25$ & 1.10&0.19 & 0.78&0.27 & 0.48&0.43 & 0.28&0.57 & 0.19&0.85 & 0.15&1.04 & 0.15&1.54 & 0.17&1.95\\
				\bottomrule
			\end{tabular}
		\end{adjustbox}
		\caption{$\epsilon_\Lambda$ and working time for the subset D: varying $k$ multiplier.}
		\label{table:app_efa_k}
	\end{table}
	
	In Tables~\ref{table:app_efa_lat}-\ref{table:app_efa_k}, $\lambda$ is a value of regularization constant for SPCA.
	
\end{appendix}
\end{document}